\documentclass[10pt,a4paper]{article}
\usepackage[english]{babel}
\usepackage[latin1]{inputenc}
\usepackage{amsfonts,amsbsy,bm,euscript,mathrsfs}
\usepackage{amssymb,stmaryrd,faktor}
\usepackage[tbtags]{amsmath}
\usepackage{bbm}
\usepackage{graphicx}
\usepackage[title,titletoc]{appendix}
\usepackage[bookmarks=true,colorlinks=true,linkcolor=blue,citecolor=blue,urlcolor=blue,bookmarksnumbered]{hyperref}

\usepackage[dvipsnames]{xcolor}
\usepackage{soul}

\usepackage{amsfonts}
\usepackage{dsfont}
\usepackage{collref}
\usepackage{lmodern}
\usepackage{mathrsfs}
\usepackage{mathtools}
\usepackage{bbm}
\usepackage{braket}
\usepackage{slashed}
\usepackage{float}
\usepackage{graphicx}
\usepackage{booktabs}
\usepackage{subfig}
\usepackage{tikz}
\usepackage[symbol]{footmisc}
\usepackage{hyperref}
\usepackage{cleveref}
\usetikzlibrary{plotmarks,calc,decorations,decorations.pathmorphing}

\usepackage{amsmath}
\usepackage{mathtools}

\makeatletter
\def\tagform@#1{\maketag@@@{\normalsize(#1)\@@italiccorr}}
\makeatother

\numberwithin{equation}{section}

\makeatletter
\renewcommand\section{\@startsection {section}{1}{\z@}
{-3.5ex \@plus -1ex \@minus -.2ex}
{2.3ex \@plus.2ex}
{\normalfont\Large\bfseries}}
\renewcommand\subsection{\@startsection{subsection}{2}{\z@}
{-3.25ex\@plus -1ex \@minus -.2ex}
{1.5ex \@plus.2ex}
{\normalfont\large\bfseries}}
\makeatother

\newcommand{\arXivlink}[1]{\href{http://arXiv.org/abs/#1}{arXiv:#1}}

\newcommand{\alg}[1]{\mathfrak{#1}}

\usepackage{blkarray}

\begin{document}

\thispagestyle{empty}
\begin{flushright}\footnotesize\ttfamily
DMUS-MP-25/01
\end{flushright}
\vspace{2em}

\begin{center}

{\Large\bf \vspace{0.2cm}
{\color{black} \large Boundary Bethe ansatz in massive $AdS_3$}} 

\vspace{4em}

\textrm{Daniele Bielli$^{a,b}$, Vasileios Moustakis$^{c}$ and Alessandro Torrielli$^{c}$ \footnote[1]{\textit{E-mail:} \texttt{d.bielli4@gmail.com, \{v.moustakis@,  a.torrielli@\}surrey.ac.uk}}
\\}

\vspace{2em}

${}^{a}$
{\it 
High Energy Physics Research Unit, Faculty of Science 
\\ 
Chulalongkorn University, Bangkok 10330, Thailand
}
\\~\\
${}^{b}$
{\it 
National Astronomical Research Institute of Thailand 
\\ 
Don Kaeo, Mae Rim, Chiang Mai 50180, Thailand
}
\\~\\
${}^{c}$ \textit{Department of Mathematics, University of Surrey        
\\
Guildford, GU2 7XH, UK}
\end{center}

\vspace{2em}

\begin{abstract}\noindent 

\noindent
In this paper we perform the boundary algebraic Bethe Ansatz for massive representations of the $AdS_3 \times S^3 \times T^4$ integrable system. This is a companion analysis to our study of massless representations \cite{Bielli:2024bve}. Our treatment is comprehensive of all possible assortments of tensor-factor polarisations which build the physical representations in the spectrum, and includes different choices of auxiliary spaces, revealing subtle differences in the procedure. We survey both singlet and vector boundaries, obtaining the auxiliary Bethe equations in very general form for all cases.  
\end{abstract}

\newpage

\overfullrule=0pt
\parskip=2pt
\parindent=12pt
\headheight=0.0in \headsep=0.0in \topmargin=0.0in \oddsidemargin=0in

\vspace{-3cm}
\thispagestyle{empty}
\vspace{-1cm}

\tableofcontents

\setcounter{footnote}{0}

\section{Introduction}

We are concerned here with the integrable model underlying the gauge-fixed string theory on $AdS_3 \times S^3 \times T^4$ \cite{Bogdan} (very nice reviews are contained in  \cite{rev3,Borsato:2016hud,Seibold:2024qkh}). The scope of this study is to test our comprehension of this unconventional integrable system, which is significantly extending the usual array of phenomena associated with AdS/CFT integrability \cite{Beisertreview}. The production associated with this sub-field is extensive \cite{OhlssonSax:2011ms,seealso3,Borsato:2012ss,Borsato:2013qpa,Borsato:2014hja,Borsato:2013hoa} and one of its novel aspects, when compared with the higher-dimensional $AdS$ counterparts, is the presence of massless modes \cite{Sax:2012jv}. A full account of these modes and their $S$-matrices \cite{Zamol2,Fendley:1993jh,DiegoBogdanAle}, see also \cite{Lloyd:2013wza} and \cite{Ben}, is a subject of continued investigation, together with their massive counterparts, which despite being better understood still reserve surprises and are themselves the object of intense study. Quantum Spectral Curve (QSC) methods have been adapted in \cite{QSC}, see also \cite{Cavaglia:2022xld}, and a full understanding of the massless sector from this perspective is emerging \cite{Ekhammar:2024kzp}. At the same time the work of \cite{AleSSergey}, see also \cite{Seibold:2022mgg}, have reconsidered the  $AdS_3 \times S^3 \times T^4$ integrable structure through the traditional sequence of steps ($S$-matrices, dressing factors, mirror TBA), and a larger and larger analysis with important input from the worldsheet \cite{Fabri:2025rok} is becoming available, showing that the picture is still dynamical and promises new interesting developments \cite{recent,riab,gaber2,frolo,sei2}.

The study of boundaries which preserve the integrable structure of a given theory \cite{GhoshalZamo} represents another established line of research, within the vast integrability literature. In \cite{Bielli:2024xuv} we have solved the boundary scattering problem for the massless modes of $AdS_3 \times S^3 \times T^4$. Together with the massive analysis which had been done in  \cite{Prinsloo:2015apa}, this completed the list of reflection matrices for this model. While exploring the massless reflection matrices, we found a wealth of new aspects which distinguishes the massless representations, particularly exploiting the difference-form variables \cite{gamma1,gamma2} which rewrite the scattering in a relativistic fashion \cite{DiegoBogdanAle,AleSSergey}. Thanks to the acquired simplicity of the difference form we could find general solutions to the boundary Yang-Baxter equation in the case of singlet boundaries. It was also especially crucial to formulate a consistent set of boundary co-ideal (sub)algebra conditions, in order to identify the appropriate symmetry relations to fix the vector-boundaries reflection matrices. The analogy of a vector boundary, with a magnon sealed at the boundary with momentum $p=\pi$, was corroborated as a result of this analysis.

In the subsequent work \cite{Bielli:2024bve} we then constructed the boundary algebraic Bethe ansatz  \cite{Bedu,Bracken:1997zww,AppelVlaar,Angela,Foerster:1993fp,Slav,Ge,Gonzalez-Ruiz:1994loj,Grab,Links,Gould,Martins:1999jbx,Shiro,Yao,Zhou,Zhou:1998joy}, restricting our analysis to singlet representations of type $\text{L}$  \cite{Borsato:2014hja} for definiteness. Inspired by the traditional treatment of \cite{Sklyanin:1988yz} we formulated the double-row monodromy matrix, the associated transfer matrix, the creation and annihilation operators and the magnon excitations built upon the pseudovacuum, obtaining the explicit RTRT algebra. Obtaining the {\it dual} equation of Sklyanin's  \cite{Sklyanin:1988yz}, in the form appropriate for our situation, was an essential step of the construction, which allowed us to demonstrate the working of the procedure adapted to our theory.  

Having proven that the procedure can work efficiently in the massless case, where it is made analytically easier by the difference form, the most urgent enterprise at this point is to complete the task of constructing the algebraic Bethe ansatz for massive representations as well, in all the variants ($\text{L}$, $\text{R}$, and tilded versions) and with boundaries in both the singlet and vector representation. This poses a series of challenges given that the expressions are instantly more complicated. In this paper we show that the procedure applies in much the same way as in the massless case, and exhibits compact exact expressions for the massive representations which allow to study every possible combination of excitations. By expressing our construction in terms of certain universal quantities, we have obtained a collective way of writing the Bethe equations, and all the other accessories of the process, in a completely general and very agile fashion. We can therefore claim to have deduced all the auxiliary (and, by extension) the non-auxiliary, Bethe equations - the next step for future work will naturally be to try to solve these equations, at least in some simple case.  The paper is structured as follows. In section \ref{sec:background} we summarise the tools at our disposal, namely the bulk and boundary representations, all the relevant bulk $R$-matrices and all the relevant boundary reflection matrices, listed with full detail for ease of reading and in order to make this paper self-contained. Section \ref{sec:Algebraic-Bethe-Ansatz} represents the main body of this work: after reviewing the dual equation in subsection \ref{subsec:dual-equation}, we construct the algebraic Bethe ansatz for singlet boundaries in subsection \ref{subsec:singlet-boundary} and conclude with vector boundaries in subsection \ref{subsec:vector-boundary}. For both singlet and vector boundaries the analysis is presented with increasing degree of complexity and for the singlet case it starts in subsection \ref{subsusbsec:singlet-all-L-reps} by displaying a complete run of the algebraic Bethe ansatz for $\text{L}$ representations only, in both the auxiliary and physical spaces. This is done with the purpose of exemplifying the method once extensively and sort out all the necessary steps, particularly when it comes to using the appropriate dual equation. In subsection \ref{subsubsec:singlet-auxiliary-L-mixed-physical-reps} the possibility of mixed $\text{L}$ and $\text{R}$ representation in the physical spaces is concisely included, with auxiliary spaces restricted to $\text{L}$. Extension of the formalism to vector boundaries starts in subsection \ref{subsubsec:vector-all-L-reps}, where all representations are for simplicity constrained once again to the $\text{L}$ sector, so as to better highlight the new features introduced by non-trivial boundaries in a concise manner. Subsection \ref{subsubsec:vector-all-mixed-reps} concludes the discussion, presenting the most general setting of a vector boundary and all possible representations allowed in either the auxiliary, physical and boundary spaces.  We finally provide some conclusions, and a conspicuous list of appendices containing the technical proofs. Amongst other things, in appendix \ref{app:6vandfake-6v} we analyse changes in the auxiliary space representations, to answer the question of what the optimal choice of auxiliary particles is to best facilitate the application of the algebraic Bethe ansatz. We find that the standard form of the exchange relations for the $\mathcal{A}$, $\mathcal{B}$, $\mathcal{C}$, $\mathcal{D}$ operators which characterise the procedure can only be derived when the two auxiliary spaces $0$ and $0'$ carry a specific subset of representations, which turn out to be associated with $R$-matrices which exhibit a genuine 6-vertex structure, as opposed to what we refer to as fake 6-vertex. This interesting pattern only affects the derivation of the exchange relations and not the applicability of the Bethe ansatz, which relies on a single auxiliary space and works equally well for any choice of auxiliary representation.

\section{Background}\label{sec:background}

We begin by summarising those definitions and results, about the pure Ramond-Ramond $AdS_{3}$ theory, which will be relevant to the rest of the paper. We will mostly adopt the conventions in \cite{Prinsloo:2015apa}, sometimes also borrowing notation from the original papers \cite{Borsato:2014hja}, where the full off-shell symmetry algebra $\mathcal{A}_{\text{full}}$ of supercharges commuting with the gauge-fixed Hamiltonian of the theory was constructed.

\subsection{Bulk theory}\label{sec:bulk-theory}
\subsubsection*{\ul{\it Symmetries and Physical Representations}}
The full symmetry algebra $\mathcal{A}_{\text{full}}$ of the worldsheet theory was found to be \cite{Borsato:2014hja}
\begin{equation}
\mathcal{A}_{\text{full}} = \mathfrak{psu}(1|1)_{\text{c.e.}}^{4} \oplus \mathfrak{so}(4)_{2} \,\, ,
\end{equation}
with the $\mathfrak{so}(4)_{2}$ part coming from the torus directions and the subsctipt $\text{c.e.}$ denoting central extension of $\mathfrak{psu}(1|1)^4$. An important property of $\mathcal{A}_{\text{full}}$ is that its representations can be constructed as tensor products of fundamental representations of the smaller algebra $\mathcal{A}=\mathfrak{su}(1|1)^2_{\text{c.e.}}$ \cite{seealso3,rev3}, which can be written as \cite{Prinsloo:2015apa}
\begin{equation}\label{algebra}
\{\mathfrak{Q}_L,\mathfrak{G}_L\}=\mathfrak{H}_L \,\, ,
\qquad \qquad 
\{\mathfrak{Q}_R,\mathfrak{G}_R\}=\mathfrak{H}_R \,\, ,
\qquad \qquad 
\{\mathfrak{Q}_L,\mathfrak{Q}_R\}=\mathfrak{P} \,\, ,
\qquad \qquad
\{\mathfrak{G}_L,\mathfrak{G}_R\}=\mathfrak{K} \,\, ,
\end{equation}
with $\mathfrak{H}_{L}, \mathfrak{H}_{R},\mathfrak{P}, \mathfrak{K}$ denoting central elements. In analogy with \cite{Prinsloo:2015apa,Bielli:2024xuv}, it will be sufficient for our purposes to ignore the $\mathfrak{so}(4)_{2}$ sector and restrict our analysis to $\mathcal{A}$. In particular, it will be sufficient to consider four inequivalent representations of $\mathcal{A}$, denoted by $\pi_{p}^{\text{L}}, \pi_{p}^{\text{R}}, \pi_{p}^{\tilde{\text{L}}}, \pi_{p}^{\tilde{\text{R}}}$ in agreement with \cite{Bielli:2024xuv} and in a mixture of notations from \cite{Borsato:2014hja} and \cite{Prinsloo:2015apa}. These representations are all based on a supersymmetric doublet involving bosons $|\phi^{a}_{p}\rangle $ and fermions $|\psi^{a}_{p} \rangle$, with $a \in \{ \text{L},\text{R},\tilde{\text{L}},\tilde{\text{R}}  \}$ and $p$ denoting the physical momentum of the excitations. The action of the generators on such a doublet can be described \cite{Prinsloo:2015apa} in terms of two-by-two supermatrices $\mathds{E}_{\alpha\beta} \in \text{End}(\mathds{C}^{1|1})$, containing 0 in all entries and 1 in the row $\alpha$, column $\beta$, with $\alpha,\beta\in \{1,2\}$. The physical representations of $\mathcal{A}_{\text{full}}$ can then be constructed \cite{Borsato:2014hja} by taking appropriate tensor products of these four building blocks.
There are two 4-dimensional massive representations and one 8-dimensional massless representation:
\begin{itemize}
\item The massive \textit{left} representation is given by the tensor product $\pi^{\text{L}}\otimes \pi^{\text{L}}$ and exhibits mass $m\!=\!1$.
\item The massive \textit{right} representation is given by the tensor product $\pi^{\text{R}}\otimes \pi^{\text{R}}$ and exhibits mass $m\!=\!-1$.
\item The massless representation exhibits mass $m=0$ and can be constructed in four different ways 
\begin{equation}
(\pi_{p}^{\text{L}} \otimes \pi_{p}^{\tilde{\text{L}}}) \oplus (\pi_{p}^{\text{L}} \otimes \pi_{p}^{\tilde{\text{L}}}) 
\,\,\, , \,\,\, 
(\pi_{p}^{\text{R}}\otimes \pi_{p}^{\tilde{\text{R}}})\oplus (\pi_{p}^{\text{R}}\otimes  \pi_{p}^{\tilde{\text{R}}}) 
\,\,\, , \,\,\,
(\pi_{p}^{\text{L}}\otimes  \pi_{p}^{\tilde{\text{L}}})\oplus (\pi_{p}^{\text{R}}\otimes  \pi_{p}^{\tilde{\text{R}}}) 
\,\,\, , \,\,\, 
(\pi_{p}^{\text{R}}\otimes  \pi_{p}^{\tilde{\text{R}}})\oplus (\pi_{p}^{\text{L}}\otimes  \pi_{p}^{\tilde{\text{L}}}) \,\, ,
\end{equation}
which turn out to be all isomorphic up to redefinitions.
\end{itemize}
All the relevant physical excitations of the model might hence be taken into account by considering the massive and massless $\text{L},\text{R}$ building blocks, together with the massless $\tilde{\text{L}}$ (or $\tilde{\text{R}}$) only. We will however proceed in a slightly different way, analysing for completeness each of the representations $\text{L},\text{R},\tilde{\text{L}},\tilde{\text{R}}$ and treating all of them as massive. Considering all possible types of interactions among these four massive building blocks will effectively lead us to study unphysical settings, since $\tilde{\text{L}},\tilde{\text{R}}$ only play a physical role in the massless case. However, the benefit of including fictitious massive $\tilde{\text{L}},\tilde{\text{R}}$ representations in our analysis is that expressing quantities associated to the boundary turns out to be somewhat less explicit but more compact, while the correct massless expressions can still be recovered by carefully taking the appropriate massless limit, as shown in \cite{Bielli:2024xuv}. This procedure does hide the difficulties arising in the massless sector due to the extra splitting of the four building block representations into further two representations, respectively associated with left and right movers, which necessarily appear in the theory when taking into account the presence of boundaries. It also masks the  difference form of the massless sector. However, since we have worked out those subtleties separately, we can allow ourselves to use this shorthand here without the fear of losing sight of the important physical effects.

\subsubsection*{\ul{\it Single-Particle and Two-Particle Representations}}
Among the four single particle representations introduced above, a special role is played by $\text{L}$ and $\text{R}$
\begin{equation}\label{L+R-rep}
\begin{aligned}
\pi^{\text{L}}_{p}(\mathfrak{Q}_{L}) &= a_{p} \, \mathbb{E}_{21} 
\qquad \quad \,\,\,\,\, 
\pi^{\text{L}}_{p}(\mathfrak{G}_{L}) = b_{p} \, \mathbb{E}_{12}
\qquad \qquad 
\pi^{\text{L}}_{p}(\mathfrak{Q}_{R}) = c_{p} \,\mathbb{E}_{12} 
\qquad \qquad
\pi^{\text{L}}_{p}(\mathfrak{G}_{R}) = d_{p} \, \mathbb{E}_{21}
\\
\pi^{\text{R}}_{p}(\mathfrak{Q}_{L}) &= c_{p} \, \mathbb{E}_{12} 
\qquad \quad\,\,\,\,\,
\pi^{\text{R}}_{p}(\mathfrak{G}_{L}) = d_{p} \, \mathbb{E}_{21}
\qquad \quad \,\,\,\,\,
\pi^{\text{R}}_{p}(\mathfrak{Q}_{R}) = a_{p} \, \mathbb{E}_{21} 
\qquad \quad\,\,\,\,\,
\pi^{\text{R}}_{p}(\mathfrak{G}_{R}) = b_{p} \, \mathbb{E}_{12} \, ,
\end{aligned}
\end{equation}
since $\tilde{\text{L}}$ and $\tilde{\text{R}}$ respectively exhibit the same coefficients as $\text{L}$ and $\text{R}$, while their matrix structure can be obtained by changing the nature of the highest weight state from bosonic to fermionic, which amounts to letting $\mathbb{E}_{\alpha\beta} \rightarrow \mathbb{E}_{\beta\alpha}$ in \eqref{L+R-rep}. The above coefficients are defined as
\begin{equation}\label{reps-coefficients}
a_{p} = b_{p} 
 = \sqrt{h} \, \eta_{p} 
\qquad \qquad 
c_{p}=-\frac{i\sqrt{h} \, \eta_{p}}{x_{p}^{-}}
\qquad \qquad 
d_{p} = \frac{i\sqrt{h} \,\eta_{p}}{x_{p}^{+}} \,\, ,
\end{equation}
which are in turn defined via
\begin{equation}\label{parameter-eta}
\eta_{p}^2 = i(x_{p}^{-}-x_{p}^{+}) \,\, 
\end{equation}
and the Zhukovski variables $x_{p}^{\pm}$. These must satisfy the following relations
\begin{equation}\label{relations-Zhukovski-variables}
\frac{x_{p}^{+}}{x_{p}^{-}}=e^{ip} =u_{p}^{2}
\qquad \qquad \text{and} \qquad \qquad 
\Biggl(x_{p}^{+}+\frac{1}{x_{p}^{+}} \Biggr)-\Biggl(x_{p}^{-}+\frac{1}{x_{p}^{-}} \Biggr)=\frac{im}{h} \,\, ,
\end{equation}
with $m \!\in\! \{ -1,0,1 \}$ the mass parameter and $h$ the coupling constant of the theory. A solution to \eqref{relations-Zhukovski-variables} is 
\begin{equation}\label{Zhukovsky-variables}
x_{p}^{\pm} = \frac{m+ \sqrt{m^2 + 16h^2 \sin^2(\tfrac{p}{2})}}{4h \sin(\tfrac{p}{2})} \, e^{\pm i \frac{p}{2}} \,\, ,
\end{equation}
but this will actually never be exploited in our analysis, which solely relies on the defining relations \eqref{relations-Zhukovski-variables}.
For the two-particle case, we will consider the coproducts defined in \cite{Prinsloo:2015apa}
\begin{equation}\label{coproducts}
\begin{aligned}
\Delta(\mathfrak{Q}_{\text{a}}) &= \mathfrak{Q}_{\text{a}} \otimes \mathds{1} + \mathfrak{U} \otimes \mathfrak{Q}_{\text{a}}
\qquad \quad 
\Delta(\mathfrak{G}_{\text{a}})= \mathfrak{G}_{\text{a}} \otimes \mathds{1} + \mathfrak{U}^{-1} \otimes \mathfrak{G}_{\text{a}}
\qquad \,\,
\Delta(\mathfrak{U}^{\alpha})= \mathfrak{U}^{\alpha} \otimes \mathfrak{U}^{\alpha}
\\
\Delta(\mathfrak{H}_{\text{a}}) &= \mathfrak{H}_{\text{a}} \otimes \mathds{1} + \mathds{1} \otimes \mathfrak{H}_{\text{a}}
\qquad \quad\,\,\,\, 
\Delta(\mathfrak{P}) = \mathfrak{P} \otimes \mathds{1} + \mathfrak{U}^{+2} \otimes \mathfrak{P}
\qquad \quad\,\,\,\,
\Delta(\mathfrak{K}) = \mathfrak{K} \!\otimes\! \mathds{1} + \mathfrak{U}^{-2} \otimes \mathfrak{K} \,\, ,
\end{aligned}
\end{equation}
where $\text{a}\in \{\text{L},\text{R}\}$, $\alpha\in \mathds{R}$ and $\mathfrak{U}$ is a central group-like generator acting as $\mathfrak{U}|\varphi^{a}_{p}\rangle = e^{i\frac{p}{2}} |\varphi^{a}_{p}\rangle $ on each excitation $\varphi \in \{ \phi, \psi \}$, for $a \in \{ \text{L},\text{R},\tilde{\text{L}},\tilde{\text{R}} \}$. For each of the above abstract generators there will also be an associated representation, denoted by $(\pi_{p}^{a}\!\otimes\!\pi_{q}^{b})\Delta(\mathfrak{a})$, for any $\mathfrak{a}\in \mathcal{A}$.

\subsubsection*{\ul{\it R-Matrices}}
Exploiting the relation $S = \Pi \circ R $, where $\Pi$ is the graded permutation operator on states $\Pi (|\alpha\rangle \otimes |\beta\rangle) = (-1)^{|\alpha||\beta|}(|\beta\rangle \otimes |\alpha\rangle)$ for $\alpha \in \{ 1,2 \}$ and $|1\rangle \equiv |\phi\rangle, |2\rangle \equiv |\psi\rangle$ respectively bosonic and fermionic excitations of any representation with gradings $|1|=0$ and $|2|=1$, one can easily extract the $R$-matrices for all types of interactions from \cite{Borsato:2014hja}. The final result exhibits of course the same structure as in \cite{Prinsloo:2015apa}, but we shall proceed exploiting a mixed notation where each matrix entry is named after \cite{Borsato:2014hja} while being explicitly defined as in \cite{Prinsloo:2015apa}. To account for the interaction of excitations in any of the four building-block representations $\text{L},\text{R},\tilde{\text{L}},\tilde{\text{R}}$, it will also be useful to introduce a restricted version of the label $a \in \{ \text{L},\text{R},\tilde{\text{L}},\tilde{\text{R}}  \}$, namely $\text{a}\in \{\text{L},\text{R}\}$, which can be conveniently dressed as $\tilde{\text{a}}\in \{ \tilde{\text{L}},\tilde{\text{R}} \}$.

From the $S$-matrices for the scattering of $\text{L}\text{L}$ and $\text{R}\text{R}$ excitations one obtains the following $R$-matrices
\begin{equation}\label{LL+RR-Rmatrices}
\begin{aligned}
R^{\text{aa}}(p,q)&=A_{pq}^{\text{aa}}\, \mathbb{E}_{11}\otimes \mathbb{E}_{11}+B_{pq}^{\text{aa}}\, \mathbb{E}_{11}\otimes \mathbb{E}_{22}+D_{pq}^{\text{aa}}\, \mathbb{E}_{22}\otimes \mathbb{E}_{11}-F_{pq}^{\text{aa}}\, \mathbb{E}_{22}\otimes \mathbb{E}_{22}
\\
& \qquad \qquad \qquad\quad \,\, -E_{pq}^{\text{aa}}\, \mathbb{E}_{12}\otimes \mathbb{E}_{21}+C_{pq}^{\text{aa}}\, \mathbb{E}_{21}\otimes \mathbb{E}_{12}
\end{aligned}
\end{equation}
with the entries defined, for both for a=L and a=R, as
\begin{equation}\label{LL+RR-functions}
\begin{aligned}
A^{\text{a}\text{a}}_{pq}&=1 
\qquad \qquad \qquad \qquad \qquad \quad \,\, 
B^{\text{a}\text{a}}_{pq}= \frac{x_{p}^{+}-x_{q}^{+}}{ u_{p} (x_{p}^{-}-x_{q}^{+}) }
\qquad \qquad \,
C^{\text{a}\text{a}}_{pq}= \frac{-iu_{q}\eta_{p}\eta_{q}}{u_{p}(x_{p}^{-}-x_{q}^{+})} 
\\
D^{\text{a}\text{a}}_{pq} &=  \frac{u_{q}( x_{p}^{-}-x_{q}^{-})}{x_{p}^{-}-x_{q}^{+} }
\qquad \qquad \qquad
E^{\text{a}\text{a}}_{pq}= \frac{-i\eta_{p}\eta_{q}}{x_{p}^{-}-x_{q}^{+}} 
\qquad \qquad \qquad
F^{\text{a}\text{a}}_{pq}= -\frac{u_{q}(x_{p}^{+}-x_{q}^{-})}{u_{p}(x_{p}^{-}-x_{q}^{+})} \,\, .
\end{aligned}
\end{equation}
From the $S$-matrix for $\text{L}\text{R}$ scattering one obtains
\begin{equation}\label{LR-Rmatrix}
\begin{aligned}
R^{\text{LR}}(p,q)&=A_{pq}^{\text{LR}}\, \mathbb{E}_{11}\otimes \mathbb{E}_{11}+C_{pq}^{\text{LR}}\, \mathbb{E}_{11}\otimes \mathbb{E}_{22}+D_{pq}^{\text{LR}}\, \mathbb{E}_{22}\otimes \mathbb{E}_{11}-E_{pq}^{\text{LR}}\, \mathbb{E}_{22}\otimes \mathbb{E}_{22}
\\
& \qquad \qquad \qquad \quad \,\,\,\, -F_{pq}^{\text{LR}}\, \mathbb{E}_{12}\otimes \mathbb{E}_{12}-B_{pq}^{\text{LR}}\, \mathbb{E}_{21}\otimes \mathbb{E}_{21}  
\end{aligned}
\end{equation}
with the entries defined as
\begin{equation}\label{LR-functions}
\begin{aligned}
A^{\text{L}\text{R}}_{pq} &=  x_{p}^{+}x_{q}^{-} -1
\qquad  \qquad \qquad \,
B^{\text{L}\text{R}}_{pq}=  - \frac{\eta_{p}\eta_{q}}{u_{q}}
\qquad \qquad \qquad \qquad \qquad \,
C^{\text{L}\text{R}}_{pq}=  u_{p}(x_{p}^{-}x_{q}^{-} - 1)
\\ 
D^{\text{L}\text{R}}_{pq} &=  \frac{x_{p}^{+}x_{q}^{+} - 1}{u_{q}}
\qquad \qquad \qquad
E^{\text{L}\text{R}}_{pq}=  - \frac{u_{p}(x_{p}^{-}x_{q}^{+} - 1)}{u_{q}}
\qquad \qquad \qquad 
F^{\text{L}\text{R}}_{pq} =  -u_{p}\eta_{p}\eta_{q} \,\, .
\end{aligned}
\end{equation}
The $R$-matrix $R^{\text{R}\text{L}}(p,q)$ has the structure of $R^{\text{L}\text{R}}(p,q)$ in \eqref{LR-Rmatrix}, with the new coefficients $A^{\text{R}\text{L}}_{pq},...,F^{\text{R}\text{L}}_{pq}$ respectively taking the same form as $A^{\text{L}\text{R}}_{pq},...,F^{\text{L}\text{R}}_{pq}$ in \eqref{LR-functions}. The $R$-matrices associated to interactions involving $\tilde{\text{L}}$ and $\tilde{\text{R}}$ excitations can then be constructed following \cite{Borsato:2014hja}: these are defined via the functions appearing in LL, RR, RL and LR, by permutation of the entries.
In particular, for $\tilde{\text{L}}\tilde{\text{L}}$ and $\tilde{\text{R}}\tilde{\text{R}}$ one has
\begin{equation}\label{LtildeLtilde+RtildeRtilde-R-matrices}
\begin{aligned}
R^{\tilde{\text{a}}\tilde{\text{a}}}(p,q)&=-F_{pq}^{\text{aa}}\, \mathbb{E}_{11}\otimes \mathbb{E}_{11}+D_{pq}^{\text{aa}}\, \mathbb{E}_{11}\otimes \mathbb{E}_{22}+B_{pq}^{\text{aa}}\, \mathbb{E}_{22}\otimes \mathbb{E}_{11}+A_{pq}^{\text{aa}}\, \mathbb{E}_{22}\otimes \mathbb{E}_{22}
\\
& \qquad \qquad \qquad\quad \,\,\,\,\,\,\, +C_{pq}^{\text{aa}}\, \mathbb{E}_{12}\otimes \mathbb{E}_{21}-E_{pq}^{\text{aa}}\, \mathbb{E}_{21}\otimes \mathbb{E}_{12}
\end{aligned}
\end{equation}
while for $\text{L}\tilde{\text{L}},\text{R}\tilde{\text{R}},\tilde{\text{L}}\text{L},\tilde{\text{R}}\text{R}$ one finds
\begin{equation}\label{LLtilde+RRtilde-Rmatrix}
\begin{aligned}
R^{\text{a}\tilde{\text{a}}}(p,q)&=B_{pq}^{\text{aa}}\, \mathbb{E}_{11}\otimes \mathbb{E}_{11}+A_{pq}^{\text{aa}}\, \mathbb{E}_{11}\otimes \mathbb{E}_{22}-F_{pq}^{\text{aa}}\, \mathbb{E}_{22}\otimes \mathbb{E}_{11}+D_{pq}^{\text{aa}}\, \mathbb{E}_{22}\otimes \mathbb{E}_{22}
\\
& \qquad \qquad \qquad \quad \,\,\, -E_{pq}^{\text{aa}}\, \mathbb{E}_{12}\otimes \mathbb{E}_{12}+C_{pq}^{\text{aa}}\, \mathbb{E}_{21}\otimes \mathbb{E}_{21}  
\\[0.5em]
R^{\tilde{\text{a}}\text{a}}(p,q)&=D_{pq}^{\text{aa}}\, \mathbb{E}_{11}\otimes \mathbb{E}_{11}-F_{pq}^{\text{aa}}\, \mathbb{E}_{11}\otimes \mathbb{E}_{22}+A_{pq}^{\text{aa}}\, \mathbb{E}_{22}\otimes \mathbb{E}_{11}+B_{pq}^{\text{aa}}\, \mathbb{E}_{22}\otimes \mathbb{E}_{22}
\\
& \qquad \qquad \qquad \quad \,\,\, +C_{pq}^{\text{aa}}\, \mathbb{E}_{12}\otimes \mathbb{E}_{12}-E_{pq}^{\text{aa}}\, \mathbb{E}_{21}\otimes \mathbb{E}_{21}  \,\, .
\end{aligned}
\end{equation}
The final bit of information is then encoded in
\begin{equation}\label{LRtilde+LtildeR+LtildeRtilde-Rmatrices}
\begin{aligned}
R^{\text{L}\tilde{\text{R}}}(p,q)&=C_{pq}^{\text{LR}}\, \mathbb{E}_{11}\otimes \mathbb{E}_{11}+A_{pq}^{\text{LR}}\, \mathbb{E}_{11}\otimes \mathbb{E}_{22}-E_{pq}^{\text{LR}}\, \mathbb{E}_{22}\otimes \mathbb{E}_{11}+D_{pq}^{\text{LR}}\, \mathbb{E}_{22}\otimes \mathbb{E}_{22}
\\
& \qquad \qquad \qquad\quad \,\,\,\, -F_{pq}^{\text{LR}}\, \mathbb{E}_{12}\otimes \mathbb{E}_{21}-B_{pq}^{\text{LR}}\, \mathbb{E}_{21}\otimes \mathbb{E}_{12}
\\[0.5em]
R^{\tilde{\text{L}}\text{R}}(p,q)&=D_{pq}^{\text{LR}}\, \mathbb{E}_{11}\otimes \mathbb{E}_{11}-E_{pq}^{\text{LR}}\, \mathbb{E}_{11}\otimes \mathbb{E}_{22}+A_{pq}^{\text{LR}}\, \mathbb{E}_{22}\otimes \mathbb{E}_{11}+C_{pq}^{\text{LR}}\, \mathbb{E}_{22}\otimes \mathbb{E}_{22}
\\
& \qquad \qquad \qquad\quad \,\,\,\, -B_{pq}^{\text{LR}}\, \mathbb{E}_{12}\otimes \mathbb{E}_{21}-F_{pq}^{\text{LR}}\, \mathbb{E}_{21}\otimes \mathbb{E}_{12}
\\[0.5em]
R^{\tilde{\text{L}}\tilde{\text{R}}}(p,q)&=E_{pq}^{\text{LR}}\, \mathbb{E}_{11}\otimes \mathbb{E}_{11}-D_{pq}^{\text{LR}}\, \mathbb{E}_{11}\otimes \mathbb{E}_{22}-C_{pq}^{\text{LR}}\, \mathbb{E}_{22}\otimes \mathbb{E}_{11}-A_{pq}^{\text{LR}}\, \mathbb{E}_{22}\otimes \mathbb{E}_{22}
\\
& \qquad \qquad \qquad \quad \,\,\,\, +B_{pq}^{\text{LR}}\, \mathbb{E}_{12}\otimes \mathbb{E}_{12}+F_{pq}^{\text{LR}}\, \mathbb{E}_{21}\otimes \mathbb{E}_{21}  
\end{aligned}
\end{equation}
from which by exchanging the L and R labels one immediately finds $\text{R}\tilde{\text{L}},\tilde{\text{R}}\text{L},\tilde{\text{R}}\tilde{\text{L}}$\footnote{The above notation could also be naturally extended to the $R^{ab}$-matrices of the mixed-flux theory \cite{Lloyd:2014bsa}, where each Zhukovski variable would inherit the appropriate $\text{a}$ or $\text{b}$ label due to the splitting of $x_{p}^{\pm}$ into $x_{p \, \text{L}}^{\pm}$ and $x_{p \, \text{R}}^{\pm}$.}. It is then straightforward to check that the above $R$-matrices correctly satisfy the bulk Yang-Baxter Equation (YBE)
\begin{equation}\label{general-bulk-YBE}
[R^{ab}(p_{1},p_{2})]_{12}[R^{ac}(p_{1},p_{3})]_{13}[R^{bc}(p_{2},p_{3})]_{23}=[R^{bc}(p_{2},p_{3})]_{23}[R^{ac}(p_{1},p_{3})]_{13}[R^{ab}(p_{1},p_{2})]_{12} 
\end{equation}
and the bulk Intertwining Equation (IE)
\begin{equation}\label{bulk-IE}
(\pi_{p}^{a}\!\otimes\! \pi_{q}^{b})\Delta^{\text{op}}(\mathfrak{a})R^{ab}(p,q) = R^{ab}(p,q)(\pi_{p}^{a}\!\otimes\! \pi_{q}^{b})\Delta(\mathfrak{a}) \,\, ,
\end{equation}
for any combination of $a,b,c \in \{ \text{L},\text{R},\tilde{\text{L}},\tilde{\text{R}} \}$ and $\mathfrak{a}\in \mathcal{A}$. Finally, they also satisfy unitarity relations
\begin{equation}\label{unitarity-1}
R^{aa}(p,q)[R^{aa}(q,p)]^{\text{op}}  \!=\! \mathds{1} \!=\! R^{\text{a}\tilde{\text{a}}}(p,q)[R^{\tilde{\text{a}}\text{a}}(q,p)]^{\text{op}} \,\, ,
\end{equation}
with usual conventions for the indices $a$ and $\text{a}$, as well as 
\begin{equation}\label{unitarity-2}
R^{ab}(p,q)[R^{ba}(q,p)]^{\text{op}}  = (x^{-}_{p}x^{-}_{q}-1)(x^{+}_{p}x^{+}_{q}-1) \,\, \mathds{1} \,\, ,
\end{equation}
for $ab \in \{ \text{L}\text{R}, \text{L}\tilde{\text{R}}, \text{R}\tilde{\text{L}}, \tilde{\text{L}}\tilde{\text{R}} \}$.

\subsection{Boundary theory}\label{sec:boundary-theory}

\begin{center}
\begin{figure}[ht]
\hspace*{2.5cm}
\includegraphics[scale=0.44]{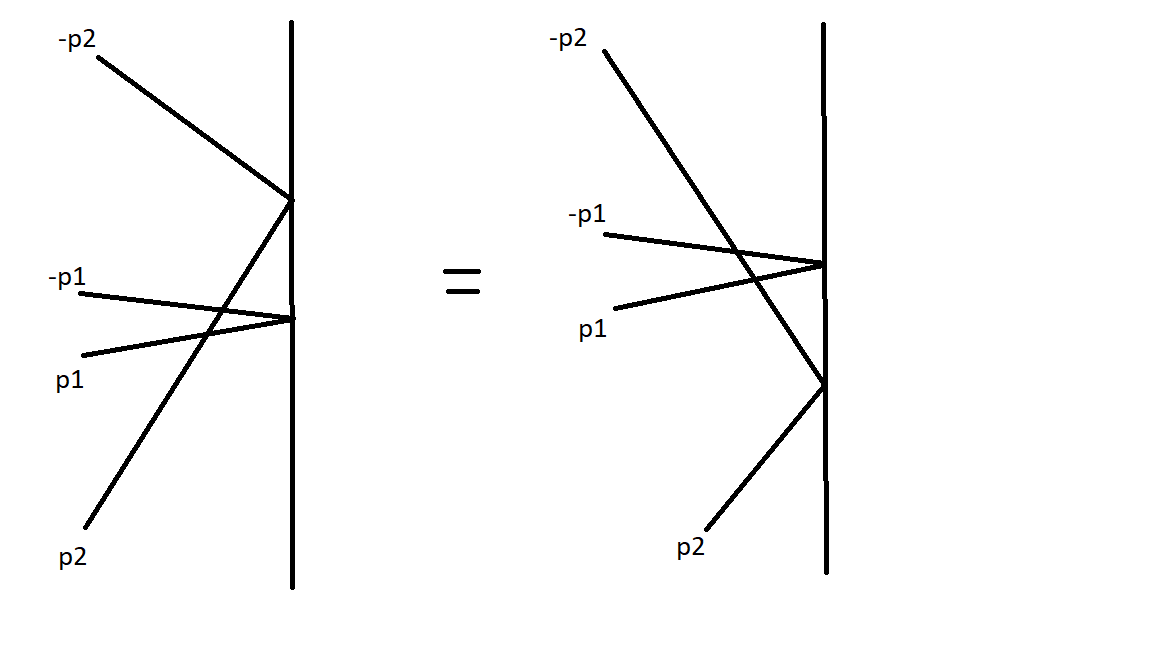}
\caption{Pictorial view of a right-wall. The bulk dynamics takes place on its left and excitations bouncing off it should respect the BYBE, encoding factorised scattering in analogy with the bulk.}
\label{fig:right-wall}
\end{figure}
\end{center}

In this section we summarise information on the interplay of boundaries with $2d$ integrability, focusing on the case of bulk excitations scattering against a right-wall, i.e. a wall placed in such a way that the dynamics takes place on its left, as depicted in figure \ref{fig:right-wall}. Two main ingredients characterise the search for boundaries which preserve, at least partially, integrability.

\subsubsection*{\ul{\it Boundary Yang-Baxter Equation (BYBE)}}
The possibility of solving such equation represents the first ingredient and a requirement for the existence of an integrable boundary. Encoding factorisation of scattering events against the wall, this is the boundary-analogue of the bulk Yang-Baxter Equation \eqref{general-bulk-YBE} and ensures compatibility of the boundary with the integrable structure underlying the bulk. Information about the bulk excitations scattering off the right wall enters the BYBE via the $R$-matrices, while the boundary information is encoded in a new set of objects, known as \textit{reflection matrices} or more simply $K$-matrices. All this is combined together in the following BYBE, which can be read off figure \ref{fig:right-wall} 
\begin{equation}\label{general-right-BYBE}
\begin{aligned}
[K^{cb}(p_{2},p_{B})]_{2B}&[R^{ca}(p_{2},-p_{1})]^{op}_{12}[K^{ab}(p_{1},p_{B})]_{1B}[R^{ac}(p_{1},p_{2})]_{12} 
\\
& = [R^{ca}(-p_{2},-p_{1})]^{op}_{12}[K^{ab}(p_{1},p_{B})]_{1B}[R^{ac}(p_{1},-p_{2})]_{12}[K^{cb}(p_{2},p_{B})]_{2B} \,\, .
\end{aligned}
\end{equation}
This describes the scattering of two excitations, respectively carrying momenta and representations $(p_{1},a)$ and $(p_{2},c)$, against a right-wall of "momentum" $p_{B}$ in the representation $b$, the meaning of which will be clarified soon. 
Like the bulk YBE, the above BYBE is an equation on the tensor product of three spaces, where the third-particle space is replaced by the boundary, such that $V_{1} \otimes V_{2} \otimes V_{B}$ respectively encode particles 1 and 2 and the boundary. Since they provide information about the bulk-dynamics, $R$-matrices only act non-trivially on $V_{1}\otimes V_{2}$. On the other hand, $K$-matrices encode information on how the boundary interacts with the bulk and hence exhibit non-trivial action in spaces $V_{1}$ or $V_{2}$ and always on $V_{B}$. In particular, one can write the action on the first space and the boundary as $[K^{ab}(p_{1},p_{B})]_{1B}=(\mathds{1}\otimes\mathds{P})(K^{ab}(p_{1},p_{B})\otimes \mathds{1})(\mathds{1}\otimes\mathds{P})$, with $\mathds{P}$ the graded permutation operator, while for the second space and the boundary one has the simpler expression $[K^{cb}(p_{2},p_{B})]_{2B}= \mathds{1} \otimes K^{cb}(p_{2},p_{B})$. Solving the BYBE means then finding a $K$-matrix which satisfies \eqref{general-right-BYBE} and ensures compatibility of bulk and boundary.

\subsubsection*{\ul{\it Coideal Subalgebras and Boundary Intertwining Equation (BIE)}}
Provided the existence of a non-trivial $K$-matrix, understanding which bulk symmetries are preserved by the boundary represents the second ingredient of the analysis. Integrability plays a very important role in this respect. At the Hopf algebra level it dictates that any subalgebra $\mathcal{B}$ of the bulk symmetry algebra $\mathcal{A}$, which is preserved by a right-boundary, should be a right-coideal subalgebra rather than a Hopf subalgebra. In particular this implies that $\mathcal{B}\subset \mathcal{A}$ can only be unbroken if the coproduct satisfies 
\begin{equation}
\Delta(\mathfrak{b}) \in \mathcal{A} \otimes \mathcal{B} \qquad \forall \,\, \mathfrak{b}\in \mathcal{B} \,\, .
\end{equation}
Additionally, at the Lie algebra level, the pair $(\mathfrak{g},\mathfrak{h})$ with the first entry associated to $\mathcal{A}$ and the second to $\mathcal{B}$, should be symmetric. This implies the vector space decomposition $\mathfrak{g} \simeq \mathfrak{h}\oplus \mathfrak{m}$ and the relations
\begin{equation}
[\mathfrak{h},\mathfrak{h}] \subseteq \mathfrak{h} 
\qquad \qquad 
[\mathfrak{h},\mathfrak{m}] \subseteq \mathfrak{m}
\qquad \qquad
[\mathfrak{m},\mathfrak{m}] \subseteq \mathfrak{h} \,\, .
\end{equation}
After identification of all possible right-coideal subalgebras, full characterisation of the boundary is obtained upon establishing which of them are effectively preserved by the boundary. This can be achieved by studying the Boundary Intertwining Equation (BIE), which represents the boundary-analogue of the bulk Intertwining Equation \eqref{bulk-IE} and should be satisfied by any generator in the preserved right-coideal subalgebra. This takes the form
\begin{equation}\label{boundary-IE}
(\pi_{-p}^{a} \otimes \pi_{p_{B}}^{b})\Delta(\mathfrak{b})K^{ab}(p,p_{B}) 
= K^{ab}(p,p_{B}) (\pi_{p}^{a} \otimes \pi_{p_{B}}^{b})\Delta(\mathfrak{b}) \qquad \forall \,\, \mathfrak{b}\in \mathcal{B} \,\, .
\end{equation}

\subsubsection*{\ul{\it Right-Wall Reflection Matrices}}
We can now specialise the study of \eqref{general-right-BYBE} and \eqref{boundary-IE} to the cases of \textit{singlet} and \textit{vector} right-walls, for different types of boundary coideal subalgebras $\mathcal{B}$, recalling known results about representations $\text{L},\text{R}$ from \cite{Prinsloo:2015apa} and extending them so as to include $\tilde{\text{L}},\tilde{\text{R}}$. Vector and singlet boundaries are of special physical interest, as they encode the scattering off elastic walls which respectively carry and do not carry extra degrees of freedom. In the vector case, these new degrees of freedom can nicely be interpreted as those of a magnon with fixed momentum attached to the wall, against which incoming excitations scatter.

\paragraph{\textbf{Singlet Boundary.}}
In this case the wall is imagined as a purely passive object, whose only action on the incoming excitations is that of changing the sign of their momenta.
For this reason it is thought of as being in a trivial representation of the symmetry algebra, from which the name singlet arises. This property implies that the boundary representation index $b$ in \eqref{general-right-BYBE} and \eqref{boundary-IE} can be completely dropped, together with the dependence of the $K$-matrix on any parameter $p_{B}$ associated to the boundary. The BYBE hence simplifies to
\begin{equation}\label{singlet-BYBE}
\begin{aligned}
[K^{c}(p_{2})]_{2}&[R^{ca}(p_{2},-p_{1})]^{op}_{12}[K^{a}(p_{1})]_{1}[R^{ac}(p_{1},p_{2})]_{12} \\
=
& [R^{ca}(-p_{2},-p_{1})]^{op}_{12}[K^{a}(p_{1})]_{1}[R^{ac}(p_{1},-p_{2})]_{12}[K^{c}(p_{2})]_{2} \,\, ,
\end{aligned}
\end{equation}
and the BIE becomes
\begin{equation}\label{singlet-intertwining}
\pi_{-p}^{a}(\mathfrak{b})K^{a}(p)=K^{a}(p)\pi_{p}^{a}(\mathfrak{b}) \,\, .
\end{equation}
In this setting, these two equations are respectively $4\times 4$ and $2\times 2$ matrix equations, which need to be solved entry-wise. Following the discussion of \cite{Prinsloo:2015apa}, there are three boundary coideal subalgebras which one may consider.
\begin{itemize}
\item \textit{Non-supersymmetric chiral}, $\mathcal{B}_{\text{NC}}=\langle \mathfrak{H}_{\text{L}},\mathfrak{H}_{\text{R}} \rangle $
\item \textit{Left half-supersymmetric}, $\mathcal{B}_{\text{L}}=\langle \mathfrak{Q}_{\text{L}}, \mathfrak{G}_{\text{L}}, \mathfrak{H}_{\text{L}},\mathfrak{H}_{\text{R}}  \rangle$
\item \textit{Right half-supersymmetric}, $\mathcal{B}_{\text{R}}=\langle \mathfrak{Q}_{\text{R}}, \mathfrak{G}_{\text{R}}, \mathfrak{H}_{\text{L}},\mathfrak{H}_{\text{R}}  \rangle$
\end{itemize}
We focus on the first one, since the $K$-matrices associated to the other two can be recovered as limiting cases. For $\text{L}$ and $\text{R}$ the result was given in \cite{Prinsloo:2015apa}, and using the extra $R$-matrices associated to $\tilde{\text{L}}$ and $\tilde{\text{R}}$ from the previous section one can complete the picture by obtaining
\begin{equation}\label{singlet-K-matrices}
\begin{aligned}
K^{a}(p)=\mathbb{E}_{11}+&g^{a}(p) \, \mathbb{E}_{22} =
\begin{pmatrix}
1 & 0
\\
0 & g^{a}(p)
\end{pmatrix} \qquad \text{with}
\\
g^{\text{L}}(p)=\frac{s-x^{+}_{p}}{s+x^{-}_{p}}
\qquad \qquad
g^{\text{R}}(p)=\frac{1+s \, x^{+}_{p}}{1-s\, x^{-}_{p}}
\qquad & \qquad 
g^{\tilde{\text{L}}}(p)=\frac{s+x^{-}_{p}}{s-x^{+}_{p}}
\qquad \qquad
g^{\tilde{\text{R}}}(p)=\frac{1-s\, x^{-}_{p}}{1+s\, x^{+}_{p}} \,\, .
\end{aligned}
\end{equation}
The BYBE \eqref{singlet-BYBE} is then satisfied for any choice of $a,c \!\in\! \{ \text{L},\text{R},\tilde{\text{L}} , \tilde{\text{R}}\}$ and the BIE \eqref{singlet-intertwining} for any representation $a$ and generator $\mathfrak{b}\!\in\! \mathcal{B}_{\text{NC}}$. The constant $s$ ($c$ in \cite{Prinsloo:2015apa}) is the same for all $K$-matrices and is left arbitrary, but it is simple to check that setting $s\!=\!\tan{\theta}$ and taking the limits $\theta\rightarrow \frac{\pi}{2}$ and $\theta\rightarrow 0$ one obtains enhancement of the preserved symmetries to the half-supersymmetric subalgebras $\mathcal{B}_{\text{L}}$ and $\mathcal{B}_{\text{R}}$, satisfying \eqref{singlet-intertwining} for all $a$.
\begin{itemize}
\item For $\theta\rightarrow \frac{\pi}{2}$ one finds $K$-matrices which intertwine the generators in $\mathcal{B}_{\text{L}}$
\begin{equation}
K^{\text{L}}(p)\!=\!K^{\tilde{\text{L}}}(p)\!=\!\mathds{1}_{2}
\qquad \qquad
K^{\text{R}}(p) = \mathbb{E}_{11}-e^{ip} \, \mathbb{E}_{22} 
\qquad \qquad 
K^{\tilde{\text{R}}}(p) = \mathbb{E}_{11}-e^{-ip} \, \mathbb{E}_{22} \,\, .
\end{equation}
\item For $\theta\rightarrow 0$ one finds $K$-matrices which intertwine the generators in $\mathcal{B}_{\text{R}}$
\begin{equation}
K^{\text{R}}(p)\!=\!K^{\tilde{\text{R}}}(p)\!=\!\mathds{1}_{2}
\qquad \qquad
K^{\text{L}}(p) = \mathbb{E}_{11}-e^{ip} \, \mathbb{E}_{22} 
\qquad \qquad 
K^{\tilde{\text{L}}}(p) = \mathbb{E}_{11}-e^{-ip} \, \mathbb{E}_{22} \,\, .
\end{equation}
\end{itemize}
Finally, changing basis of $\mathcal{A}$ as in (4.24) and (4.25) of \cite{Prinsloo:2015apa} it is straightforward to check the existence of a hidden \textit{diagonally supersymmetric} boundary algebra $\mathcal{B}_{\text{D}}= \langle \mathfrak{q}_{+},\mathfrak{q}_{-},\mathfrak{d},\tilde{\mathfrak{d}}  \rangle$, which solves the BIE in all representations $\text{L},\text{R},\tilde{\text{L}},\tilde{\text{R}}$ for the set of matrices \eqref{singlet-K-matrices}, provided that $s^2\neq -1$.

\paragraph{\textbf{Vector Boundary.}}
Picturing the elastic wall as an object which is able to interact with bulk excitations while possibly carrying some extra degrees of freedom leads to the concept of vector boundary. In this context the boundary index $b$ in \eqref{general-right-BYBE} and \eqref{boundary-IE} is non-trivial and a choice of boundary representation for the symmetry algebra and associated parameter $p_{\text{B}}$ should be made. In the spirit of the previous sections it is very natural to proceed here with the choice made in \cite{Prinsloo:2015apa}, i.e. extending the definitions \eqref{L+R-rep} to the boundary by letting $p \rightarrow p_{\text{B}}$, with the interpretation that $\pi_{p_{\text{B}}}^{\text{L}}$ and $\pi_{p_{\text{B}}}^{\text{R}}$ (as well as their tilde versions) arise as special choices of the bulk counterparts for which the coupling has been rescaled as $h \rightarrow h/2$ and the momentum fixed to $p\equiv p_{\text{B}} \equiv \pi$. This leads to the physical interpretation of a vector boundary as of a magnon with fixed momentum attached to a wall. The representation coefficients read
\begin{equation}\label{boundary-reps-coefficients}
a_{p_{\text{B}}}= b_{p_{\text{B}}} 
=\sqrt{h} \, \eta_{p_{\text{B}}}
\qquad \qquad 
c_{p_{\text{B}}}= d_{p_{\text{B}}} = \frac{i\sqrt{h}\,\eta_{p_{\text{B}}}}{x_{p_{\text{B}}}}
\qquad \qquad 
\eta_{p_{\text{B}}}^2=-i\,x_{p_{\text{B}}} \,\, ,
\end{equation}
with $x_{p_{B}}$ defined again as in \cite{Prinsloo:2015apa}
\begin{equation}
x_{p_{\text{B}}}=x_{p=\pi}^{+}=-x_{p=\pi}^{-} \,\, .
\end{equation}
We can now proceed in considering the only boundary coideal subalgebra allowed by the vector case, namely the \textit{totally supersymmetric} algebra \cite{Prinsloo:2015apa}
\begin{equation}\label{vector-coideal-subalgebra}
\mathcal{B}_{\text{T}}= \langle \mathfrak{Q}_{\text{L}},\mathfrak{G}_{\text{L}}, \mathfrak{H}_{\text{L}}, \mathfrak{Q}_{\text{R}},\mathfrak{G}_{\text{R}}, \mathfrak{H}_{\text{R}}, \mathfrak{P}, \mathfrak{K}  \rangle \,\, .
\end{equation}
As for the singlet case, the $K$-matrices associated to $\text{L}$ and $\text{R}$ representations satisfying BYBE \eqref{general-right-BYBE} and BIE \eqref{boundary-IE} were found in \cite{Prinsloo:2015apa} and their extension to $\tilde{\text{L}},\tilde{\text{R}}$ can be obtained using the $R$-matrices provided in section \ref{sec:bulk-theory}. Exploiting an analogous notation, with the addition of hats, for the labelling of the coefficient function which appear in each of the relevant $K$-matrices, one finds
\begin{equation}\label{LL+RR-Kmatrices}
\begin{aligned}
K^{\text{aa}}(p,p_{\text{B}})&=\hat{A}_{pp_{\text{B}}}^{\text{aa}}\, \mathbb{E}_{11}\otimes \mathbb{E}_{11}+\hat{B}_{pp_{\text{B}}}^{\text{aa}}\, \mathbb{E}_{11}\otimes \mathbb{E}_{22}+\hat{D}_{pp_{\text{B}}}^{\text{aa}}\, \mathbb{E}_{22}\otimes \mathbb{E}_{11}-\hat{F}_{pp_{\text{B}}}^{\text{aa}}\, \mathbb{E}_{22}\otimes \mathbb{E}_{22}
\\
& \qquad \qquad \qquad\quad \,\,\,\,\, -\hat{E}_{pp_{\text{B}}}^{\text{aa}}\, \mathbb{E}_{12}\otimes \mathbb{E}_{21}+\hat{C}_{pp_{\text{B}}}^{\text{aa}}\, \mathbb{E}_{21}\otimes \mathbb{E}_{12}
\end{aligned}
\end{equation}
for a=L or a=R, with the entries defined as
\begin{equation}\label{K-matrix-LL+RR-functions}
\begin{aligned}
\hat{A}^{\text{a}\text{a}}_{pp_{\text{B}}}=1 
\qquad \qquad \qquad \qquad \quad
\hat{B}^{\text{a}\text{a}}_{pp_{\text{B}}}&= \frac{x_{p}^{+}-u_{p}^{-2} \, x_{p_{\text{B}}}}{x_{p}^{-}-x_{p_{\text{B}}} }
\qquad \qquad \qquad\,\,\,
\hat{C}^{\text{a}\text{a}}_{pp_{\text{B}}}= \frac{-i(u_{p}+u_{p}^{-1})\eta_{p}\,\eta_{p_{\text{B}}}}{x_{p}^{-}-x_{p_{\text{B}}} } 
\\
\hat{D}^{\text{a}\text{a}}_{pp_{\text{B}}}=  \frac{x_{p}^{-}+u_{p}^{2} \, x_{p_{\text{B}}}}{x_{p}^{-}-x_{p_{\text{B}}} }
\qquad \qquad
\hat{E}^{\text{a}\text{a}}_{pp_{\text{B}}} &= \frac{-i(u_{p}+u_{p}^{-1})\eta_{p} \, \eta_{p_{\text{B}}}}{x_{p}^{-}-x_{p_{\text{B}}}}  
\qquad \qquad
\hat{F}^{\text{a}\text{a}}_{pp_{\text{B}}}= -\frac{x_{p}^{+}+x_{p_{\text{B}}} }{x_{p}^{-}-x_{p_{\text{B}}} } \,\, .
\end{aligned}
\end{equation}
For the mixed case one has
\begin{equation}\label{LR-Kmatrix}
\begin{aligned}
K^{\text{LR}}(p,p_{\text{B}})&=\hat{A}_{pp_{\text{B}}}^{\text{LR}}\, \mathbb{E}_{11}\otimes \mathbb{E}_{11}+\hat{C}_{pp_{\text{B}}}^{\text{LR}}\, \mathbb{E}_{11}\otimes \mathbb{E}_{22}+\hat{D}_{pp_{\text{B}}}^{\text{LR}}\, \mathbb{E}_{22}\otimes \mathbb{E}_{11}-\hat{E}_{pp_{\text{B}}}^{\text{LR}}\, \mathbb{E}_{22}\otimes \mathbb{E}_{22}
\\
& \qquad \qquad \qquad \quad \,\,\,\,\,\, -\hat{F}_{pp_{\text{B}}}^{\text{LR}}\, \mathbb{E}_{12}\otimes \mathbb{E}_{12}-\hat{B}_{pp_{\text{B}}}^{\text{LR}}\, \mathbb{E}_{21}\otimes \mathbb{E}_{21}  
\end{aligned}
\end{equation}
with the entries defined as
\begin{equation}\label{K-matrix-LR-functions}
\begin{aligned}
\hat{A}^{\text{L}\text{R}}_{pp_{\text{B}}}=  x_{p}^{+}x_{p_{\text{B}}} +u_{p}^{-2}
\qquad \qquad 
\hat{B}^{\text{L}\text{R}}_{pp_{\text{B}}} &=  -(u_{p}+u_{p}^{- 1}) \eta_{p}\eta_{p_{\text{B}}}
\qquad \qquad
\hat{C}^{\text{L}\text{R}}_{pp_{\text{B}}}=  x_{p}^{-}x_{p_{\text{B}}} +1
\\
\hat{D}^{\text{L}\text{R}}_{pp_{\text{B}}}=  x_{p}^{+}x_{p_{\text{B}}} -1
\qquad \qquad  \quad \,
\hat{E}^{\text{L}\text{R}}_{pp_{\text{B}}} &=  - (x_{p}^{-}x_{p_{\text{B}}} -u_{p}^{2})
\qquad \qquad \quad \,\,
\hat{F}^{\text{L}\text{R}}_{pp_{\text{B}}}=  (u_{p}+u_{p}^{- 1}) \eta_{p}\eta_{p_{\text{B}}} \,\, ,
\end{aligned}
\end{equation}
and $K^{\text{R}\text{L}}(p,p_{\text{B}})$ having the same structure as $K^{\text{L}\text{R}}(p,p_{\text{B}})$, with the new coefficients $\hat{A}^{\text{R}\text{L}}_{pq},...,\hat{F}^{\text{R}\text{L}}_{pq}$ respectively taking the same form as $\hat{A}^{\text{L}\text{R}}_{pq},...,\hat{F}^{\text{L}\text{R}}_{pq}$ in \eqref{K-matrix-LR-functions}.

Notice that the above $K$-matrices can be written by formally putting a hat on the coefficients appearing in the $R$-matrices \eqref{LL+RR-Rmatrices} and \eqref{LR-Rmatrix}, using then the explicit functions found in \cite{Prinsloo:2015apa}. This choice of notation for the $K$-matrices allows to readily exploit the fact that the $R$-matrices for $\tilde{\text{L}},\tilde{\text{R}}$ can be written by simply permuting the coefficients in $\text{LL}, \text{RR}, \text{LR}, \text{RL}$: the same mapping can indeed be straightforwardly applied to the respective $K$-matrices, leading to the desired result. For $\tilde{\text{L}}\tilde{\text{L}}$ and $\tilde{\text{R}}\tilde{\text{R}}$ one obtains
\begin{equation}\label{LtildeLtilde+RtildeRtilde-K-matrices}
\begin{aligned}
K^{\tilde{\text{a}}\tilde{\text{a}}}(p,p_{\text{B}})&=-\hat{F}_{pp_{\text{B}}}^{\text{aa}}\, \mathbb{E}_{11}\otimes \mathbb{E}_{11}+\hat{D}_{pp_{\text{B}}}^{\text{aa}}\, \mathbb{E}_{11}\otimes \mathbb{E}_{22}+\hat{B}_{pp_{\text{B}}}^{\text{aa}}\, \mathbb{E}_{22}\otimes \mathbb{E}_{11}+\hat{A}_{pp_{\text{B}}}^{\text{aa}}\, \mathbb{E}_{22}\otimes \mathbb{E}_{22}
\\
& \qquad \qquad \qquad\quad \,\,\,\,\,\,\,\,\, +\hat{C}_{pp_{\text{B}}}^{\text{aa}}\, \mathbb{E}_{12}\otimes \mathbb{E}_{21}-\hat{E}_{pp_{\text{B}}}^{\text{aa}}\, \mathbb{E}_{21}\otimes \mathbb{E}_{12} \,\, ,
\end{aligned}
\end{equation}
while for $\text{L}\tilde{\text{L}},\text{R}\tilde{\text{R}},\tilde{\text{L}}\text{L},\tilde{\text{R}}\text{R}$ one finds
\begin{equation}\label{LLtilde+RRtilde-Kmatrix}
\begin{aligned}
K^{\text{a}\tilde{\text{a}}}(p,p_{\text{B}})&=\hat{B}_{pp_{\text{B}}}^{\text{aa}}\, \mathbb{E}_{11}\otimes \mathbb{E}_{11}+\hat{A}_{pp_{\text{B}}}^{\text{aa}}\, \mathbb{E}_{11}\otimes \mathbb{E}_{22}-\hat{F}_{pp_{\text{B}}}^{\text{aa}}\, \mathbb{E}_{22}\otimes \mathbb{E}_{11}+\hat{D}_{pp_{\text{B}}}^{\text{aa}}\, \mathbb{E}_{22}\otimes \mathbb{E}_{22}
\\
& \qquad \qquad \qquad \quad \,\,\,\,\, -\hat{E}_{pp_{\text{B}}}^{\text{aa}}\, \mathbb{E}_{12}\otimes \mathbb{E}_{12}+\hat{C}_{pp_{\text{B}}}^{\text{aa}}\, \mathbb{E}_{21}\otimes \mathbb{E}_{21}  
\\[0.5em]
K^{\tilde{\text{a}}\text{a}}(p,p_{\text{B}})&=\hat{D}_{pp_{\text{B}}}^{\text{aa}}\, \mathbb{E}_{11}\otimes \mathbb{E}_{11}-\hat{F}_{pp_{\text{B}}}^{\text{aa}}\, \mathbb{E}_{11}\otimes \mathbb{E}_{22}+\hat{A}_{pp_{\text{B}}}^{\text{aa}}\, \mathbb{E}_{22}\otimes \mathbb{E}_{11}+\hat{B}_{pp_{\text{B}}}^{\text{aa}}\, \mathbb{E}_{22}\otimes \mathbb{E}_{22}
\\
& \qquad \qquad \qquad \quad \,\,\,\,\,\, +\hat{C}_{pp_{\text{B}}}^{\text{aa}}\, \mathbb{E}_{12}\otimes \mathbb{E}_{12}-\hat{E}_{pp_{\text{B}}}^{\text{aa}}\, \mathbb{E}_{21}\otimes \mathbb{E}_{21}  \,\, .
\end{aligned}
\end{equation}
Finally, the remaining $K$-matrices read
\begin{equation}\label{LRtilde+LtildeR+LtildeRtilde-Kmatrices}
\begin{aligned}
K^{\text{L}\tilde{\text{R}}}(p,p_{\text{B}})&=\hat{C}_{pp_{\text{B}}}^{\text{LR}}\, \mathbb{E}_{11}\otimes \mathbb{E}_{11}+\hat{A}_{pp_{\text{B}}}^{\text{LR}}\, \mathbb{E}_{11}\otimes \mathbb{E}_{22}-\hat{E}_{pp_{\text{B}}}^{\text{LR}}\, \mathbb{E}_{22}\otimes \mathbb{E}_{11}+\hat{D}_{pp_{\text{B}}}^{\text{LR}}\, \mathbb{E}_{22}\otimes \mathbb{E}_{22}
\\
& \qquad \qquad \qquad\quad \,\,\,\,\,\, -\hat{F}_{pp_{\text{B}}}^{\text{LR}}\, \mathbb{E}_{12}\otimes \mathbb{E}_{21}-\hat{B}_{pp_{\text{B}}}^{\text{LR}}\, \mathbb{E}_{21}\otimes \mathbb{E}_{12}
\\[0.5em]
K^{\tilde{\text{L}}\text{R}}(p,p_{\text{B}})&=\hat{D}_{pp_{\text{B}}}^{\text{LR}}\, \mathbb{E}_{11}\otimes \mathbb{E}_{11}-\hat{E}_{pp_{\text{B}}}^{\text{LR}}\, \mathbb{E}_{11}\otimes \mathbb{E}_{22}+\hat{A}_{pp_{\text{B}}}^{\text{LR}}\, \mathbb{E}_{22}\otimes \mathbb{E}_{11}+\hat{C}_{pp_{\text{B}}}^{\text{LR}}\, \mathbb{E}_{22}\otimes \mathbb{E}_{22}
\\
& \qquad \qquad \qquad\quad \,\,\,\,\,\, -\hat{B}_{pp_{\text{B}}}^{\text{LR}}\, \mathbb{E}_{12}\otimes \mathbb{E}_{21}-\hat{F}_{pp_{\text{B}}}^{\text{LR}}\, \mathbb{E}_{21}\otimes \mathbb{E}_{12}
\\[0.5em]
K^{\tilde{\text{L}}\tilde{\text{R}}}(p,p_{\text{B}})&=\hat{E}_{pp_{\text{B}}}^{\text{LR}}\, \mathbb{E}_{11}\otimes \mathbb{E}_{11}-\hat{D}_{pp_{\text{B}}}^{\text{LR}}\, \mathbb{E}_{11}\otimes \mathbb{E}_{22}-\hat{C}_{pp_{\text{B}}}^{\text{LR}}\, \mathbb{E}_{22}\otimes \mathbb{E}_{11}-\hat{A}_{pp_{\text{B}}}^{\text{LR}}\, \mathbb{E}_{22}\otimes \mathbb{E}_{22}
\\
& \qquad \qquad \qquad \quad \,\,\,\,\,\, +\hat{B}_{pp_{\text{B}}}^{\text{LR}}\, \mathbb{E}_{12}\otimes \mathbb{E}_{12}+\hat{F}_{pp_{\text{B}}}^{\text{LR}}\, \mathbb{E}_{21}\otimes \mathbb{E}_{21}  
\end{aligned}
\end{equation}
from which by exchanging the L and R labels one can again recover $\text{R}\tilde{\text{L}},\tilde{\text{L}}\text{R},\tilde{\text{R}}\tilde{\text{L}}$.
The BYBE \eqref{general-right-BYBE} and BIE \eqref{boundary-IE} are then straightforwardly solved for any combination of $a,b,c \in \{ \text{L},\text{R},\tilde{\text{L}}, \tilde{\text{R}} \}$ using the above $K$-matrices and substituting the boundary parameters \eqref{boundary-reps-coefficients} in the bulk representations \eqref{L+R-rep}.

\section{The boundary algebraic Bethe ansatz}\label{sec:Algebraic-Bethe-Ansatz}
In this section we apply the boundary algebraic Bethe ansatz method of \cite{Bielli:2024bve} to the massive excitations described in section \ref{sec:background}, exhibiting a few new interesting features which did not appear in the massless case.
We start by considering the usual setup for the application of the Bethe ansatz, which requires a system with $N$ physical excitations, placed in the bulk to the left of a right-wall and standing at fixed locations without interacting. The key step consists then in introducing a so-called \textit{auxiliary} excitation, able to move throughout the system and sequentially interacting with all the physical elements which characterise it, namely the $N$ physical excitations and in this case also the wall. This interaction condenses in a single fundamental object, the \textit{double-row monodromy matrix}, which crucially satisfies the same boundary Yang-Baxter equation (BYBE) as the reflection matrix associated to the wall\footnote{ See appendix \ref{app:Tminus-satisfies-BYBE} for a general proof encompassing both singlet and vector boundaries, for any combination of representations in either the auxiliary or physical spaces.} and encodes (almost) all the necessary information about the underlying integrable structure. In the absence of boundaries, e.g. in the possibly more familiar setting of a spin chain on a circle, this object does indeed encode all the necessary information, which can be extracted in a purely algebraic fashion and results in the Bethe equations. When boundaries are added to the system, the double-row monodromy is not sufficient to encode all the data required to extract the Bethe equations, and the construction of a second fundamental object is needed, the \textit{dual monodromy}. This satisfies the so-called \textit{dual boundary Yang-Baxter equation} \eqref{dual_BYBE} and completes the information contained in the double-row monodromy by ensuring that the trace of a full-fledged monodromy, build out as the product of the two partial ones, exhibits vanishing commutator and can thus be used as a generating function for conserved charges. 

In section \ref{subsec:dual-equation} the dual equation is introduced, and its solution in terms of a dual monodromy is provided. The rest of the section describes the completion of the boundary algebraic Bethe ansatz method. In particular, since the possible excitations and boundaries appearing in the system are characterised by various types of representations, we will present the method in several stages of increasing complexity, hopefully making the whole analysis more transparent and accessible.
\begin{itemize}
\item In section \ref{subsec:singlet-boundary} we focus on the case of boundaries carrying a trivial singlet representation. As a first step in this direction, subsection \ref{subsusbsec:singlet-all-L-reps} further specialises to the case of massive $\text{L}$ representations only, for both the auxiliary and the physical particles. Subsection \ref{subsubsec:singlet-auxiliary-L-mixed-physical-reps} refines then the analysis by taking into account the possibility of mixed $\text{L}$ and $\text{R}$ representations in the physical spaces.
\item In section \ref{subsec:vector-boundary} we extend the procedure to the case of boundaries carrying a non-trivial vector representation. In analogy with the singlet, we first specialise subsection \ref{subsubsec:vector-all-L-reps} to the case of all-$\text{L}$ massive representations for auxiliary, physical and boundary spaces. This allows to highlight the new ingredients introduced by vector boundaries, while still retaining a simple notation. We close the analysis in subsection \ref{subsubsec:vector-all-mixed-reps}, finally jumping to the most general setting involving fully mixed representations in either auxiliary, physical or boundary space.
\end{itemize} 

\subsection{The dual equation}\label{subsec:dual-equation}
As discussed above, the transfer matrix which we aim to diagonalize through the algebraic Bethe ansatz is defined in terms of two boundary monodromy matrices, the double-row monodromy $\left[T_{-}^{ab}(p_{0},p_{B})\right]_{0B}$ and the dual monodromy $\left[T_{+}^{ab}(p_{0},p_{B})\right]_{0B}$, by tracing along the auxiliary space $0$:
\begin{equation}\label{transfer-matrix-def}
\tau^{ab}(p_0,p_{B})=\text{str}_0\Bigl(\left[T_{+}^{ab}(p_{0},p_{B})\right]_{0B}\left[T_{-}^{ab}(p_{0},p_{B})\right]_{0B}\Bigr) \,\, .
\end{equation}
The labels $a,b$ refer to the representations carried by the auxiliary space $0$ and boundary space $B$, on which the two monodromies act. More details about the setup are provided in the appendices \ref{app:notation-and-conventions},\ref{app:useful-identities},\ref{app:Tminus-satisfies-BYBE},\ref{app:commuting-monodromies}.
The requirement that the transfer matrix $\tau^{ab}(p_{0},p_{B})$ commutes with itself for arbitrary momenta 
\begin{equation}
\tau^{ab}(p_0,p_{B})\tau^{ab}(p_{0'},p_{B})=\tau^{ab}(p_{0'},p_{B})\tau^{ab}(p_0,p_{B}) \,\, ,
\end{equation}
guarantees that the infinite charges it generates are in involution, and is met when each of the two monodromies satisfies a BYBE. More specifically, $T_-$ needs to be a solution to the right-wall equation \eqref{general-right-BYBE}, namely it has to satisfy \eqref{general-right-BYBE} like the $K$-matrix $K^{ab}(p_{0},p_{B})$ associated to the right-wall, while $T_+$ needs to satisfy the dual BYBE \eqref{dual_BYBE_appendix}, namely it has to satisfy \eqref{dual_BYBE_appendix} like the dual $K$-matrix $K^{ab}_{D}(p_{0},p_{B})$.  Here we directly express the dual BYBE in terms of $T_{+}$ and provide its derivation at the beginning of appendix \ref{app:dual-BYBE}. The proof of commutativity of the transfer matrix is given at the end of the same appendix and the construction of the monodromies in terms of the corresponding $K$-matrices is discussed below.

The appropriate choice of dual equation for our setting is \begin{equation}\label{dual_BYBE}
\begin{aligned}
\left[T_+^{cb}(p_{0'},p_{B})\right]^{ist_{0'}}_{0'B}&\left[X^{ac}(p_0,p_{0'})\right]_{00'}\left[T_+^{ab}(p_0,p_{B})\right]_{0B}^{st_{0}}\left[Y^{ac}(p_0,p_{0'})\right]_{00'}=
\\
&\left[R^{ac}(-p_0,-p_{0'})\right]_{00'}^{st_{0}ist_{0'}}\left[T^{ab}_+(p_0,p_{B})\right]_{0B}^{st_{0}}\left[Z^{ac}(p_0,p_{0'})\right]_{00'}\left[T_+^{cb}(p_{0'},p_{B})\right]^{ist_{0'}}_{0'B} \, ,
\end{aligned}
\end{equation}
where $X,Y,Z$ are defined in appendix \ref{app:dual-BYBE} and $(i)st_j$ denotes (inverse) supertransposition in the space $j$. Like the right-wall BYBE, it is an equation of matrices acting on $V_0\otimes V_{0'}\otimes V_B$, where $V_B$ is the boundary space carrying representation $b$, and $V_0,V_{0'}$ are auxiliary spaces carrying representations $a,c$. 

This dual BYBE is applicable to both the vector and singlet boundary cases. In the latter case, where $V_B$ is trivial, one can simply drop the $B$ (and $b$) indexes with no other changes to \eqref{dual_BYBE}. In fact, along the lines of \cite{Sklyanin:1988yz} one usually associates to the monodromy matrix $T_+$ the solution $K_D$ of the dual equation which exhibits trivial action on the physical spaces
\begin{equation}
[T_{+}^{ab}(p_{0},p_{B})]_{0B}:=[K_{D}^{ab}(p_{0},p_{B})]_{0B} \,\, .
\end{equation}
This is meant to guarantee the commutativity of $T_{-}$ and $T_{+}$, which is a crucial assumption on which commutativity of the transfer matrices $\tau$ relies.

In the case of singlet boundaries, the general solutions $K_D^a$ can be obtained by brute force, in the same way that the singlet right wall equations \eqref{general-right-BYBE} can be solved for $K^a$, giving 
\begin{equation}\label{singlet-dual-K-matrices}
\begin{aligned}
K_D^{a}(p)&=\mathbb{E}_{11}+g_D^{a}(p) \, \mathbb{E}_{22} =
\begin{pmatrix}
1 & 0
\\
0 & g_D^{a}(p)
\end{pmatrix}\,, \qquad\qquad  \qquad\qquad \text{with}
\\
g_D^{\text{a}}(p)&=\frac{s^{\text{a}}_D-x^-_{p}}{s^{\text{a}}_D + x^+_{p}}\,, 
\qquad  \qquad 
g_D^{\tilde{\text{a}}}(p)=\frac{s^{\tilde{\text{a}}}_D + x^+_{p}}{s^{\tilde{\text{a}}}_D-x^-_{p}} \qquad \text{for} \qquad \text{a}\in\{\text{L},\text{R}\} \,\, .
\end{aligned}
\end{equation}
The case of vector boundaries turns out to be more delicate and, as discussed in appendix \ref{app:commuting-monodromies}, the assumption of commutativity strongly constraints the form of the dual $K$-matrix, effectively forcing it to coincide with the singlet ones, up to a trivial extension acting on the boundary space
\begin{equation}\label{vector-K-plus-intro}
[K_{D}^{ab}(p_{0},p_{B})]_{0B}=K_{D}^{a}(p_{0})\otimes \mathds{1}_{B} \,\, .
\end{equation}
In particular, each of the four homogeneous BYBEs \eqref{dual_BYBE_appendix}, where both vector spaces $V_0,V_{0'}$ carry the same representation, is satisfied by the corresponding $K^a_D$-matrix in equation \eqref{singlet-dual-K-matrices} for any value of its complex constant $s_D^a\in\mathbb{C}$ for $a\in\{\text{L},\text{R},\tilde{\text{L}},\tilde{\text{R}}\}$. The various mixed BYBEs, where the two spaces carry different representations, then require these constants to be related in the following way
\begin{align}
s_{D}\equiv s_D^{L}=s_D^{\tilde{L}}=-\frac{1}{s^R_D}=-\frac{1}{s^{\tilde{R}}_D} \,\, .
\end{align}
This also allows to bring the dual $K$-matrices to a form similar to the right-wall ones in \eqref{singlet-K-matrices}:
\begin{equation}\label{dual-singlet-K-matrices}
\begin{aligned}
K_D^{\text{L}}(p) = \mathbb{E}_{11}+\frac{s_D-x^{-}_{p}}{s_D+x^{+}_{p}} \, \mathbb{E}_{22}
\qquad & \qquad 
K_D^{\text{R}}(p) = \mathbb{E}_{11}+\frac{1+s_D\,x^{-}_{p}}{1-s_D\,x^{+}_{p}} \, \mathbb{E}_{22}
\\
K_D^{\tilde{\text{L}}}(p) = \mathbb{E}_{11}+\frac{s_D+x^{+}_{p}}{s_D-x^{-}_{p}} \, \mathbb{E}_{22}
\qquad & \qquad 
K_D^{\tilde{\text{R}}}(p) = \mathbb{E}_{11}+ \frac{1-s_D\,x^{+}_{p}}{1+s_D\,x^{-}_{p}} \, \mathbb{E}_{22} \,\, .
\end{aligned}
\end{equation}

\subsection{Singlet boundary}\label{subsec:singlet-boundary}
As previously anticipated, we start from the simplest setting of boundaries carrying a trivial singlet representation, implying that the associated reflection matrix satisfies the BYBE \eqref{singlet-BYBE}. The relevant boundary coideal subalgebra in this context is the \textit{non-supersymmetric chiral} one, since the remaining two can be recovered for appropriate limiting values of the constant $s$. This characterises the singlet $K$-matrices in equation \eqref{singlet-K-matrices}, which correctly satisfy the BIE \eqref{boundary-IE} for any value of $s$.

The first step of the boundary Bethe ansatz technique is the construction of the double-row monodromy: we begin by placing the auxiliary excitation to the far left of the wall, away from all the $N$ physical particles, we then let it move to the right towards the wall, bounce off it and move back towards the left, finally reaching its original position. In this process the auxiliary excitation also scatters against all the physical ones, in reversed orders when first moving towards the right and then towards the left. This translates into the following expression for the double-row monodromy matrix\footnote{When comparing with \cite{Bielli:2024bve}, we have here adopted a different and more standard ordering of the $R$-matrices in the definition of the double-row monodromy. In the massless case considered in \cite{Bielli:2024bve} it turns out that either order
produces a good description of the conserved charges and both produce two consistent versions of the algebraic Bethe ansatz, as detailed in \cite{Bielli:2024bve}. We thank again the referee of \cite{Bielli:2024bve} for a series of remarks which have helped us appreciate this point.}
\begin{equation}\label{singlet-monodromy}
[T_{-}^{a}(p_{0})]_{0} \! := \! [R^{d_{N}a}(p_N,-p_0)]^{op}_{0N} ... [R^{d_{1}a}(p_1,-p_0)]^{op}_{01} \, [K^{a}(p_0)]_{0} \,  [R^{a d_{1}}(p_0,p_1)]_{01} ... [R^{ad_{N}}(p_0,p_N)]_{0N},   
\end{equation}
with $K^{a}(p_{0})$ denoting one of the $K$-matrices in \eqref{singlet-K-matrices}, satisfying equations \eqref{singlet-BYBE} and \eqref{boundary-IE}.
These matrices only act non-trivially on the auxiliary space $0$, where the auxiliary particle is taken in a generic representations $a \in \{\text{L},\text{R},\tilde{\text{L}},\tilde{\text{R}}\}$, which appear as a label on both $K$ and the monodromy itself. The $R$-matrices (and their $op$) act on the same auxiliary space as $K$ and only one physical space. The labels $d_i\in \{\text{L},\text{R},\tilde{\text{L}},\tilde{\text{R}}\}$ refer to the representation acting on the physical space $i$. They do not explicitly appear on the left hand side of \eqref{singlet-monodromy} because, as shown in appendix \ref{app:Tminus-satisfies-BYBE} for the most general case of a vector boundary, the monodromy satisfies the BYBE for any collection $\{d_{1}, d_{2},...,d_{N} \}$ and in this sense is not sensitive to such information.

The second step of the boundary Bethe ansatz procedure is the construction of the dual monodromy, which as anticipated is taken to coincide with the dual $K$-matrix $K_{D}$ in \eqref{singlet-dual-K-matrices}
\begin{equation}\label{dual-monodromy-def}
[T_{+}^{a}(p_{0})]_{0}:=[K_{D}^{a}(p_{0})]_{0} \,\, ,
\end{equation}
which solves by definition the dual equation \eqref{dual_BYBE}. Like their cousins \eqref{singlet-K-matrices}, these matrices only act non-trivially on the auxiliary space $0$, which carries representation $a$.

\subsubsection{Auxiliary-$\text{L}$ and physical-$\text{L}$  representations}\label{subsusbsec:singlet-all-L-reps}
In a homogeneous setting where both the auxiliary and the physical excitations are taken in the massive $\text{L}$ representation, the double-row monodromy \eqref{singlet-monodromy} becomes
\begin{equation}\label{usea}
[T_-(p_{0})]_{0} = [R(p_N,-p_0)]^{op}_{0N} ... [R(p_1,-p_0)]^{op}_{01} \, [K(p_0)]_{0} \,  [R(p_0,p_1)]_{01} ... [R(p_0,p_N)]_{0N},   
\end{equation}
where $K(p_{0})$ is the singlet reflection matrix $K^{\text{L}}(p)$ in \eqref{singlet-K-matrices}, the $R$-matrices are the $R^{\text{L}\text{L}}(p,q)$ in \eqref{LL+RR-Rmatrices} (and their \textit{op} version) and we dropped all the $\text{L}$ labels to make the presentation as simple as possible. Applying the same simplification to the dual monodromy \eqref{dual-monodromy-def}, we can construct the full monodromy
\begin{eqnarray}
T(p_{0})= T_+(p_0) T_-(p_0).  \label{double_row_monodromy}  
\end{eqnarray}
As discussed in appendix \ref{app:dual-BYBE}, the very construction of $T_+$ ensures that the supertrace of $T$ commutes at distinct values of $p_0$, providing a generating function for the conserved charges in involution:
\begin{eqnarray}
[\mbox{str}_0 T(p_0), \mbox{str}_{0'}T(p_{0'})] = 0 \qquad \forall \, \, \, p_0, p_{0'} \,\, .
\end{eqnarray}
The RTRT relations can then be written in the same abstract form as in the massless case \cite{Bielli:2024bve}, namely 
\begin{eqnarray}\label{operators-1}
[T_-(p_0)]_{0} = \begin{pmatrix}
\mathcal{A}(p_0)&\mathcal{B}(p_0)\\\mathcal{C}(p_0)&\mathcal{D}(p_0)\end{pmatrix}_0 = \mathbb{E}_{11} \otimes \mathcal{A}(p_0) +\mathbb{E}_{12} \otimes \mathcal{B}(p_0) +\mathbb{E}_{21} \otimes \mathcal{C}(p_0) +\mathbb{E}_{22} \otimes \mathcal{D}(p_0),  \label{tmi}
\end{eqnarray}
and consequently, combining the $R
$-matrix expansion
\begin{equation}
R(p,q) =  r^{\alpha\beta\gamma\delta}_{pq} \, \mathbb{E}_{\alpha\beta} \otimes \mathbb{E}_{\gamma\delta}
\end{equation}
with the definitions
\begin{equation}\label{operators-2}
\begin{aligned}
\mathcal{A}(p_0) &\equiv  \mathcal{A}_{11}(p_0), \qquad \mathcal{B}(p_0) \equiv  \mathcal{A}_{21}(p_0), \qquad \mathcal{C}(p_0) \equiv  \mathcal{A}_{12}(p_0), \qquad \mathcal{D}(p_0) \equiv  \mathcal{A}_{22}(p_0), 
\end{aligned}
\end{equation}
we again have from the reflection equation for $T_-$ 
\begin{equation}
\begin{aligned}\label{indices}
& (-1)^{(\mu+\delta)(\gamma+\mu)+(\rho+\beta)(\sigma+\mu)+(\nu+\beta+\sigma+\delta)(\nu+\rho)} \, r_{p_{0'}\, -p_{0}}^{\alpha \rho \mu\sigma}  \, r_{p_{0}p_{0'}}^{\nu \beta\sigma \delta}  \,  \mathbb{E}_{\alpha \beta} \otimes \mathbb{E}_{\gamma \delta} \otimes \mathcal{A}_{\mu\gamma}(p_{0'}) \mathcal{A}_{\nu\rho}(p_0) =
\\
&    (-1)^{(\mu+\beta)(\gamma+\sigma+\nu+\mu)+(\gamma+\delta)(\nu+\mu)} \, r_{-p_{0'}\, -p_{0}}^{\alpha \nu\gamma\sigma} \, r_{p_{0}\, -p_{0'}}^{\mu\beta\sigma\rho} \mathbb{E}_{\alpha \beta} \otimes \mathbb{E}_{\gamma \delta} \otimes \mathcal{A}_{\mu \nu}(p_{0}) \mathcal{A}_{\delta\rho}(p_{0'}) \,\, .
\end{aligned}
\end{equation}
When written down in full detail, we can extract the relations which are most useful. These are the relations between two $\mathcal{B}$ operators:
\begin{eqnarray}
\mathcal{B}(p_{0})\mathcal{B}(p_{0'})= \frac{x^{-}_{p_{0}}x^{+}_{p_{0'}}\big[(x^{-}_{p_{0'}})^2-(x^{+}_{p_{0}})^2\big]}{x^{+}_{p_{0}} x^{-}_{p_{0'}}  \big[(x^{-}_{p_{0}})^2-(x^{+}_{p_{0'}})^2\big]} \, \mathcal{B}(p_{0'})\mathcal{B}(p_{0}).    
\label{exch}
\end{eqnarray}
The relation between two $\mathcal{B}$'s is suggestive of braided statistics - in particular, two $\mathcal{B}$'s do NOT commute. As a check, it is easy to verify that they tend to commuting operators in the limit of massless kinematic.

The relations to commute $\mathcal{A}$ and $\mathcal{D}$ through a $\mathcal{B}$ are next. To produce clean formulas for the latter we need  to perform some manipulations, exactly the same as in the massless case. First, we rewrite a selected relation involving  $\mathcal{AB}$ products with the auxiliary momenta swapped, and solve both versions of this  relation simultaneously - this allows to separate $\mathcal{A}(p_{0})\mathcal{B}(p_{0'})$ from $\mathcal{A}(p_{0'})\mathcal{B}(p_{0})$. We do the same with $\mathcal{DB}$. Second, we combine the resulting $\mathcal{AB}$ and $\mathcal{DB}$ relations, achieving everywhere having only products of the type $\mathcal{BA}$ and $\mathcal{DA}$ on the right-hand-side. The net result is as follows:
\begin{align}
\mathcal{A}(p_{0}) \mathcal{B}(p_{0'}) & =  \frac{x^{-}_{p_{0}}[(x^{+}_{p_{0}})^2 - (x^{-}_{p_{0'}})^2]}{x^{+}_{p_{0}}[(x^{-}_{p_{0}})^2 - (x^{-}_{p_{0'}})^2]} \, \mathcal{B}(p_{0'})\mathcal{A}(p_{0}) 
\notag \\[1em]
& + \frac{i \,u_{p_{0}}^{-1} \, u_{p_{0'}}^{-1} \,  \eta_{p_{0}} \, \eta_{p_{0'}} \, (x^{-}_{p_{0}}+x^{+}_{p_{0'}})}{ [(x^{-}_{p_{0'}})^2 - (x^{-}_{p_{0}})^2]} \, \mathcal{B}(p_{0})\mathcal{A}(p_{0'})
\notag \\[1em]
& +\frac{i \,u_{p_{0}}^{-1} \, u_{p_{0'}}^{-1} \, \eta_{p_{0}} \, \eta_{p_{0'}}}{(x^{-}_{p_{0'}} + x^{-}_{p_{0}})} \, \mathcal{B}(p_{0})\mathcal{D}(p_{0'})
\label{AB-commutator-LL-reps}
\\[2em]
\mathcal{D}(p_{0}) \mathcal{B}(p_{0'}) &=  \frac{x^{-}_{p_{0}}[(x^{+}_{p_{0}})^2 - (x^{-}_{p_{0'}})^2]}{x^{+}_{p_{0}}[(x^{-}_{p_{0}})^2 - (x^{-}_{p_{0'}})^2]} \, \mathcal{B}(p_{0'})\mathcal{D}(p_{0})
\notag \\[1em]
& - \frac{i \,u_{p_{0}}^{-1} \, u_{p_{0'}}^{-1} \, \eta_{p_{0}} \, \eta_{p_{0'}} \, (x^{-}_{p_{0'}}+x^{+}_{p_{0}})}{[(x^{-}_{p_{0}})^2 - (x^{-}_{p_{0'}})^2]} \, \mathcal{B}(p_{0})\mathcal{D}(p_{0'}) 
\notag \\[1em]
& + \frac{i \,u_{p_{0}}^{-1} \, u_{p_{0'}}^{-1} \, \eta_{p_{0}} \, \eta_{p_{0'}} \, (x^{+}_{p_{0}}-x^{+}_{p_{0'}})}{[(x^{-}_{p_{0}})^2 - (x^{-}_{p_{0'}})^2]} \, \mathcal{B}(p_{0})\mathcal{A}(p_{0'})
\label{DB-commutators-LL-reps}
\end{align}
We have checked that these relations, obtained abstractly, are satisfied in particular by the explicit matrices for the case $N=2$.

We can immediately obtain the first two eigenstates by means of the algebraic Bethe ansatz, using these relations which we have just found. The pseudo-vacuum can be taken to be
\begin{eqnarray}
|0\rangle_{N} = |\phi\rangle_{1} \otimes |\phi\rangle_{2} \otimes ... \otimes |\phi\rangle_{N}  \,\, ,
\end{eqnarray}
obtained by tensoring $N$ copies of the fundamental boson $|\phi\rangle$, and it can easily be seen as an eigenstate of $\mathcal{A}$ and $\mathcal{D}$ separately:
\begin{eqnarray}\label{fore}
\mathcal{A}(p_0)|0\rangle_{N} = |0\rangle_{N}, \qquad \qquad
\mathcal{D}(p_0)|0\rangle_{N} = \lambda_{2(\!N\!)} (p_0;p_1,...,p_N)|0\rangle_{N} \,\, ,
\end{eqnarray}
with $\!\lambda_{2(\!N\!)}\!$ a function to be shortly determined. 
Let us schematically rewrite the RTRT relations above 
\begin{eqnarray}
&&\mathcal{A}_{0} \mathcal{B}_{0'} = \alpha \, \mathcal{B}_{0'} \mathcal{A}_0 + \beta \, \mathcal{B}_0 \mathcal{A}_{0'} + \gamma \, \mathcal{B}_0 \mathcal{D}_{0'}, \nonumber\\
&&\mathcal{D}_0 \mathcal{B}_{0'} = \alpha \, \mathcal{B}_{0'} \mathcal{D}_0 - \tilde{\beta} \, \mathcal{B}_0 \mathcal{D}_{0'} + \tilde{\gamma} \, \mathcal{B}_0 \mathcal{A}_{0'},\label{alp}
\end{eqnarray}
where the coefficients $\alpha, \beta,...$ depend on $p_{0},p_{0'}$, namely $\alpha(p_0,p_{0'}), \beta(p_0,p_{0'}),...$. The one-magnon state 
\begin{eqnarray}
\mathcal{B}_{0'} |0\rangle_{N}
\end{eqnarray}
is an eigenstate of the transfer matrix if and only if
\begin{eqnarray}
\!\!\!\tau_0 \mathcal{B}_{0'} |0\rangle_{N} \!=\!
\bigl(\mathcal{A}_0 - g^L_{D}(p_{0}) \mathcal{D}_0\bigr)\mathcal{B}_{0'}|0\rangle
\!=\!  \alpha \mathcal{B}_{0'} |0\rangle_{N} +   \Bigl[\beta - g^L_{D}(p_{0}) \tilde{\gamma} + \bigl(\gamma+ g^L_{D}(p_{0}) \tilde{\beta}\bigr)  \lambda_{2(\!N\!)}(p_{0'})   \Bigr] \mathcal{B}_{0'}|0\rangle_{N}   
\end{eqnarray}
is an eigenstate, namely if
\begin{eqnarray}
\beta - g^L_{D}(p_{0}) \tilde{\gamma}+ \bigl(\gamma+ g^L_{D}(p_{0}) \tilde{\beta}\bigr)  \lambda_{2(\!N\!)}(p_{0'};p_1,...,p_N)=0 \,\, .
\end{eqnarray}
Simplifying this expression shows that this is equivalent to imposing
\begin{eqnarray}
-s_D - x^{+}_{p_{0'}} +\big(s_D - x^{-}_{p_{0'}}\big)\lambda_{2(\!N\!)}(p_{0'};p_1,...,p_N)=0, 
\end{eqnarray}
which shows that the dependence on $p_0$ rightfully has dropped out of the Bethe equation, hence of the eigenvector. The eigenvalue is therefore proportional to the pseudo-vacuum eigenvalue, that is 
\begin{eqnarray}
\tau_0 \mathcal{B}_{0'}|0\rangle_{N} =  \alpha \big[1 - g_{D}^{L}(p_{0}) \lambda_{2(\!N\!)}(p_0)\big] \mathcal{B}_{0'}|0\rangle_{N} \,\, .
\end{eqnarray}
From this we can very simply
deduce the formulas pertaining to the one-magnon state, written now in the more traditional way as $\mathcal{B}(q_1)|0\rangle_{N}$, by specialising our results to $p_{0'} = q_1$.

When it comes to multi-magnon states, we need to consider that the $\mathcal{B}$ operators do not commute. Fortunately the relatively simple exchange factor in (\ref{exch}) means that the normalised version of the (rays of the) eigenvectors
\begin{eqnarray}
    \mathcal{B}(q_1) \mathcal{B}(q_2)...\mathcal{B}(q_M)|0\rangle_{N}
\end{eqnarray}
are independent on the ordering of the string of $\mathcal{B}$'s. However,  the exchange factor could matter in deriving the Bethe equations, and in fact its existence appears essential to obtain the correct answer. To convince ourselves of this fact, we have repeated the procedure for two magnons: $\bigl(\mathcal{A}_0 - g_{D}^{L}(p_{0}) \mathcal{D}_0\bigr) \mathcal{B}(q_1) \mathcal{B}(q_2)|0\rangle_{N}$. We have commuted twice the $\mathcal{A}$ and $\mathcal{D}$ operators through the two $\mathcal{B}$s with the help of  symbolic programming, and then rearranged the result using the relations for swapping two $\mathcal{B}$'s. After a massive cancellation, and implementing the exchange factors collected rearranging the $\mathcal{B}$'s, we find that we can cancel the two unwanted terms (proportional to $\mathcal{B}_{q_1} \mathcal{B}_0$ and $\mathcal{B}_{q_2}\mathcal{B}_0$) by setting
\begin{eqnarray}
-s_D - x^{+}_{q_{i}} +\big(s_D - x^{-}_{q_{i}}\big)\lambda_{2(\!N\!)}(q_i;p_1,...,p_N)=0,   \qquad i=1,2, 
\label{BAE}
\end{eqnarray}
which is perfectly consistent and suggest the generalisation to arbitrary magnon number $i=1,...,M$. A more compact way to rewrite the Bethe equations is actually
\begin{eqnarray}
g_{D}^{L}(q_i) \lambda_{2(\!N\!)}(q_i;p_1,...,p_N) = 1, \qquad i=1,...,M,    
\end{eqnarray}
where the information about the physical sites $N$ is buried in the function $\lambda_{2(\!N\!)}$. With a little calculation it is actually possible to see that the massless equations \cite{Bielli:2024bve} have exactly the same form. It is possible to adapt the proof in appendix D of \cite{Bielli:2024bve} to prove the above statement as well. In fact, one can argue in the same way that a first swap of arguments followed by all remaining diagonal actions returns the same Bethe equation as we have found for one magnon times an immaterial global factor. Likewise, choosing any other than the first momentum $q_1$ just amounts to a global exchange factor, which again is immaterial if we set to zero the unwanted term. We discover therefore that taking into account the exchange factor in rearranging the $\mathcal{B}$'s is necessary for consistency of the procedure, but its precise value cancels out in the Bethe equations, as we actually realise when revisiting the brute force calculation.   

As far as the eigenvalue of the transfer matrix is concerned, it is straightforwardly obtained by focusing only on the diagonal term, hence we always collect $\alpha$-type terms in (\ref{alp}). Therefore we have
\begin{eqnarray}\label{transfer-eigenvalue-LL}
\tau_0 \mathcal{B}_1 ... \mathcal{B}_M |0\rangle_{N} = \big[1 - g^L_{D}(p_{0}) \lambda_{2(\!N\!)}(p_0;p_1,...,p_N)\big]\prod_{m=1}^M \alpha(p_0,q_m) \,\,  \mathcal{B}_1 ... \mathcal{B}_M |0\rangle_{N}   
\end{eqnarray}
where 
\begin{eqnarray}
\alpha(p_0,p_{0'}) =  \frac{x_{p_{0}}^{-}[({x^+_{p_{0}}})^2 -({x^-_{p_{o'}}})^2 ]}{x_{p_{0}}^{+}[({x^-_{p_{0}}})^2 -({x^-_{p_{0'}}})^2]}.
\end{eqnarray}

When it comes to computing the eigenvalue $\lambda_{2(\!N\!)}$ for any number of physical excitations, we employ an inductive procedure, exactly as in \cite{Bielli:2024bve}, by expressing the $(N+1)$-site eigenvalue $\lambda_{2(\!N+1\!)}$ in terms of the $N$-site one. The trick is to rewrite the $R$-matrix acting on the $(N+1)^{\text{th}}$ physical space as
\begin{align}
\left[R(p_0,p_{N+1})\right]_{0,N+1}= r_{p_0,p_{N+1}}^{\alpha\beta\gamma\delta} \,  \mathbb{E}_{\alpha\beta}\otimes \mathbb{E}_{\gamma\delta}=\mathbb{E}_{\alpha\beta}\otimes \nu_{\beta\alpha}
\end{align}
and the opposite one in the same way
\begin{align}
\left[R(p_{N+1},-p_0)\right]_{0,N+1}^{op}=\mathbb{E}_{\gamma\delta}\otimes \mu_{\delta\gamma}.
\end{align}
The explicit $\mu_{\delta\gamma}$ and $\nu_{\beta\alpha}$ matrices appearing here are
\begin{equation}
\begin{aligned}
&\mu_{11} = A^{\text{LL}}_{p_{N+1},-p_0}\mathbb{E}_{11} +D^{\text{LL}}_{p_{N+1},-p_0}\mathbb{E}_{22} \,,
\qquad  \qquad  
\mu_{12} = E^{LL}_{p_{N+1},-p_0}\mathbb{E}_{12} \,\, ,
\\
&\mu_{22} = B^{\text{LL}}_{p_{N+1},-p_0}\mathbb{E}_{11} -F^{\text{LL}}_{p_{N+1},-p_0}\mathbb{E}_{22}\, , 
\qquad  \qquad \,
\mu_{21} = -C^{\text{LL}}_{p_{N+1},-p_0}\mathbb{E}_{21},
\end{aligned}
\end{equation}
and
\begin{equation}
\begin{aligned}
&\nu_{11} = A^{\text{LL}}_{p_0,p_{N+1}}\mathbb{E}_{11} +B^{\text{LL}}_{p_0,p_{N+1}}\mathbb{E}_{22}\, ,
\qquad  \qquad
\nu_{12} = C^{\text{LL}}_{p_0,p_{N+1}}\mathbb{E}_{12} \,\, 
\\
&\nu_{22} = D^{\text{LL}}_{p_0,p_{N+1}}\mathbb{E}_{11} -F^{\text{LL}}_{p_0,p_{N+1}}\mathbb{E}_{22}\, ,
\qquad  \qquad 
\nu_{21} = -E^{\text{LL}}_{p_0,p_{N+1}}\mathbb{E}_{21}.
\end{aligned}
\end{equation}
Using this form of the $R$-matrix it is straightforward to express the $(N+1)$-site monodromy $T_{-(\!N+1\!)}$ in terms of the $N$-site one, by dressing the latter with a pair of $R$-matrices, as in \eqref{usea}. Adding extra labels $(N+1)$ and $(N)$ to the monodromy and the operators to remind ourselves of how many sites they are meant to act on, we find 
\begin{equation}
\begin{aligned}
    T_{-(\!N+1\!)}=&\left(\mathbb{E}_{\gamma\delta}\otimes\mathds{1}_N\otimes\mu_{\delta \gamma}\right)\left(\mathbb{E}_{\rho\sigma}\otimes \mathcal{A}_{\sigma\rho(\!N\!)}\otimes \mathds{1}\right)\left(\mathbb{E}_{\alpha\beta}\otimes\mathds{1}_N\otimes \nu_{\beta\alpha}\right)\\
    =&(-1)^{(|\alpha|+|\beta|)(|\alpha|+|\gamma|)}\mathbb{E}_{\gamma \beta}\otimes \mathcal{A}_{\alpha\delta(\!N\!)}\otimes \mu_{\delta \gamma}\nu_{\beta \alpha}
\end{aligned}
\end{equation}
where we use \eqref{operators-1} and \eqref{operators-2} for $T_{-(\!N\!)}$. Picking out the $\mathbb{E}_{22}$ term in this relation, we can extract the action of the $\mathcal{D}_{\!(N+1\!)}$ on the pseudovacuum
\begin{align}
    \mathcal{D}_{\!(N+1\!)}|0\rangle_{N+1}=(-1)^{(|\alpha|+1)}\mathcal{A}_{\alpha\delta(\!N\!)}|0\rangle_N\otimes\mu_{\delta2}\nu_{2\alpha}|\phi\rangle \,\, .
\end{align}
Out of the four terms in the above expression, two vanish identically, namely
\begin{align}    \mathcal{A}_{12(\!N\!)}|0\rangle_N\otimes\mu_{22}\nu_{21}|0\rangle_1=\mathcal{C}_{(\!N\!)}|0\rangle_N\otimes\mu_{22}\nu_{21}|\phi\rangle=0 \,\, ,
\end{align}
as $\mathcal{C}_{(\!N\!)}$ annihilates $|0\rangle_N$ by construction, and
\begin{align} \mathcal{A}_{21(\!N\!)}|0\rangle_N\otimes\mu_{12}\nu_{22}|0\rangle_1=\mathcal{B}_{(\!N\!)}|0\rangle_N\otimes\mu_{12}\nu_{22}|\phi\rangle=0 \,\, ,
\end{align}
because $\mu_{12}\nu_{22}$ annihilates $|\phi\rangle$ due to its matrix structure.
The relevant terms are therefore the ones including $\mathcal{A}_{(\!N\!)}$ and $\mathcal{D}_{\!(N\!)}$, both of which act diagonally on $|0\rangle_N$ according to \eqref{fore}. Combining this with
\begin{equation}
\begin{aligned}
\mu_{12} \nu_{21}|\phi\rangle &= -E^{\text{LL}}_{p_0,p_{N+1}} E^{\text{LL}}_{p_{N+1},-p_0}|\phi\rangle \equiv -s_{N+1} |\phi\rangle 
\\
\mu_{22} \nu_{22}|\phi\rangle &=  D^{\text{LL}}_{p_0,p_{N+1}} B^{\text{LL}}_{p_{N+1},-p_0}|\phi\rangle \equiv - t_{N+1} |\phi\rangle \,\, ,  
\end{aligned}
\end{equation}
we obtain the following recursive relation for the eigenvalues of the $\mathcal{D}$-operator
\begin{eqnarray}
\lambda_{2(\!N+1\!)} = E^{\text{LL}}_{p_0,p_{N+1}} E^{\text{LL}}_{p_{N+1},-p_0} + D^{\text{LL}}_{p_0,p_{N+1}} B^{\text{LL}}_{p_{N+1},-p_0} \, \lambda_{2(\!N\!)}  = s_{N+1} - t_{N+1} x_N\nonumber \,\, .
\end{eqnarray}
The solution is the same as the one in \cite{Bielli:2024bve} {\it mutatis mutandis}, that is
\begin{eqnarray}
x_N = \lambda_{2(\!N\!)} = s_N + \sum_{n=1}^{N-1} (-1)^n s_{N-n} \prod_{k=N-n+1}^N t_k - (-1)^N c_0 \prod_{k=1}^N t_k,\label{same}
\end{eqnarray}
with 
\begin{eqnarray}
c_0 \equiv - g(p_0) =  -\frac{s - x^{+}_{p_{0}}}{s + x^{-}_{p_{0}}} \,\, , 
\end{eqnarray}
where $g(p_{0})$ is the function appearing in the singlet $K^{\text{L}}$-matrix in equation \eqref{singlet-K-matrices}.

The value of $c_0$ can be fixed from a simple calculation, involving the explicit computation of the operator $\mathcal{D}$ for $N=1$, which can be done very straightforwardly using (\ref{usea}).
We have then tested this formula explicitly for $N=2$, finding perfect agreement with the available expression we have obtained of the eigenvalue $\lambda_{2(\!N\!)}$ by direct matrix action.

\subsubsection{Auxiliary-$\text{L}$ and mixed physical representations}\label{subsubsec:singlet-auxiliary-L-mixed-physical-reps}
If we have physical particles in mixed $\text{L}$ and $\text{R}$ representations, which we need to include for completeness of the physical spectrum, we can still proceed as follows. The auxiliary particle, as traditionally the case, merely provides an organising principle and can therefore be kept in the $\text{L}$ representation, as in the previous section\footnote{We will consider the possibility of different auxiliary representations in the final subsection \ref{subsubsec:vector-all-mixed-reps}.}. Therfore the RTRT relations are always the same as we had before, and only the representation of $T_{-}$ changes.

This modification in the representation of $T_-$ changes in turn the structure of the $\mathcal{C}$ operator, such that a different choice of pseudovacuum $|0\rangle_N$ needs to be made for every combination of physical representations. Explicitly computing $\mathcal{C}$ for different cases, we find that the state it annihilates consists of a bosonic basis vector $|\phi\rangle$ on every physical space that carries an $\text{L}$ or $\tilde{\text{R}}$ representation and a fermionic one $|\psi\rangle$ in the spaces that carry an $\text{R}$ or $\tilde{\text{L}}$ representation. That is, denoting $|\rho^{\text{L}}\rangle=|\rho^{\tilde{\text{R}}}\rangle=|\phi\rangle$ and $|\rho^{\text{R}}\rangle=|\rho^{\tilde{\text{L}}}\rangle=|\psi\rangle$, the appropriate pseudovacuum for the physical space $V_{d_1}\otimes\cdots\otimes V_{d_N}$ is
\begin{align}
|0\rangle_N=\bigotimes_{i=0}^N |\rho^{d_i}\rangle\,,
\end{align}
where $d_i\in\{\text{L},\text{R},\tilde{\text{L}},\tilde{\text{R}}\}$. This form of $|0\rangle_N$ looks somewhat connected to the similar structures which respectively appear in the $\text{L},\tilde{\text{R}}$ and $\text{R},\tilde{\text{L}}$ representations, as in particular the state annihilated by each generator is the same in each of these pairs of representations. For instance, from \eqref{L+R-rep} follows that
\begin{equation}
\begin{aligned}
&\pi^\text{L}(\mathfrak{Q}_R)|\phi\rangle=\pi^{\tilde{\text{R}}}(\mathfrak{Q}_R)|\phi\rangle=0 \,\, , \qquad \text{while} \qquad\pi^\text{R}(\mathfrak{Q}_R)|\psi\rangle=\pi^{\tilde{\text{L}}}(\mathfrak{Q}_R)|\psi\rangle=0
\end{aligned}
\end{equation}
and there is a similar pattern for the remaining fermionic generators. This might be interpreted as a hint of the fact that in the purely algebraic form of the monodromy $T_-$, the local generators that appear in $\mathcal{C}$, including the $\mathfrak{psu}(1|1)^2_{c.e.}$ ones and the rest of the infinite generators of the Hopf algebra, are those that act like $\mathfrak{G}_L\,\text{and}\,\mathfrak{Q}_R$, which annihilate $|\phi\rangle$ and $|\psi\rangle$ in the $\text{L},\tilde{\text{R}}$ and $\text{R},\tilde{\text{L}}$ representations respectively. To verify this intuition one would need to explicitly construct the algebraic element corresponding to $T_-$ using the universal $R$-matrix corresponding to our system, which remains unknown.

Since the RTRT relations remain unchanged compared to the previous subsection, the exchange relations \eqref{exch}-\eqref{DB-commutators-LL-reps} still hold. Then, the forms of the Bethe equations \eqref{BAE} and transfer matrix eigenvalues \eqref{transfer-eigenvalue-LL}, which are completely fixed by those exchange relations, are also the same in the case of mixed physical representations. What remains to complete the study of the auxiliary Bethe equations in this case is to calculate the eigenvalue $\lambda_{2(\!N\!)}$ of $\mathcal{D}$, which should be different from the previous case, as the explicit form of $\mathcal{D}$ depends on the physical representations.

A recursion relation for $\lambda_{2(\!N\!)}$ can be derived in an almost identical way to the one shown above, after adjusting the normalisation of $R^{\text{RL}}$ and $R^{\text{LR}}$. For example (\ref{fore}) holds with an appropriate normalisation\footnote{For instance, if we have all the physical particles in the $\text{R}$ representation, it is sufficient to normalise the associated $T_-$ such that every $R^{\text{LR}}(p,q)$ is multiplied by $\Phi^{\text{LR}}(p,q) = \frac{e^{-i p}}{x^{-}_{p} x^{-}_{q}-1}$ and every $R^{\text{RL}}(p,q)$ is multiplied by $\Phi^{\text{RL}}(p,q) = \frac{1}{x^{+}_{p} x^{+}_{q}-1}$.} and with the new pseudovacuum. 
Now, we can again proceed exactly as in \cite{Bielli:2024bve} - section 3.2.3. Let us for example consider the case of all the physical particles in representation $\text{R}$ (the auxiliary particle always in representation $\text{L}$). The relevant terms in this particular case are given by
\begin{equation}
\begin{aligned}
\mu_{11} &= A^{\text{RL}}_{p_{N+1},-p_0}\mathbb{E}_{11} +D^{\text{RL}}_{p_{N+1},-p_0}\mathbb{E}_{22}\,,
\qquad  \qquad
\mu_{12} = B^{\text{RL}}_{p_{N+1},-p_0}\mathbb{E}_{21}
\\
\mu_{22} &= C^{\text{RL}}_{p_{N+1},-p_0}\mathbb{E}_{11} -E^{\text{RL}}_{p_{N+1},-p_0}\mathbb{E}_{22}\,, \qquad  \qquad
\mu_{12} = F^{\text{RL}}_{p_{N+1},-p_0}\mathbb{E}_{12},
\end{aligned}
\end{equation}
and
\begin{equation}
\begin{aligned}
\nu_{11} &= A^{LR}_{p_0,p_{N+1}}\mathbb{E}_{11} +C^{LR}_{p_0,p_{N+1}}\mathbb{E}_{22}\,,
\qquad  \qquad 
\nu_{12} = -B^{LR}_{p_0,p_{N+1}}\mathbb{E}_{21}\\
\nu_{22} &= D^{LR}_{p_0,p_{N+1}}\mathbb{E}_{11} -E^{LR}_{p_0,p_{N+1}}\mathbb{E}_{22}\,,
\qquad  \qquad 
\nu_{21} = -F^{LR}_{p_0,p_{N+1}}\mathbb{E}_{12},
\end{aligned}
\end{equation}
(please notice the different assortment of indices given the different matrix structure of the mixed $R$-matrices)
hence
\begin{equation}
\begin{aligned}
\mu_{12} \nu_{21}|\psi\rangle &= - F^{\text{LR}}_{p_0,p_{N+1}} B^{\text{RL}}_{p_{N+1},-p_0}|\psi\rangle \equiv -s_{N+1} |\psi\rangle
\\
\mu_{22} \nu_{22}|\psi\rangle &=  E^{\text{LR}}_{p_0,p_{N+1}} E^{\text{RL}}_{p_{N+1},-p_0}|\psi\rangle \equiv - t_{N+1} |\psi\rangle.  
\end{aligned}
\end{equation}
The recursion is therefore again of the same type $x_{N+1} = s_{N+1} - t_{N+1} x_N$ (even taking into account of the fact that the fermionic degree of $|0\rangle_{N}$ is now $(-1)^N$), and so the solution is again the same (\ref{same}) {\it mutatis mutandis}. We have checked $N=2$ explicitly, namely $\lambda_{2(\!N=2\!)} = s_2 - t_2 s_1 - c_0 t_1 t_2$, finding again perfect agreement with the explicit expressions from the direct matrix action.

If we now have a non-zero numbers of $\text{L}$ particle and a non-zero number of $\text{R}$ particles, the recursion is such that at every step $N \to N+1$ the very form of the next term depends whether the representation of the $N+1$ physical particle is $\text{L}$ or $\text{R}$. We can collect the results which we have already obtained by saying that 
\begin{equation}
\begin{aligned}
\text{if site} \quad  N+1 &= \text{L}  \qquad \text{then} \qquad    x_{N+1} = s^{\text{L}}_{N+1} - t^{\text{L}}_{N+1} x_N \,\, ,
\\
\text{if site} \quad N+1 &= \text{R} \qquad \text{then} \qquad  x_{N+1} = s^{\text{R}}_{N+1} - t^{\text{R}}_{N+1} x_N \,\, ,
\end{aligned}
\end{equation}
where
\begin{equation}
\begin{aligned}
s^{\text{L}}_n &= E^{\text{LL}}_{p_0,p_n}E^{\text{LL}}_{p_n,-p_0}, \qquad \qquad     t^{\text{L}}_n = - D^{\text{LL}}_{p_0,p_n}B^{\text{LL}}_{p_n,-p_0},\nonumber\\
s^{\text{R}}_n &= F^{\text{LR}}_{p_0,p_n}B^{\text{RL}}_{p_n,-p_0}, \qquad \qquad     t^{\text{R}}_n = - E^{\text{LR}}_{p_0,p_n}E^{\text{RL}}_{p_n,-p_0}.
\end{aligned}
\end{equation}
We can then achieve a solution by the following steps. First, let us fix an assortment of physical representations $\vec{\alpha} = (\text{L},\text{L},\text{R},\text{L},\text{R},\text{L},...,\text{L},\text{R})$ {\it etc.}, where in general $\vec{\alpha} = (\alpha_1,...,\alpha_{N+1})$, and $\alpha_i = \text{L},\text{R}$. For any fixed $\vec{\alpha}$, we define 
\begin{eqnarray}
    \sigma_m \equiv s_m^{\alpha_m}, \qquad \tau_m \equiv t_m^{\alpha_m}.\label{extra}
\end{eqnarray}
This means that we have reduced the problem to the standard recursion if we keep track of the extra indices as in (\ref{extra}), which we therefore solve as
\begin{eqnarray}
x_N = \lambda_{2(\!N\!)} = \sigma_N + \sum_{n=1}^{N-1} (-1)^n \sigma_{N-n} \prod_{k=N-n+1}^N \tau_k - (-1)^N c_0 \prod_{k=1}^N \tau_k,\label{samen}
\end{eqnarray}
We have explicitly checked this formula against direct diagonalisation  for all the combinations at $N=2$ (that is, $\text{LL}$, $\text{LR}$, $\text{RL}$ and $\text{RR}$ in the physical spaces).

\subsection{Vector boundary}\label{subsec:vector-boundary}
We are now ready to extend the procedure to the more subtle case of a non-trivial boundary, with associated reflection matrices satisfying the BYBE \eqref{general-right-BYBE}. In this setting, the only relevant boundary coideal subalgebra is the \textit{totally supersymmetric} algebra \eqref{vector-coideal-subalgebra}, solving the BIE \eqref{boundary-IE} for the $K$-matrices listed below equation \eqref{vector-coideal-subalgebra}. Also in this case, the first step of the Bethe ansatz procedure is the construction of the double-row monodromy, which follows the same reasoning described for the singlet at the beginning of section \ref{subsec:singlet-boundary} and leads to the expression
\begin{align}\label{vector-monodromy}
[T_{-}^{ab}&(p_{0},p_{B})]_{0}  := 
\\
& =[R^{d_{N}a}(p_N,-p_0)]^{op}_{0N} ... [R^{d_{1}a}(p_1,-p_0)]^{op}_{01} \, [K^{ab}(p_0,p_{B})]_{0B} \,  [R^{a d_{1}}(p_0,p_1)]_{01} ... [R^{ad_{N}}(p_0,p_N)]_{0N}, 
\notag
\end{align}
which simply carries an extra boundary-representation index $b\in \{ \text{L},\text{R},\tilde{\text{L}},\tilde{\text{R}} \}$ with respect to \eqref{singlet-monodromy}. We stress again that \eqref{vector-monodromy} satisfies the same BYBE \eqref{general-right-BYBE} as the $K$-matrices associated to the right-wall, which is shown explicitly in appendix \ref{app:Tminus-satisfies-BYBE}. As already anticipated in section \ref{subsec:dual-equation}, the construction of a dual monodromy can then be formally addressed, as for the singlet case, by choosing 
\begin{equation}\label{dual-vector-monodromy}
[T_{+}^{ab}(p_{0},p_{B})]_{0B}:=[K_{D}^{ab}(p_{0},p_{B})]_{0B} \,\, ,
\end{equation}
where $K_{D}^{ab}(p_{0},p_{B})$ is the solution to the dual boundary Yang-Baxter equation \eqref{dual_BYBE_appendix} which exhibits a trivial action on the physical spaces $1,...,N$. While this may a priori seem like an innocuous generalisation, it turns out to be actually quite subtle and to require a careful look back at the dual equation constructed in section \ref{subsec:dual-equation}. Indeed, a crucial step in the derivation of the dual equation relies on the assumption that the two monodromies $T_{-}$ and $T_{+}$ should commute: while for the singlet case this was immediately ensured by the fact that $K_{D}^{a}(p_{0})$, and in turn the dual monodromy $T_{+}^{a}(p_{0})$, only acted non-trivially on the auxiliary space, the same is not true for vector boundaries. Indeed, its action on the boundary is now made quite intricate by the mixing with the action on physical spaces and it is a priori non-straightforward to see that the two monodromies should still commute. This question has been carefully analysed in appendix \ref{app:commuting-monodromies}, leading to the conclusion that requiring commutativity under the assumption of fermion-number conservation, enforces a solution to the dual equation of the form 
\begin{equation}\label{vector-K-plus}
[K_{D}^{ab}(p_{0},p_{B})]_{0B}:=[K_{D}^{a}(p_{0})]_{0}\otimes \mathds{1}_{B} \,\, ,
\end{equation}
with $K_{D}^{a}(p_{0})$ being of the same diagonal form as for the singlet boundary. Altogether, including a trivial action on the physical spaces, this leads to a dual monodromy $T_{+}$ of the following explicit form
\begin{equation}\label{explicit-dual-monodromy}
[T_{+}^{ab}(p_{0},p_{B})]_{0B}:=K_{D}^{a}(p_{0})\otimes \mathds{1}_{N}\otimes \mathds{1}_{B} \,\, .
\end{equation}

\subsubsection{Auxiliary-$\text{L}$, physical-$\text{L}$ and boundary-$\text{L}$ representations}\label{subsubsec:vector-all-L-reps} Let us initially focus on the case where the auxiliary and physical excitations are L and the boundary also carries an L-type representation with a rescaling $h\rightarrow \frac{h}{2}$ and at fixed momentum $p_B=\pi$. This choice allows us to interpret the boundary degree of freedom as an L-type particle fixed at the boundary.\\
Now the right-wall reflection matrix $K^{\text{LL}}(p)$ acts in $V_0\otimes V_B$, it is reported in \eqref{LL+RR-Kmatrices} and explicitly reads
\begin{align}
\label{frome}
K^{\text{LL}}(p,p_{B})=&\mathbb{E}_{11}\otimes \mathbb{E}_{11}+\frac{x_p^+-e^{-i p}x_B}{x^-_p-x_B}\mathbb{E}_{11}\otimes \mathbb{E}_{22}+\frac{x_p^- +e^{i p}x_B}{x^-_p-x_B}\mathbb{E}_{22}\otimes \mathbb{E}_{11}+\frac{x_p^+ +x_B}{x^-_p-x_B}\mathbb{E}_{22}\otimes \mathbb{E}_{22}\nonumber\\
&+\frac{i(e^{i\frac{p}{2}}-e^{-i\frac{p}{2}})\eta_p\eta_B}{x^-_p-x_B} \mathbb{E}_{12}\otimes \mathbb{E}_{21}-\frac{i(e^{i\frac{p}{2}}-e^{-i\frac{p}{2}})\eta_p\eta_B}{x^-_p-x_B} \mathbb{E}_{21}\otimes \mathbb{E}_{12} \,\, .
\end{align}
Using the solution to the dual equation already found for the singlet boundary, the derivation of the Bethe equations is almost identical to what is discussed in the previous subsections. The pseudovacuum for $N$ physical excitations is now
\begin{align}
|0 \rangle_{NB}= |0\rangle_{N} \otimes |\phi\rangle_{B}=|\phi \rangle_1\otimes\cdots\otimes |\phi \rangle_N \otimes|\phi \rangle_B\, ,
\end{align}
i.e. the singlet pseudovacuum with an added bosonic boundary state. The reflection equation that $T_-^{ab}$ satisfies is identical to the previous cases, since the auxiliary space carries yet again the L representation, therefore the exchange relations (\ref{indices}) still hold. Also, due to our choice of $T_+^{ab}$, the general form of the transfer matrix in terms of $\mathcal{A}$ and $\mathcal{D}$ is the same as in the previous cases, namely
\begin{align}
\tau(p_0)=\mathcal{A}(p_0)-g^L_D(p_0) \mathcal{D}(p_0) \,\, .
\end{align}
Due to these two facts, acting with $\tau(p_0)$ on a state $\mathcal{B}(q_1)\cdots \mathcal{B}(q_2)|0 \rangle_{NB}$ and requiring that the spurious terms vanish yields (\ref{BAE}). The auxiliary Bethe equations are thus identical to the ones we obtained in the singlet case when expressed abstractly, in terms of the vacuum eigenvalues of the operators $\mathcal{A}$ and $\mathcal{D}$. The difference now is in the explicit form of $\lambda_{2(\!N\!)}$, which depends on the choice of boundary.
In turn, the recursion to fix $\lambda_{2(\!N\!)}$ (the eigenvalue of $\mathcal{A}$ is $1$ in the normalisation which we insist on) is again the same as we encountered many times before, but it has now a different initial condition. The solution is 
\begin{eqnarray}
x_N = \lambda_{2(\!N\!)} = s_N + \sum_{n=1}^{N-1} (-1)^n s_{N-n} \prod_{i=N-n+1}^N t_i - (-1)^N (-k^{2211}) \prod_{i=1}^N t_i,\label{samel}
\end{eqnarray}
having written the vector boundary reflection matrix as
\begin{eqnarray}
K^{\text{LL}}(p,p_{B}) = k^{\alpha\beta\gamma\delta} \, \mathbb{E}_{\alpha\beta} \otimes \mathbb{E}_{\gamma\delta},
\qquad \text{with} \qquad k^{1111}=1 \,\, .
\end{eqnarray}
This means that (from \ref{frome})
\begin{eqnarray}
    k^{2211} =\frac{x_p^- + e^{ip}x_B}{x_p^- - x_B}.
\end{eqnarray}
The initial condition is again ascertained by performing a simple calculation at $N=1$. We have also checked this against the explicit calculation at $N=2$.
\begin{figure}
\centerline{\includegraphics[scale=0.7]{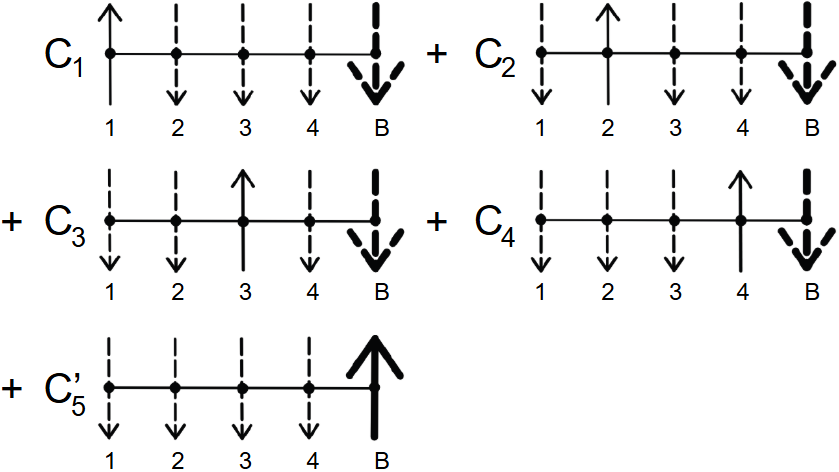}}
\caption{The schematic structure of the eigenstate in the form of a travelling excitation. The disturbance (schematically depicted here as a solid line up arrow in a sea of dashed line down arrows) travels toward the right and when it hits the wall, depicted as a thick dashed arrow, it excites that. This form  of the eigenvectors is observed in the explicit expression, which is however coded in a symbolic computer programme and is not particularly illuminating to report. The $C$'s (and $C'$ for when the boundary is excited) coefficients are certain complicated expressions which are also not particularly instructive.}
\label{fig1}
\end{figure}
See now figure \ref{fig1} for a pictorial representation of the form of the eigenstates in the case of a vector-boundary.

\subsubsection{Fully mixed representations}\label{subsubsec:vector-all-mixed-reps}
We conclude here our analysis of the boundary Bethe ansatz. We started in subsection \ref{subsusbsec:singlet-all-L-reps} by spelling out all the details of the computation and in the subsequent subsections \ref{subsubsec:singlet-auxiliary-L-mixed-physical-reps} and \ref{subsubsec:vector-all-L-reps} we gradually introduced complications, such as possible physical R-representations and a vector boundary, keeping the discussion compact. Here we will be dealing with the most general setup where the auxiliary particle $0$, the physical particles $1,...,N$ and the boundary $B$ are allowed to be in arbitrary representations, and for this reason we will go back to a more detailed style. We will also proceed in a way which is equivalent to the previous discussions, but slightly different in flavour: aiming at complete freedom in the choice of representations which characterise the system, it will be somewhat natural to directly exploit the definition \eqref{vector-monodromy} of double-row monodromy to explicitly construct its coefficients for any number $N$ of physical excitations and use this result to algebraically derive the correct structure of vacua, its eigenvalues and Bethe equations. While necessarily allowing to recover the results of previous sections, this slight change in perspective will provide a more systematic tool, at the price of keeping track of a somewhat heavier notation. We will begin by analysing the structure of pseudo-vacua for $N$ physical particles, successively determining the explicit form of the eigenvalues of the double-row monodromy and finally deriving the Bethe equations and eigenvalues of the transfer matrix \eqref{transfer-matrix-def} via the exchange relations in appendix \ref{app:6vandfake-6v}. The appendix studies the possibility of constructing the exchange relations via arbitrary representations $a$ and $c$ associated to the auxiliary spaces $0$ and $0'$, findings that this is possible when $(ac)$ have associated $R$-matrices of the 6-vertex type, namely $(ac)\in\{ (\text{L}\text{L}),(\text{R}\text{R}),(\tilde{\text{L}}\tilde{\text{L}}),(\tilde{\text{R}}\tilde{\text{R}}),(\text{L}\tilde{\text{R}}),(\text{R}\tilde{\text{L}}),(\tilde{\text{L}}\text{R}),(\tilde{\text{R}}\text{L}) \}$, but is not immediately duable when $R$-matrices with fake-6-vertex type enter the game, namely when $(ac)\in\{ (\text{L}\text{R}),(\text{R}\text{L}),(\text{L}\tilde{\text{L}}),(\text{R}\tilde{\text{R}}),(\tilde{\text{L}}\text{L}),(\tilde{\text{R}}\text{R}),(\tilde{\text{L}}\tilde{\text{R}}),(\tilde{\text{R}}\tilde{\text{L}}) \}$.

\subsubsection*{\ul{\it Structure of $N$-particle pseudo-vacua}}
Aiming at an explicit characterisation of the double-row monodromy matrix, it is natural to start looking at the simplest case of $N=1$, for which \eqref{vector-monodromy} can be written as
\begin{equation}\label{N=1-monodrmy}
[T_{-(\!N\!=\!1\!)}^{ab}(p_{0},p_{B})]_{0B} = \Bigl(t_{-(\!N\!=\!1\!)}^{ab}(p_{0},p_{B}) \Bigr)^{\alpha_{1}\beta_{1}\gamma_{1}\delta_{1}\rho\sigma} \, \mathbb{E}_{\alpha_{1}\beta_{1}}\otimes \mathbb{E}_{\gamma_{1}\delta_{1}}\otimes \mathbb{E}_{\rho\sigma} \,\, ,
\end{equation}
with coefficients
\begin{equation}\label{N=1-mnonodromy-coefficient}
\Bigl(t_{-(\!N\!=\!1\!)}^{ab}(p_{0},p_{B}) \Bigr)^{\alpha_{1}\beta_{1}\gamma_{1}\delta_{1}\rho\sigma} \!\!=\! (-1)^{(\alpha_{1}+\beta_{1})(\gamma_{1}+\lambda_{1})}
\Bigl(r^{d_{1} \, a}_{p_{1}  -p_{0}}\Bigr)^{\gamma_{1}\lambda_{1}\alpha_{1}\alpha_{0}}
\Bigl(r^{a \, d_{1}}_{p_{0} \, p_{1}}\Bigr)^{\beta_{0}\beta_{1}\lambda_{1}\delta_{1}}
\Bigl( k_{p_{0}p_{B}}^{ab} \Bigr)^{\alpha_{0}\beta_{0}\rho\sigma} \,\, .
\end{equation}
Exploiting then the matrix structure encoded in the tensor products of \eqref{N=1-monodrmy} and the explicit coefficients \eqref{N=1-mnonodromy-coefficient}, one can construct the action of the $N=1$ entries of the 
monodromy on a generic state of the form $|\theta_{1}\rangle \otimes |\pi_{B}\rangle$, where $\theta_{1}  \in \{ 1,2\}$ represents the physical state and $\pi_{B} \in \{1,2\}$ a vector boundary state:
\begin{align}\label{N=1-monodromy-action-on-state}
[T_{-(\!N\!=\!1\!)}^{ab}&(p_{0},p_{B})]_{0B}^{\alpha_{1}\beta_{1}}|\theta_{1}\rangle \otimes |\pi_{B}\rangle = 
\\
&= (-1)^{(\alpha_{1}+\beta_{1})(\gamma_{1}+\lambda_{1})+\theta_{1}(\rho+\pi_{B})}\Bigl(r^{d_{1} \, a}_{p_{1}  -p_{0}}\Bigr)^{\gamma_{1}\lambda_{1}\alpha_{1}\alpha_{0}}
\Bigl(r^{a \, d_{1}}_{p_{0} \, p_{1}}\Bigr)^{\beta_{0}\beta_{1}\lambda_{1}\theta_{1}}
\Bigl( k_{p_{0}p_{B}}^{ab} \Bigr)^{\alpha_{0}\beta_{0}\rho\pi_{B}} |\gamma_{1}\rangle \otimes |\rho\rangle \,\, .
\notag
\end{align}
The correct $N=1$ vacuum $|\theta_{1}\rangle \otimes |\pi_{B}\rangle$ can then be determined by using, for all $N$, the identifications 
\begin{equation}\label{monodromy-identifications}
\begin{aligned}
[T_{-(\!N\!)}^{ab}(p_{0},p_{B})]_{0B}^{11}=\mathcal{A}_{(\!N\!)}^{ab}
\qquad & \qquad
[T_{-(\!N\!)}^{ab}(p_{0},p_{B})]_{0B}^{12}=\mathcal{B}_{(\!N\!)}^{ab}
\\
[T_{-(\!N\!)}^{ab}(p_{0},p_{B})]_{0B}^{21}=\mathcal{C}_{(\!N\!)}^{ab}
\qquad & \qquad
[T_{-(\!N\!)}^{ab}(p_{0},p_{B})]_{0B}^{22}=\mathcal{D}_{(\!N\!)}^{ab} \,\, ,
\end{aligned}
\end{equation}
and requiring satisfaction of the standard Bethe ansatz relations
\begin{equation}\label{N=1-Bethe-actions}
\mathcal{A}_{(\!1\!)}^{ab} |\theta_{1}\rangle \otimes |\pi_{B}\rangle = \lambda_{1(\!N\!=\!1\!)}^{ab}|\theta_{1}\rangle \otimes |\pi_{B}\rangle
\qquad 
\mathcal{D}_{(\!1\!)}^{ab} |\theta_{1}\rangle \otimes |\pi_{B}\rangle = \lambda_{2(\!N\!=\!1\!)}^{ab}|\theta_{1}\rangle \otimes |\pi_{B}\rangle
\qquad 
\mathcal{C}_{(\!1\!)}^{ab} |\theta_{1}\rangle \otimes |\pi_{B}\rangle =0 \,\, .
\end{equation}
Unfortunately, spelling out the relations encoded in \eqref{N=1-monodromy-action-on-state} and imposing \eqref{N=1-Bethe-actions} is not yet sufficient to fully determine the correct vacuum, as many non-vanishing and non-eigenstate contributions appear for generic $R$ and $K$. However, having noticed that all such matrices fall in the two main categories \textit{6-vertex} and \textit{fake-6-vertex} discussed in appendix \ref{app:6vandfake-6v}, one can systematically determine the vacuum.
\begin{enumerate}
\item When both $R$ and $K$ are of the 6-vertex type, the correct vacuum is
\begin{equation}
|0\rangle_{1B} := |1\rangle \otimes |1\rangle_{B}
\end{equation}
with eigenvalues given by 
\begin{align}\label{N=1-eigenvalues-case1}
\mathcal{A}_{(\!1\!)}^{ab} |0\rangle_{1B} &= \Bigl( k^{ab}_{p_{0}\, p_{B}} \Bigr)^{1111} \Bigl( r^{a \, d_{1}}_{p_{0} \, p_{1}} \Bigr)^{1111}
\Bigl( r^{d_{1} \, a}_{p_{1} \, -p_{0}} \Bigr)^{1111} |0\rangle_{1B}
\notag \\
\\
\mathcal{D}_{(\!1\!)}^{ab} |0\rangle_{1B} &= \Bigl[ \Bigl( k^{ab}_{p_{0}\, p_{B}} \Bigr)^{1111} \Bigl( r^{a \, d_{1}}_{p_{0} \, p_{1}} \Bigr)^{1221}
\Bigl( r^{d_{1} \, a}_{p_{1} \, -p_{0}} \Bigr)^{1221} +\Bigl( k^{ab}_{p_{0}\, p_{B}} \Bigr)^{2211} \Bigl( r^{a \, d_{1}}_{p_{0} \, p_{1}} \Bigr)^{2211}
\Bigl( r^{d_{1} \, a}_{p_{1} \, -p_{0}} \Bigr)^{1122}\Bigr]|0\rangle_{1B}
\notag
\end{align}
\item When $R$ is 6-vertex and $K$ fake-6-vertex the vacuum is
\begin{equation}
|0\rangle_{1B} := |1\rangle \otimes |2\rangle_{B}
\end{equation}
with eigenvalues given by 
\begin{align}\label{N=1-eigenvalues-case2}
\mathcal{A}_{(\!1\!)}^{ab} |0\rangle_{1B} &= \Bigl( k^{ab}_{p_{0}\, p_{B}} \Bigr)^{1122} \Bigl( r^{a \, d_{1}}_{p_{0} \, p_{1}} \Bigr)^{1111}
\Bigl( r^{d_{1} \, a}_{p_{1} \, -p_{0}} \Bigr)^{1111} |0\rangle_{1B}
\notag \\
\\
\mathcal{D}_{(\!1\!)}^{ab} |0\rangle_{1B} &= \Bigl[ \Bigl( k^{ab}_{p_{0}\, p_{B}} \Bigr)^{1122} \Bigl( r^{a \, d_{1}}_{p_{0} \, p_{1}} \Bigr)^{1221}
\Bigl( r^{d_{1} \, a}_{p_{1} \, -p_{0}} \Bigr)^{1221} +\Bigl( k^{ab}_{p_{0}\, p_{B}} \Bigr)^{2222} \Bigl( r^{a \, d_{1}}_{p_{0} \, p_{1}} \Bigr)^{2211}
\Bigl( r^{d_{1} \, a}_{p_{1} \, -p_{0}} \Bigr)^{1122}\Bigr]|0\rangle_{1B}
\notag
\end{align}
\item When $R$ is fake-6-vertex and $K$ 6-vertex the vacuum is
\begin{equation}
|0\rangle_{1B} := |2\rangle \otimes |1\rangle_{B}
\end{equation}
with eigenvalues given by 
\begin{align}\label{N=1-eigenvalues-case3}
\mathcal{A}_{(\!1\!)}^{ab} |0\rangle_{1B} &= \Bigl( k^{ab}_{p_{0}\, p_{B}} \Bigr)^{1111} \Bigl( r^{a \, d_{1}}_{p_{0} \, p_{1}} \Bigr)^{1122}
\Bigl( r^{d_{1} \, a}_{p_{1} \, -p_{0}} \Bigr)^{2211} |0\rangle_{1B}
\\
\mathcal{D}_{(\!1\!)}^{ab} |0\rangle_{1B} &= \Bigl[ \Bigl( k^{ab}_{p_{0}\, p_{B}} \Bigr)^{1111} \Bigl( r^{a \, d_{1}}_{p_{0} \, p_{1}} \Bigr)^{1212}
\Bigl( r^{d_{1} \, a}_{p_{1} \, -p_{0}} \Bigr)^{2121} +\Bigl( k^{ab}_{p_{0}\, p_{B}} \Bigr)^{2211} \Bigl( r^{a \, d_{1}}_{p_{0} \, p_{1}} \Bigr)^{2222}
\Bigl( r^{d_{1} \, a}_{p_{1} \, -p_{0}} \Bigr)^{2222}\Bigr]|0\rangle_{1B}
\notag 
\end{align}
\item When both $R$ and $K$ are of the fake-6-vertex type the vacuum is
\begin{equation}
|0\rangle_{1B} := |2\rangle \otimes |2\rangle_{B}
\end{equation}
with eigenvalues given by 
\begin{align}\label{N=1-eigenvalues-case4}
\mathcal{A}_{(\!1\!)}^{ab} |0\rangle_{1B} &= \Bigl( k^{ab}_{p_{0}\, p_{B}} \Bigr)^{1122} \Bigl( r^{a \, d_{1}}_{p_{0} \, p_{1}} \Bigr)^{1122}
\Bigl( r^{d_{1} \, a}_{p_{1} \, -p_{0}} \Bigr)^{2211} |0\rangle_{1B}
\notag \\
\\
\mathcal{D}_{(\!1\!)}^{ab} |0\rangle_{1B} &= \Bigl[ \Bigl( k^{ab}_{p_{0}\, p_{B}} \Bigr)^{1122} \Bigl( r^{a \, d_{1}}_{p_{0} \, p_{1}} \Bigr)^{1212}
\Bigl( r^{d_{1} \, a}_{p_{1} \, -p_{0}} \Bigr)^{2121} +\Bigl( k^{ab}_{p_{0}\, p_{B}} \Bigr)^{2222} \Bigl( r^{a \, d_{1}}_{p_{0} \, p_{1}} \Bigr)^{2222}
\Bigl( r^{d_{1} \, a}_{p_{1} \, -p_{0}} \Bigr)^{2222}\Bigr]|0\rangle_{1B}
\notag
\end{align}
\end{enumerate}
Given the above results, we would like to determine the generic vacuum structure of an $N+1$ particle state by using a similar approach and to this aim it is not hard to explicitly compute the monodromy \eqref{vector-monodromy} for $N=1,2,3$, finding the following pattern\footnote{For the sake of clarity we should stress that in the expansion \eqref{N-particle-expansion-T-}, the label $N$ in $\alpha_{N},\beta_{N}$ refers to the fact that we are looking at the $N$-particles monodromy. This notation is exploited at any $N$, hence also for the case of $N=1$ in \eqref{N=1-monodrmy}, and helps in finding the recursion \eqref{monodromy-on-N+1-particle-states}. The matrices acting on physical spaces are on the other hand always denoted by indices $\gamma_{1}\delta_{1}$...$\gamma_{N}\delta_{N}$, while $\rho\sigma$ are used for the action on the boundary state.}
\begin{equation}\label{N-particle-expansion-T-}
[T_{-(\!N\!)}^{ab}(p_{0},p_{B})]_{0B} = \Bigl(t_{-(\!N\!)}^{ab}(p_{0},p_{B}) \Bigr)^{\alpha_{N}\beta_{N}\gamma_{1}\delta_{1}...\gamma_{N}\delta_{N}
\rho\sigma} \, \mathbb{E}_{\alpha_{N}\beta_{N}}\otimes \mathbb{E}_{\gamma_{1}\delta_{1}} \otimes ... \otimes \mathbb{E}_{\gamma_{N}\delta_{N}}\otimes \mathbb{E}_{\rho\sigma} \,\, ,
\end{equation}
with coefficients
\begin{equation}
\begin{aligned}
\Bigl(t_{-(\!N\!)}^{ab}&(p_{0},p_{B}) \Bigr)^{\alpha_{N}\beta_{N}\gamma_{1}\delta_{1}...\gamma_{N}\delta_{N}\rho\sigma} = 
\\
= &(-1)^{(\alpha_{N}+\beta_{N})(\gamma_{N}+\lambda_{N})+(\gamma_{N}+\delta_{N})(1+\rho+\sigma+\alpha_{N}+\beta_{N})}
\\
& 
\Bigl(r^{d_{N} \, a}_{p_{N}  -p_{0}}\Bigr)^{\gamma_{N}\lambda_{N}\alpha_{N}\alpha_{N-1}}
\Bigl(r^{a \, d_{N}}_{p_{0} \, p_{N}}\Bigr)^{\beta_{N-1}\beta_{N}\lambda_{N}\delta_{N}}
\Bigl(t_{-(\!N-1\!)}^{ab}(p_{0},p_{B}) \Bigr)^{\alpha_{N-1}\beta_{N-1}\gamma_{1}\delta_{1}...\gamma_{N-1}\delta_{N-1}\rho\sigma} \,\, .
\end{aligned}
\end{equation}
The general action on a generic $N+1$ particle state can then be computed, in an analogy with $N=1$,
\begin{equation}\label{monodromy-on-N+1-particle-states}
\begin{aligned}
[T_{-(\!N+1\!)}^{ab}&(p_{0},p_{B})]_{0B}^{\alpha_{N+1}\beta_{N+1}}  |\theta_{1}\rangle \otimes ... \otimes |\theta_{N}\rangle \otimes |\theta_{N+1}\rangle \otimes |\pi_{B}\rangle =
\\
&(-1)^{(\alpha_{N+1}+\beta_{N+1})(\gamma_{N+1}+\lambda_{N+1})+\theta_{N+1}(\alpha_{N}+\beta_{N}+\sum_{i=1}^{N}\theta_{i})} 
\\
&\Bigl( r^{d_{N+1} \, a}_{p_{N+1} \, -p_{0}} \Bigr)^{\gamma_{N+1}\lambda_{N+1}\alpha_{N+1}\alpha_{N}}
\Bigl( r^{a \, d_{N+1}}_{p_{0} \, p_{N+1}} \Bigr)^{\beta_{N}\beta_{N+1}\lambda_{N+1}\theta_{N+1}}
\\
& P_{s}^{(N+1)N}...P_{s}^{21}\Biggl( |\gamma_{N+1}\rangle \otimes \Bigl( [T_{-(\!N\!)}^{ab}(p_{0},p_{B})]_{0B}^{\alpha_{N}\beta_{N}} |\theta_{1}\rangle \otimes ... \otimes |\theta_{N}\rangle \otimes |\pi_{B}\rangle \Bigr) \Biggr) \,\, ,
\end{aligned}
\end{equation}
after having introduced permutation operators
\begin{equation}
P_{s}\!:=\!(-1)^{\nu} \mathbb{E}_{\mu\nu}\otimes\mathbb{E}_{\nu\mu}
\quad \text{acting as} \quad
P_{s}(|\alpha\rangle \otimes |\beta\rangle)\! =\! (-1)^{\alpha+\beta} |\beta\rangle \otimes |\alpha \rangle
\quad \text{and satisfying} \quad
P_{s}\circ P_{s} \!=\! \mathds{1} \,\, .
\end{equation}
These enjoy an obvious extension to the case of $N$ physical spaces with a boundary
\begin{equation}
P_{s}^{(i+1)i}  := (-1)^{\nu} \underbrace{\mathds{1} \otimes... \otimes\mathds{1}}_{\text{spaces $1$ to $i-1$}}\otimes \underbrace{\mathbb{E}_{\mu\nu}\otimes\mathbb{E}_{\nu\mu}}_{\text{spaces $i$ and $i+1$}} \otimes \underbrace{\mathds{1}\otimes ...\otimes \mathds{1}\otimes \mathds{1}_{B}}_{\text{spaces $i+2$ to $N$ and boundary} }  \,\, ,
\end{equation}
in particular allowing to derive the relations
\begin{align}
&\mathbb{E}_{\gamma_{1}\delta_{1}} \otimes ... \otimes \mathbb{E}_{\gamma_{N}\delta_{N}}\otimes \mathbb{E}_{\rho\sigma} =
\notag \\
&=(-1)^{(\gamma_{N}+\delta_{N})\sum_{i=1}^{N-1}(\gamma_{i}+\delta_{i})}P_{s}^{N(N-1)}...P_{s}^{21}\Bigl( \mathbb{E}_{\gamma_{N}\delta_{N}}\otimes\mathbb{E}_{\gamma_{1}\delta_{1}} \otimes ... \otimes\mathbb{E}_{\gamma_{N-1}\delta_{N-1}}\otimes \mathbb{E}_{\rho\sigma}\Bigr)P_{s}^{21}...P_{s}^{N(N-1)}
\notag \\
\\
&P_{s}^{21}...P_{s}^{N(N-1)}\Bigl( |\theta_{1}\rangle \otimes ...\otimes |\theta_{N}\rangle \otimes |\pi_{B}\rangle \Bigr) = (-1)^{\theta_{N}\sum_{i=1}^{N-1}\theta_{i} }\Bigl( |\theta_{N}\rangle \otimes |\theta_{1}\rangle \otimes... \otimes |\theta_{N-1}\rangle \otimes |\pi_{B}\rangle \Bigr) \,\, ,
\notag 
\end{align}
needed to obtain the right hand side of \eqref{monodromy-on-N+1-particle-states}. 
At this point, since the action on a $N+1$ particle state can be written in terms of the action on a $N$-particle state, we can study the structure of the $N+1$ vacuum by using once again the identifications \eqref{monodromy-identifications} and making the induction hypothesis
\begin{equation}\label{induction-assumption}
\mathcal{A}_{(\!N\!)}^{ab} |0\rangle_{NB} = \lambda_{1(\!N\!)}^{ab} |0\rangle _{NB}
\qquad \qquad
\mathcal{D}_{(\!N\!)}^{ab} |0\rangle_{NB} = \lambda_{2(\!N\!)}^{ab} |0\rangle _{NB}
\qquad \qquad
\mathcal{C}_{(\!N\!)}^{ab} |0\rangle_{NB} = 0 \,\, ,
\end{equation}
where to shorten the notation we have denoted the $N$-particle vacuum by
\begin{equation}
|0\rangle_{NB}:=|\theta_{1}\rangle \otimes ... \otimes |\theta_{N}\rangle \otimes |\pi_{B}\rangle \,\, ,
\end{equation}
for an unknown set of single particle states $\theta_{1},...,\theta_{N}$ and boundary state $\pi_{B}$ to be determined.

Spelling out the equations  in \eqref{monodromy-on-N+1-particle-states} one finds that the action of $\mathcal{A}_{(\!N+1\!)}^{ab},\mathcal{B}_{(\!N+1\!)}^{ab},\mathcal{C}_{(\!N+1\!)}^{ab},\mathcal{D}_{(\!N+1\!)}^{ab}$ contains all contributions from $\mathcal{A}_{(\!N\!)}^{ab},\mathcal{B}_{(\!N\!)}^{ab},\mathcal{C}_{(\!N\!)}^{ab},\mathcal{D}_{(\!N\!)}^{ab}$ and using the induction assumption \eqref{induction-assumption} is not sufficient to find the correct vacuum. However, noting that the $N+1$ particle action is proportional to the $R$-matrix which acts on spaces $0$ and $N+1$, one can again systematically find out the correct vacuum structure by distinguishing between the two main categories of $R$-matrices in the game.
\begin{itemize}
\item For $R^{a d_{N+1}}(p_{0},p_{N+1})$ of the 6-vertex type the correct vacuum is
\begin{equation}
\mathcal{C}_{(\!N+1\!)}^{ab} |\theta_{1}\rangle \otimes...\otimes|\theta_{N-1}\rangle \otimes |1\rangle \otimes |\pi_{B}\rangle  :=\mathcal{C}_{(\!N+1\!)}^{ab}|0\rangle_{N+1}= 0
\end{equation}
with eigenvalues
\begin{align}\label{6v-eigenvalues}
\mathcal{A}_{(\!N+1\!)}^{ab} |0\rangle _{N+1} &= \lambda_{1(\!N\!)}^{ab} \Bigl( r^{a \, d_{N+1}}_{p_{0} \, p_{N+1}} \Bigr)^{1111}  \Bigl( r^{d_{N+1} \, a}_{p_{N+1} \, -p_{0}} \Bigr)^{1111} |0\rangle _{N+1}
\\
\mathcal{D}_{(\!N+1\!)}^{ab} |0\rangle _{N+1} & = \Bigl[ \lambda_{1(\!N\!)}^{ab}\Bigl( r^{a \, d_{N+1}}_{p_{0} \, p_{N+1}} \Bigr)^{1221}  \Bigl( r^{d_{N+1} \, a}_{p_{N+1} \, -p_{0}} \Bigr)^{1221}+\lambda_{2(\!N\!)}^{ab}\Bigl( r^{a \, d_{N+1}}_{p_{0} \, p_{N+1}} \Bigr)^{2211}  \Bigl( r^{d_{N+1} \, a}_{p_{N+1} \, -p_{0}} \Bigr)^{1122} \Bigr] |0\rangle _{N+1}
\notag
\end{align}
\item For $R^{a d_{N+1}}(p_{0},p_{N+1})$ of the fake-6-vertex type the correct vacuum is
\begin{equation}
\mathcal{C}_{(\!N+1\!)}^{ab} |\theta_{1}\rangle \otimes...\otimes|\theta_{N-1}\rangle \otimes |2\rangle \otimes |\pi_{B}\rangle  :=\mathcal{C}_{(\!N+1\!)}^{ab}|0\rangle_{N+1}= 0
\end{equation}
with eigenvalues
\begin{align}\label{fake-6v-eigenvalues}
\mathcal{A}_{(\!N+1\!)}^{ab} |0\rangle _{N+1} &= \lambda_{(\!N\!)}^{ab} \Bigl( r^{a \, d_{N+1}}_{p_{0} \, p_{N+1}} \Bigr)^{1122}  \Bigl( r^{d_{N+1} \, a}_{p_{N+1} \, -p_{0}} \Bigr)^{2211} |0\rangle _{N+1}
\\
\mathcal{D}_{(\!N+1\!)}^{ab}|0\rangle _{N+1} & = \Bigl[ \lambda_{1(\!N\!)}^{ab}\Bigl( r^{a \, d_{N+1}}_{p_{0} \, p_{N+1}} \Bigr)^{1212}  \Bigl( r^{d_{N+1} \, a}_{p_{N+1} \, -p_{0}} \Bigr)^{2121}+\lambda_{2(\!N\!)}^{ab}\Bigl( r^{a \, d_{N+1}}_{p_{0} \, p_{N+1}} \Bigr)^{2222}  \Bigl( r^{d_{N+1} \, a}_{p_{N+1} \, -p_{0}} \Bigr)^{2222} \Bigr] |0\rangle _{N+1}
\notag
\end{align}
\end{itemize}
The structure of the generic $N+1$ particle vacuum is morally in agreement with the one found for the single particle case and we can thus summarise our findings by establishing the following pattern: the vacuum is fully determined by the \textit{6-vertex} vs \textit{fake-6-vertex} nature of the $R$ and $K$ matrices appearing in \eqref{vector-monodromy}, which respectively act on the auxiliary space $0$ plus one of the physical spaces $1,...N$ and the auxiliary space $0$ plus the boundary. In particular, whenever one of these matrices is 6-vertex the physical/boundary state on which they act must be a boson, while whenever either of them is fake-6-vertex, the corresponding physical/boundary state must be a fermion. This implies that given a string of representations $(a,d_{1},...,d_{N},b)$, the composition of the vacuum is fully determined by looking at the auxiliary space representation $a$ and noting whether the couples $(a,d_{i})$ and $(a,b)$ fall in the 6-vertex or fake-6-vertex case. As we also recalled at the beginning of this subsection, in appendix \ref{app:6vandfake-6v} we noticed that $R$ and $K$ matrices of the 6-vertex type carry representations $\{ (\text{L}\text{L}),(\text{R}\text{R}),(\tilde{\text{L}}\tilde{\text{L}}),(\tilde{\text{R}}\tilde{\text{R}}),(\text{L}\tilde{\text{R}}),(\text{R}\tilde{\text{L}}),(\tilde{\text{L}}\text{R}),(\tilde{\text{R}}\text{L}) \}$, while those of the fake-6-vertex type carry representations $\{ (\text{L}\text{R}),(\text{R}\text{L}),(\text{L}\tilde{\text{L}}),(\text{R}\tilde{\text{R}}),(\tilde{\text{L}}\text{L}),(\tilde{\text{R}}\text{R}),(\tilde{\text{L}}\tilde{\text{R}}),(\tilde{\text{R}}\tilde{\text{L}}) \}$.

\subsubsection*{\ul{\it Eigenvalues of the double-row monodromy}}
Having established the existence of a pseudo-vacuum state for each choice of representation in the auxiliary, physical and boundary spaces, and having fully characterised its structure, we are now in the position to explicitly construct the eigenvalues $\lambda_{1(\!N\!)}^{ab}$ and $\lambda_{2(\!N\!)}^{ab}$ on any $N$-particle pseudo-vacuum. Looking back at the expressions \eqref{6v-eigenvalues} and \eqref{fake-6v-eigenvalues} for the eigenvalues of $\mathcal{A}_{(\!N+1\!)}^{ab}$ and $\mathcal{D}_{(\!N+1\!)}^{ab}$ in cases where the $(N+1)$-th particle respectively involves 6-vertex and fake-6-vertex $R$-matrices, we can immediately notice the following recursive structures
\begin{equation}\label{recursion-1}
\lambda_{1(\!N+1\!)}^{ab}=\alpha_{p_{0} p_{N+1}}^{ad_{N+1}} \lambda_{1(\!N\!)}^{ab}
\qquad \text{and} \qquad
\lambda_{2(\!N+1\!)}^{ab}=\beta_{p_{0} p_{N+1}}^{ad_{N+1}} \lambda_{1(\!N\!)}^{ab}+\delta_{p_{0} p_{N+1}}^{ad_{N+1}} \lambda_{2(\!N\!)}^{ab}
\end{equation}
with
\begin{equation}
\alpha_{p_{0} p_{N+1}}^{ad_{N+1}} = 
\begin{cases}
\Bigl( r^{a \, d_{N+1}}_{p_{0} \, p_{N+1}} \Bigr)^{1111}  \Bigl( r^{d_{N+1} \, a}_{p_{N+1} \, -p_{0}} \Bigr)^{1111}
\quad \quad \text{for} \quad \quad R^{ad_{N+1}}(p_{0},p_{N+1}) \in \text{6-vertex}
\\
\\
\Bigl( r^{a \, d_{N+1}}_{p_{0} \, p_{N+1}} \Bigr)^{1122}  \Bigl( r^{d_{N+1} \, a}_{p_{N+1} \, -p_{0}} \Bigr)^{2211}
\quad \quad \text{for} \quad \quad R^{ad_{N+1}}(p_{0},p_{N+1}) \in \text{fake-6-vertex}
\end{cases}
\end{equation}
\begin{equation}
\beta_{p_{0} p_{N+1}}^{ad_{N+1}} = 
\begin{cases}
\Bigl( r^{a \, d_{N+1}}_{p_{0} \, p_{N+1}} \Bigr)^{1221}  \Bigl( r^{d_{N+1} \, a}_{p_{N+1} \, -p_{0}} \Bigr)^{1221}
\quad \quad \text{for} \quad \quad R^{ad_{N+1}}(p_{0},p_{N+1}) \in \text{6-vertex}
\\
\\
\Bigl( r^{a \, d_{N+1}}_{p_{0} \, p_{N+1}} \Bigr)^{1212}  \Bigl( r^{d_{N+1} \, a}_{p_{N+1} \, -p_{0}} \Bigr)^{2121}
\quad \quad \text{for} \quad \quad R^{ad_{N+1}}(p_{0},p_{N+1}) \in \text{fake-6-vertex}
\end{cases}
\end{equation}
\begin{equation}
\delta_{p_{0} p_{N+1}}^{ad_{N+1}} = 
\begin{cases}
\Bigl( r^{a \, d_{N+1}}_{p_{0} \, p_{N+1}} \Bigr)^{2211}  \Bigl( r^{d_{N+1} \, a}_{p_{N+1} \, -p_{0}} \Bigr)^{1122}
\quad \quad \text{for} \quad \quad R^{ad_{N+1}}(p_{0},p_{N+1}) \in \text{6-vertex}
\\
\\
\Bigl( r^{a \, d_{N+1}}_{p_{0} \, p_{N+1}} \Bigr)^{2222}  \Bigl( r^{d_{N+1} \, a}_{p_{N+1} \, -p_{0}} \Bigr)^{2222}
\quad \quad \text{for} \quad \quad R^{ad_{N+1}}(p_{0},p_{N+1}) \in \text{fake-6-vertex}
\end{cases}
\end{equation}
The above relations are true for any $N\geq 1$, but must also be supplemented with some stopping conditions, for the recursion to eventually terminate, which can be determined from the eigenvalues at $N=1$. Looking at \eqref{N=1-eigenvalues-case1}, \eqref{N=1-eigenvalues-case2}, \eqref{N=1-eigenvalues-case3}, \eqref{N=1-eigenvalues-case4} one indeed finds that
\begin{equation}\label{recursion-2}
\lambda_{1(\!N\!=\!1\!)}^{ab}=\kappa^{ab} \alpha_{p_{0}p_{1}}^{ad_{1}} 
\qquad \text{and} \qquad
\lambda_{2(\!N\!=\!1\!)}^{ab}=\kappa^{ab}\beta_{p_{0}p_{1}}^{ad_{1}}+\omega^{ab}\delta_{p_{0}p_{1}}^{ad_{1}}
\end{equation}
with
\begin{equation}
\kappa^{ab} = 
\begin{cases}
\Bigl( k^{ab}_{p_{0} \, p_{B}} \Bigr)^{1111}  
\quad \quad \text{for} \quad \quad K^{ab}(p_{0},p_{B}) \in \text{6-vertex}
\\
\\
\Bigl( k^{ab}_{p_{0} \, p_{B}} \Bigr)^{1122} 
\quad \quad \text{for} \quad \quad K^{ab}(p_{0},p_{B}) \in \text{fake-6-vertex}
\end{cases}
\end{equation}
\begin{equation}
\omega^{ab} = 
\begin{cases}
\Bigl( k^{ab}_{p_{0} \, p_{B}} \Bigr)^{2211}  
\quad \quad \text{for} \quad \quad K^{ab}(p_{0},p_{B}) \in \text{6-vertex}
\\
\\
\Bigl( k^{ab}_{p_{0} \, p_{B}} \Bigr)^{2222} 
\quad \quad \text{for} \quad \quad K^{ab}(p_{0},p_{B}) \in \text{fake-6-vertex}
\end{cases}
\end{equation}
Altogether, combining \eqref{recursion-1} and \eqref{recursion-2}, one finally obtains the following eigenvalues\footnote{The summations and products are understood to vanish for values of $N$ leading to negative ranges of the indices.}
\begin{align}\label{general-eigenvalues}
\lambda_{1(\!N+1\!)}^{ab}&\!=\!\kappa^{ab}\!\prod_{i=1}^{N+1}\alpha_{p_{0}p_{i}}^{ad_{i}}
\notag\\
\\
\lambda_{2(\!N+1\!)}^{ab}&\!=\!\omega^{ab}\!\prod_{i=1}^{N+1}\delta_{p_{0}p_{i}}^{ad_{i}}+\kappa^{ab}\Bigl[ \beta_{p_{0}p_{N+1}}^{ad_{N+1}}\prod_{j=1}^{N}\alpha_{p_{0}p_{j}}^{ad_{j}}+\sum_{i=1}^{N-1}\beta_{p_{0}p_{i+1}}^{ad_{i+1}}\Bigl(\prod_{j=1}^{i}\alpha_{p_{0}p_{j}}^{ad_{j}}\Bigr)\Bigl(\prod_{k=i+2}^{N+1}\delta_{p_{0}p_{k}}^{ad_{k}} \Bigr)+\beta_{p_{0}p_{1}}^{ad_{1}}\prod_{j=2}^{N+1}\delta_{p_{0}p_{j}}^{ad_{j}}\Bigr]  .
\notag
\end{align}
As one should reasonably expect, these expressions include and generalise those obtained in the previous sections for either the vector or singlet case. For example, the expression \eqref{samel} is easily recovered by noting that for $\text{L}$ representations only, the factors in \eqref{general-eigenvalues} become
\begin{equation}
\kappa^{\text{LL}}\!=\!1
\quad , \quad 
\omega^{\text{LL}}\!=\!\Bigl( k^{LL}_{p_{0} \, p_{B}} \Bigr)^{2211}
\quad , \quad 
\alpha_{p_{0}p_{i}}^{\text{LL}}\!=\!1 
\quad , \quad
\beta_{p_{0}p_{i}}^{\text{LL}}\!=\!E^{\text{LL}}_{p_{0}p_{i}}E^{\text{LL}}_{p_{i} \, -p_{0}}
\quad , \quad 
\delta_{p_{0}p_{i}}^{\text{LL}}\!=\!D^{\text{LL}}_{p_{0}p_{i}}B^{\text{LL}}_{p_{i} \, -p_{0}} \,\, .
\notag
\end{equation}
Similarly, the result \eqref{same} for the singlet case in the presence of $\text{L}$ representations is recovered by trivialising the information associated to the boundary, which corresponds to dropping the index $b$ as
\begin{equation}
\kappa^{ab} \rightarrow
\kappa^{a}\equiv \Bigl(k^{a}_{p_0} \Bigr)^{11} = 1 \qquad \forall a\in \{{\text{L},\text{R},\tilde{\text{L}},\tilde{\text{R}}}\} 
\qquad \text{and}
\qquad
\omega^{ab} \rightarrow
\omega^{a}\equiv \Bigl(k^{a}_{p_0} \Bigr)^{22}=g^{a}(p_{0}) \,\, .
\end{equation}
In fact equation \eqref{same} is precisely recovered by also noting the following mapping 
\begin{equation}
\beta_{p_{0}p_{i}}^{\text{LL}}\!=\!E^{\text{LL}}_{p_{0}p_{i}}E^{\text{LL}}_{p_{i} \, -p_{0}}\equiv s_{i}
\qquad , \qquad 
\delta_{p_{0}p_{i}}^{\text{LL}}\!=\!D^{\text{LL}}_{p_{0}p_{i}}B^{\text{LL}}_{p_{i} \, -p_{0}}\equiv -t_{i} \,\, .
\end{equation}

\subsubsection*{\ul{\it Eigenvalues of the transfer matrix and Bethe equations}}
Having established the structure of pseudo-vacua and determined the eigenvalues of the double-row monodromy $T_{-(\!N\!)}^{ab}$ for any $N$, we can finally conclude by deriving the eigenvalues of the transfer matrix and the Bethe equations. We begin by recalling the transfer matrix \eqref{transfer-matrix-def}
\begin{equation}
\tau^{ab}(p_{0},p_{B}):=\text{str}_{0}\Bigl([T_{+}^{ab}(p_{0},p_{B})]_{0B} [T_{-}^{ab}(p_{0},p_{B})]_{0B} \Bigr)=\mathcal{A}_{(\!N\!)}^{ab}(p_{0},p_{B})-g^{a}_{D}(p_{0})\mathcal{D}_{(\!N\!)}^{ab}(p_{0},p_{B}) \,\, ,
\end{equation}
which as discussed above, even in the case of a vector boundary, must exhibit a singlet-like dual monodromy with \eqref{singlet-dual-K-matrices}. The next step in the Bethe ansatz is the action with the transfer matrix on $M$-magnon states built out of a $N$-particle vacuum and the extraction of the eigenvalue part after cancellation of unwanted terms, which give rise to the Bethe equations. Starting from a 1-magnon state, one finds
\begin{equation}
\begin{aligned}
\tau^{ab}&(p_{0},p_{B})\mathcal{B}_{(\!N\!)}^{ab}(q_{1},p_{B}) |0\rangle_{NB} = 
\\
& =\mathcal{A}_{(\!N\!)}^{ab}(p_{0},p_{B})\mathcal{B}_{(\!N\!)}^{ab}(q_{1},p_{B}) |0\rangle_{NB}-g_{D}^{a}(p_{0})\mathcal{D}_{(\!N\!)}^{ab}(p_{0},p_{B})\mathcal{B}_{(\!N\!)}^{ab}(q_{1},p_{B}) |0\rangle_{NB}=
\\
&=\tilde{\varphi}^{aa}(p_{0},q_{1})\Bigl[ \lambda_{1(\!N\!)}^{ab}(p_{0};p_{1},...,p_{N})-g_{D}^{a}(p_{0})\lambda_{2(\!N\!)}^{ab}(p_{0};p_{1},...,p_{N}) \Bigr]\mathcal{B}_{(\!N\!)}^{ab}(q_{1},p_{B}) |0\rangle_{NB}+
\\
& \,\,\,\,\,\, +\Bigl[ \lambda_{1(\!N\!)}^{ab}(q_{1};p_{1},...,p_{N})\Bigl( \hat{\chi}^{aa}(p_{0},q_{1})-g_{D}^{a}(p_{0})\tilde{\psi}^{aa}(p_{0},q_{1}) \Bigr) +
\\
& \,\,\,\,\,\,\,\,\, +\lambda_{2(\!N\!)}^{ab}(q_{1};p_{1},...,p_{N})\Bigl( \hat{\psi}^{aa}(p_{0},q_{1})-g_{D}^{a}(p_{0})\tilde{\chi}^{aa}(p_{0},q_{1}) \Bigr) 
\Bigr]\mathcal{B}_{(\!N\!)}^{ab}(p_{0},p_{B}) |0\rangle_{NB} \,\, ,
\end{aligned}
\end{equation}
after having exploited the exchange relations in appendix \ref{app:genuine-6-vertex-subsection}, adapted to the case with one single auxiliary space $0$, in which the representation indices $a$ and $c$ coincide, namely $a=c$. The first term on the right hand side of the above expression represents the eigenvalue of the transfer matrix, while the second bracket represents unwanted terms and should hence vanish. Using the explicit expressions for the functions $\hat{\chi},\hat{\psi},\tilde{\chi},\tilde{\psi}$ in appendix \ref{app:genuine-6-vertex-subsection}, it is not hard to realise that the dependence on the auxiliary momentum $p_{0}$ correctly factorises and the vanishing of the large square bracket reduces to the condition
\begin{equation}
\begin{aligned}
-\lambda_{1(\!N\!)}^{ab}(q_{1};p_{1},...,p_{N})[s_{D}^{a}+x_{q_{1}}^{+}]+\lambda_{2(\!N\!)}^{ab}(q_{1};p_{1},...,p_{N})[s_{D}^{a}-x_{q_{1}}^{-}] &\equiv 0 \qquad \text{for} \qquad a=\text{L},\text{R}
\\
-\lambda_{1(\!N\!)}^{ab}(q_{1};p_{1},...,p_{N})[s_{D}^{a}-x_{q_{1}}^{-}]+\lambda_{2(\!N\!)}^{ab}(q_{1};p_{1},...,p_{N})[s_{D}^{a}+x_{q_{1}}^{+}] &\equiv 0 \qquad \text{for} \qquad a=\tilde{\text{L}},\tilde{\text{R}}
\end{aligned}
\end{equation}
which can also be combined into the single expression
\begin{equation}\label{first-Bethe-eq}
g_{D}^{a}(q_{1})\lambda_{2(\!N\!)}^{ab}(q_{1};p_{1},...,p_{N})=\lambda_{1(\!N\!)}^{ab}(q_{1};p_{1},...,p_{N}) 
\quad \text{for} \quad a,b\in\{\text{L},\text{R},\tilde{\text{L}},\tilde{\text{R}}\} \,\, ,
\end{equation}
with the functions $g_{D}^{a}$ given in \eqref{singlet-dual-K-matrices}. If \eqref{first-Bethe-eq} is satisfied, then the 1-magnon state is an eigenstate of the transfer matrix
\begin{equation}
\begin{aligned}
\tau^{ab}&(p_{0},p_{B})\mathcal{B}_{(\!N\!)}^{ab}(q_{1},p_{B}) |0\rangle_{NB} 
=
\\
& =\tilde{\varphi}^{aa}(p_{0},q_{1})\Bigl[ \lambda_{1(\!N\!)}^{ab}(p_{0};p_{1},...,p_{N})-g_{D}^{a}(p_{0})\lambda_{2(\!N\!)}^{ab}(p_{0};p_{1},...,p_{N}) \Bigr]\mathcal{B}_{(\!N\!)}^{ab}(q_{1},p_{B}) |0\rangle_{NB} \,\, .
\end{aligned}
\end{equation}
At this point one needs to proceed with the 2-magnon case: the computation is a bit more involved, but in the end one obtains two Bethe equations, the first being exactly as \eqref{first-Bethe-eq} and the second one reading
\begin{align}
0&\equiv  \lambda_{1(\!N\!)}^{ab}(q_{2};p_{1},...,p_{N}) \Bigl( \varphi^{aa}(q_{1},p_{0})\tilde{\varphi}^{aa}(p_{0},q_{1})[\hat{\chi}^{aa}(p_{0},q_{2})-g_{D}^{a}(p_{0})\tilde{\psi}^{aa}(p_{0},q_{2})]+
\notag \\
& \quad \quad \qquad \qquad \qquad \qquad \,\,\,\,\, +\hat{\chi}^{aa}(q_{1},q_{2})[\hat{\chi}^{aa}(p_{0},q_{1})-g_{D}^{a}(p_{0})\tilde{\psi}^{aa}(p_{0},q_{1})]+
\notag \\
&  \quad \quad \qquad \qquad \qquad \qquad \,\,\,\,\, +\tilde{\psi}^{aa}(q_{1},q_{2})[\hat{\psi}^{aa}(p_{0},q_{1})-g_{D}^{a}(p_{0})\tilde{\chi}^{aa}(p_{0},q_{1})]\Bigr)+
\notag \\
& + \lambda_{2(\!N\!)}^{ab}(q_{2};p_{1},...,p_{N}) \Bigl( \varphi^{aa}(q_{1},p_{0})\tilde{\varphi}^{aa}(p_{0},q_{1})[\hat{\psi}^{aa}(p_{0},q_{2})-g_{D}^{a}(p_{0})\tilde{\chi}^{aa}(p_{0},q_{2})]+
\notag \\
& \quad \quad \qquad \qquad \qquad \qquad \,\,\,\,\, +\hat{\psi}^{aa}(q_{1},q_{2})[\hat{\chi}^{aa}(p_{0},q_{1})-g_{D}^{a}(p_{0})\tilde{\psi}^{aa}(p_{0},q_{1})]+
\notag \\
&  \quad \quad \qquad \qquad \qquad \qquad \,\,\,\,\, +\tilde{\chi}^{aa}(q_{1},q_{2})[\hat{\psi}^{aa}(p_{0},q_{1})-g_{D}^{a}(p_{0})\tilde{\chi}^{aa}(p_{0},q_{1})]\Bigr) \,\, .
\end{align}
Using the explicit expressions in \ref{app:genuine-6-vertex-subsection} one can however recover again a factorised dependence on $p_{0}$, leaving 
\begin{equation}
\begin{aligned}
-\lambda_{1(\!N\!)}^{ab}(q_{1};p_{1},...,p_{N})[s_{D}^{a}+x_{q_{2}}^{+}]+\lambda_{2(\!N\!)}^{ab}(q_{2};p_{1},...,p_{N})[s_{D}^{a}-x_{q_{2}}^{-}] &\equiv 0 \qquad \text{for} \qquad a=\text{L},\text{R}
\\
-\lambda_{1(\!N\!)}^{ab}(q_{2};p_{1},...,p_{N})[s_{D}^{a}-x_{q_{2}}^{-}]+\lambda_{2(\!N\!)}^{ab}(q_{2};p_{1},...,p_{N})[s_{D}^{a}+x_{q_{2}}^{+}] &\equiv 0 \qquad \text{for} \qquad a=\tilde{\text{L}},\tilde{\text{R}}
\end{aligned}
\end{equation}
which can be combined into the single expression
\begin{equation}\label{second-Bethe-eq}
g_{D}^{a}(q_{2})\lambda_{2(\!N\!)}^{ab}(q_{2};p_{1},...,p_{N})=\lambda_{1(\!N\!)}^{ab}(q_{2};p_{1},...,p_{N}) 
\quad \text{for} \quad a,b\in\{\text{L},\text{R},\tilde{\text{L}},\tilde{\text{R}}\} \,\, ,
\end{equation}
finally leading to the 2-magnon eigenstate
\begin{align}
&\tau^{ab}(p_{0},p_{B})\mathcal{B}_{(\!N\!)}^{ab}(q_{1},p_{B})\mathcal{B}_{(\!N\!)}^{ab}(q_{2},p_{B}) |0\rangle_{NB}=
\\
&\tilde{\varphi}^{aa}(p_{0},q_{1})\tilde{\varphi}^{aa}(p_{0},q_{1})\Bigl[ \lambda_{1(\!N\!)}^{ab}(p_{0};p_{1},...,p_{N})-g_{D}^{a}(p_{0})\lambda_{2(\!N\!)}^{ab}(p_{0};p_{1},...,p_{N}) \Bigr]\mathcal{B}_{(\!N\!)}^{ab}(q_{1},p_{B})\mathcal{B}_{(\!N\!)}^{ab}(q_{2},p_{B}) |0\rangle_{NB} \,\, .
\notag
\end{align}
In the spirit of the previous sections we can then conclude by writing down, for generic $M$-magnon states, the eigenvalue of the transfer matrix
\begin{align}
&\tau^{ab}(p_{0},p_{B})\mathcal{B}_{(\!N\!)}^{ab}(q_{1},p_{B})...\mathcal{B}_{(\!N\!)}^{ab}(q_{M},p_{B}) |0\rangle_{NB}=
\\
&\prod_{i=1}^{M}\tilde{\varphi}^{aa}(p_{0},q_{i})\Bigl[ \lambda_{1(\!N\!)}^{ab}(p_{0};p_{1},...,p_{N})-g_{D}^{a}(p_{0})\lambda_{2(\!N\!)}^{ab}(p_{0};p_{1},...,p_{N}) \Bigr]\mathcal{B}_{(\!N\!)}^{ab}(q_{1},p_{B})...\mathcal{B}_{(\!N\!)}^{ab}(q_{M},p_{B}) |0\rangle_{NB}
\notag
\end{align}
and the Bethe equations
\begin{equation}
g_{D}^{a}(q_{i})\lambda_{2(\!N\!)}^{ab}(q_{i};p_{1},...,p_{N})\!=\!\lambda_{1(\!N\!)}^{ab}(q_{i};p_{1},...,p_{N}) 
\quad \text{for} \quad a,b\in\{\text{L},\text{R},\tilde{\text{L}},\tilde{\text{R}}\} \quad \text{and} \quad i=1,...,M \,\, .
\end{equation}

\section{Conclusion}

In this paper we have extended the boundary Bethe ansatz programme to the case of massive representations in the integrable model associated with strings propagating in an $AdS_3 \times S^3 \times T^4$ geometry. We have employed the standard framework, in the adaptation which we have constructed starting with the massless $AdS_3 \times S^3 \times T^4$ representations \cite{Bielli:2024bve}. This time the complication is significantly higher due to the lack of difference-form of the $R$-matrix in the massive sector, and so we had to work much harder to perform all the steps. All of these steps turned out to be controllable, and it has been possible to write all of our results in universal form. Another distinction of this paper with respect to the massless case is that we have encompassed all {\it possible} massive representations, leading us to obtain the complete result for the massive sector.   

Our main results are a complete array of explicit fully worked out Bethe equations and spectrum- generating creation operators to populate the transfer-matrix (energy) eigenvalues for any assortment of allowed representations.  We  included those representations which can generate the massless sector in suitable careful limits, which we have however treated in detail in an earlier publication. In the appendix we allow for all possible assignments of representations in both the auxiliary and   the physical spaces, we have understood which of these assignments lead to a functioning procedure  and which do not. Another result we have proven is a constraint on the form of the vector-boundary $T_+$ monodromy matrix which is allowed for our procedure to apply. Within our analysis we have therefore found ample context for a general treatment of the boundary algebraic Bethe ansatz for the $AdS_3 \times S^3 \times T^4$ scattering theory.

We list here a number of possible ideas for future work
\begin{enumerate}
\item It will be necessary to repeat this analysis for mixed massive-massless cases. Using the Zukhowsky variables could be a way of compactly combining massive and massless representations, however at the cost of having explicit formulas for the various assignments of chiralities. Since in mixed-mass cases the difference-form of the $R$-matrix is also lost, this might not be such a big cost to pay in this case. 

\item It will be important to obtain some simple solutions of the Bethe ansatz, which we recall are the auxiliary system and therefore are only determining the algebraic form of the eigenvectors without quantising the momenta on a circle.

\item Having identified the dual equation for all the representations concerned, we should now investigate the issue of crossing symmetry, and obtain dressing factors for all the reflection matrices which we have found here (and in \cite{Bielli:2024xuv}). This is a major task but an absolute  necessary step to proceed with the thermodynamics.

\item We should then proceed formulating the momentum-carrying equations, which should be relatively straighforward given that we do have the explicit formulas for the eigenvectors, and then we should derive the boundary thermodynamic Bethe ansatz.

\item It will be interesting to study possible connections and potential $AdS_3 \times S^3 \times T^4$ analogies with the work of \cite{Cava}\footnote{We thank A. Cavagli\'a for related remarks.}.  
\end{enumerate}

\section*{Acknowledgments}

A.T. thanks Davide Polvara and Alessandro Sfondrini for the very useful email exchange, and Andrea Cavagli\'a, Davide Fioravanti, Chiara Paletta and Robert Weston for very helpful comments in the occasion of the {\it In Tropea 2024} workshop. D.\,B. thanks Sibylle Driezen and Davide Polvara for interesting discussions connected to boundary scattering in occasion of the \textit{2025 Favignana Workshop on Higher-d integrability}, as well as all the participants of the workshop for stimulating conversations on integrability-related topics. D.\,B. is supported by Thailand NSRF via PMU-B, grant number B13F680083. V.\,M. is supported by the STFC under the grant ST/X508809/1.

\begin{appendix}
\addtocontents{toc}{\protect\setcounter{tocdepth}{1}}

\section{Notation and conventions}\label{app:notation-and-conventions}
We begin by considering a setup with two auxiliary spaces (respectively labelled as $0$ and $0'$), $N$ physical spaces (labeled by integers $1,...,N$) and one boundary space (labeled by the subscript $B$).
\begin{equation}\label{setup-spaces}
V_{0}\otimes V_{0'}\otimes V_{1}\otimes ... \otimes V_{N}  \otimes V_{B}
\end{equation}
Our discussion concerns various $4\times4$ matrices of two variables which can be written in the form
\begin{equation}\label{generic-M-decomposition}
M(p,q) = \sum_{\mu,\nu,\rho,\sigma=1}^{2} m_{pq}^{\mu\nu\rho\sigma} \,\, \mathbb{E}_{\mu\nu}\otimes \mathbb{E}_{\rho\sigma} \,\, .
\end{equation}
All such matrices are invertible, with the inverse $M^{-1}(p,q)$ enjoying a similar decomposition, and respect the conservation of the fermion number. This means non-vanishing coefficients satisfy the relations
\begin{equation}\label{fermion-number-conservation}
|\mu|+|\nu|+|\rho|+|\sigma|=0 \,\,\, \text{mod 2} 
\qquad 
\text{and}
\qquad 
|\mu|+|\nu|=|\rho|+|\sigma| \,\, \text{mod 2} 
\qquad
\text{with} 
\qquad 
|1| = 0 \,\, , \,\, |2|=1 \,\, .
\end{equation}
In the form \eqref{generic-M-decomposition}, the matrices act on two copies of $V$ as $M(p,q): \,\, V_{p}\otimes V_{q} \rightarrow V_{p}\otimes V_{q}$, which also makes clear that the two variables $p$ and $q$ are respectively associated to the first and the second copy of $V$.
To deal with the setup \eqref{setup-spaces}, we extend the matrices \eqref{generic-M-decomposition} in the usual fashion, introducing a trivial action on the spaces where $M(p,q)$ is not meant to act, and explicitly highlighting by extra subscripts the spaces where the action is non-trivial. In particular, since we will be dealing with non-trivial actions on spaces $(V_{0},V_{i})$, $(V_{0'}, V_{i})$, with $i \in \{1,...,N\}$, and $(V_{0}, V_{B})$, $(V_{0'}, V_{B})$, we will consider matrices like
\begin{align}\label{matrices-M-and-N}
[M(p_{0},p_{i})]_{0i} \! &:= \!
m_{p_{0}p_{i}}^{\mu\nu\rho\sigma} \! \,\, \mathbb{E}_{\mu\nu} \otimes \mathds{1}\otimes \mathbb{E}_{\rho\sigma}^{(i)}\otimes \mathds{1}
\qquad \quad \quad \,\,
[M(p_{0'},p_{i})]_{0'i} \!:=\! 
m_{p_{0'}p_{i}}^{\mu\nu\rho\sigma} \,\, \mathds{1} \otimes \mathbb{E}_{\mu\nu}\otimes \mathbb{E}_{\rho\sigma}^{(i)}\otimes \mathds{1} \,\, ,
\notag \\
\\
[N(p_{0},p_{B})]_{0B} \! & :=\! 
n_{p_{0}p_{B}}^{\alpha\beta\gamma\delta} \,\, \mathbb{E}_{\alpha\beta} \otimes \mathds{1} \otimes \mathds{1}_{N} \otimes \mathbb{E}_{\gamma\delta}
\qquad \quad
[N(p_{0'},p_{B})]_{0'B} \! :=\!
n_{p_{0'}p_{B}}^{\alpha\beta\gamma\delta} \,\, \mathds{1}\otimes \mathbb{E}_{\alpha\beta} \otimes \mathds{1}_{N} \otimes \mathbb{E}_{\gamma\delta}
\,\, ,
\notag
\end{align}
where we suppressed summations making use of Einstein's convention and introduced shorthand notations for the trivial action on $N$ physical spaces and the non-trivial action on the physical space $i$
\begin{equation}
\mathds{1}_{N} \!:=\! \underbrace{\mathds{1}\otimes \mathds{1} \otimes ... \otimes \mathds{1}}_{\text{$N$-times}} 
\qquad \,\, , \qquad \,\,
\mathbb{E}^{(i)}_{\gamma\delta} \!:=\! \underbrace{\mathds{1} \otimes \mathds{1} \otimes ... }_{\text{spaces $1$ to $i-1$}}\otimes \underbrace{\mathbb{E}_{\gamma\delta}}_{\text{space $i$}} \otimes \underbrace{\mathds{1} \otimes ... \otimes \mathds{1}}_{\text{spaces $i+1$ to $N$}} \quad \forall \, i \in \{1,...,N\} \,\, .
\end{equation}

\section{Useful identities}\label{app:useful-identities}
Here we collect relations appearing in the proofs reported in later appendices.
\subsection{Commuting matrices on auxiliary, physical and boundary spaces}
The matrices \eqref{matrices-M-and-N}, which often appear throughout the text, satisfy the following useful commutators\footnote{These commutators of matrices acting on different spaces would be trivial in a non-supersymmetric theory. In this work, where the tensor product is graded leading to fermion signs, we find their vanishing to be not immediately obvious.}
\begin{equation}\label{identity-MN-commutator}
\Bigl[ [M(p_{0},p_{i})]_{0i} \, , \,[N(p_{0'},p_{B})]_{0'B} \Bigr] = 0 
\qquad \qquad \qquad 
\Bigl[ [M(p_{0'},p_{i})]_{0'i} \, , \, [N(p_{0},p_{B})]_{0B} \Bigr]=0 \,\, .
\end{equation}
Both can immediately be checked by direct computation. For example in the first case one has
\begin{equation}
\begin{aligned}
[M(p_{0},p_{i})]_{0i}[N(p_{0'},p_{B})]_{0'B} & = m_{p_{0}p_{i}}^{\mu\nu\rho\sigma} \,\, n_{p_{0'}p_{B}}^{\alpha\beta\gamma\delta} \,\,(\mathbb{E}_{\mu\nu}\otimes \mathds{1} \otimes \mathbb{E}_{\rho\sigma}^{(i)}\otimes \mathds{1})(\mathds{1}\otimes \mathbb{E}_{\alpha\beta}\otimes \mathds{1}_{N}\otimes \mathbb{E}_{\gamma\delta})
\\
& =
(-1)^{(\rho+\sigma)(\alpha+\beta)} m_{p_{0}p_{i}}^{\mu\nu\rho\sigma} \,\, n_{p_{0'}p_{B}}^{\alpha\beta\gamma\delta} \,\, \mathbb{E}_{\mu\nu}\otimes \mathbb{E}_{\alpha\beta} \otimes \mathbb{E}_{\rho\sigma}^{(i)} \otimes \mathbb{E}_{\gamma\delta}
\end{aligned}
\end{equation}
and
\begin{align}
[N(p_{0'},p_{B})]_{0'B} [M(p_{0},p_{i})]_{0i} & = m_{p_{0}p_{i}}^{\mu\nu\rho\sigma} \,\, n_{p_{0'}p_{B}}^{\alpha\beta\gamma\delta} \,\,(\mathds{1}\otimes \mathbb{E}_{\alpha\beta}\otimes \mathds{1}_{N}\otimes \mathbb{E}_{\gamma\delta})(\mathbb{E}_{\mu\nu}\otimes \mathds{1} \otimes \mathbb{E}_{\rho\sigma}^{(i)}\otimes \mathds{1})
\\
& =
(\!-1\!)^{(\mu+\nu+\rho+\sigma)(\gamma+\delta)} (\!-1\!)^{(\mu+\nu)(\alpha+\beta)} m_{p_{0}p_{i}}^{\mu\nu\rho\sigma} \,\, n_{p_{0'}p_{B}}^{\alpha\beta\gamma\delta} \,\, \mathbb{E}_{\mu\nu}\otimes \mathbb{E}_{\alpha\beta} \otimes \mathbb{E}_{\rho\sigma}^{(i)} \otimes \mathbb{E}_{\gamma\delta} \,\, ,
\notag
\end{align}
where for simplicity we omitted the symbol $|\cdot|$ in the sign factors. The two expressions agree if
\begin{equation}
(-1)^{(\rho+\sigma)(\alpha+\beta)}=\underbrace{(-1)^{(\mu+\nu+\rho+\sigma)(\gamma+\delta)}}_{\text{always $+1$}} (-1)^{(\mu+\nu)(\alpha+\beta)} \,\, ,
\end{equation}
which is always true due to the fermion number conservation \eqref{fermion-number-conservation}.
\subsection{Yang-Baxter equation and unitarity of bulk $R$-matrices}
The $R$-matrices describing interaction of bulk excitations satisfy the Yang-Baxter equation \eqref{general-bulk-YBE}
\begin{equation}\label{YBE-appendix}
[R^{ab}(p_{1},p_{2})]_{12}[R^{ac}(p_{1},p_{3})]_{13}[R^{bc}(p_{2},p_{3})]_{23}=[R^{bc}(p_{2},p_{3})]_{23}[R^{ac}(p_{1},p_{3})]_{13}[R^{ab}(p_{1},p_{2})]_{12} \,\, ,
\end{equation}
acting on any triplet $V_{1}\otimes V_{2}\otimes V_{3}$ (possibly involving both auxiliary and physical spaces), with generic representations $a,b,c$ respectively associated to spaces $1,2,3$ and being any combination of $\text{L},\text{R},\tilde{\text{L}},\tilde{\text{R}}$.
Additionally, all such $R$-matrices satisfy unitarity relations \eqref{unitarity-1} and \eqref{unitarity-2}, which can be rewritten as
\begin{equation}\label{unitarity-general}
[R^{ab}(p_{1},p_{2})]_{12}[R^{ba}(p_{2},p_{1})]^{op}_{12}= f^{ab}(p_{1},p_{2}) \mathds{1}\otimes \mathds{1}
\qquad \qquad
\forall \,\, \{ a,b\} \in \{\text{L},\text{R},\tilde{\text{L}},\tilde{\text{R}}\} \,\, ,
\end{equation}
with $f^{ab}(p_{1},p_{2})=f^{ab}(p_{2},p_{1})$ irrespectively of the specific choice of representations $a,b$. These relations immediately allow to obtain
\begin{align}
[R^{bc}(p_{2},p_{3})]_{23}[R^{ba}(p_{2},p_{1})]^{op}_{12}[R^{ca}(p_{3},p_{1})]^{op}_{13} & = [R^{ca}(p_{3},p_{1})]^{op}_{13} [R^{ba}(p_{2},p_{1})]^{op}_{12} [R^{bc}(p_{2},p_{3})]_{23}
\label{op-YBE-1}\\
[R^{cb}(p_{3},p_{2})]_{23}^{op}[R^{ca}(p_{3},p_{1})]^{op}_{13}[R^{ba}(p_{2},p_{1})]^{op}_{12} & = [R^{ba}(p_{2},p_{1})]^{op}_{12} [R^{ca}(p_{3},p_{1})]^{op}_{13} [R^{cb}(p_{3},p_{2})]_{23}^{op}
\label{op-YBE-2}\\
[R^{cb}(p_{3},p_{2})]_{23}^{op}[R^{ab}(p_{1},p_{2})]_{12}[R^{ac}(p_{1},p_{3})]_{13} & = [R^{ac}(p_{1},p_{3})]_{13}[R^{ab}(p_{1},p_{2})]_{12}[R^{cb}(p_{3},p_{2})]_{23}^{op} \,\, .
\label{op-YBE-3}
\end{align}
In particular, all the above relations are obtained by acting on the YBE \eqref{YBE-appendix} as follows
\begin{itemize}
\item \eqref{op-YBE-1}: multiply by $[R^{ba}(p_{2},p_{1})]_{12}^{op}[R^{ca}(p_{3},p_{1})]_{13}^{op}$ from the right 

and $[R^{ca}(p_{3},p_{1})]_{13}^{op}[R^{ba}(p_{2},p_{1})]_{12}^{op}$ from the left.
\item \eqref{op-YBE-2}: multiply by $[R^{cb}(p_{3},p_{2})]_{23}^{op}[R^{ca}(p_{3},p_{1})]^{op}_{13}[R^{ba}(p_{2},p_{1})]^{op}_{12}$ from the right

and $[R^{ba}(p_{2},p_{1})]^{op}_{12}[R^{ca}(p_{3},p_{1})]^{op}_{13}[R^{cb}(p_{3},p_{2})]_{23}^{op} $ from the left.
\item \eqref{op-YBE-3}: multiply by $[R^{cb}(p_{3},p_{2})]_{23}^{op}$ from both the left and the right.
\end{itemize}

\section{$T_{-}^{ab}$ satisfies BYBE}\label{app:Tminus-satisfies-BYBE}
In this appendix we show that the double-row monodromy $T_{-}^{ab}$ for a vector boundary satisfies the BYBE \eqref{general-right-BYBE}, like the reflection matrices $K^{ab}(p,p_{B})$, independently of the specific representations $a,b$. Replacing the spaces $1,2$ with auxiliary spaces $0,0'$ and all instances of $K$ with $T_{-}$ in \eqref{general-right-BYBE}, one obtains
\begin{equation}\label{BYBE-appendix}
\begin{aligned}
[T_{-}^{cb}(p_{0'},p_{B})]_{0'B}&[R^{ca}(p_{0'},-p_{0})]_{00'}^{op} [T_{-}^{ab}(p_{0},p_{B})]_{0B}[R^{ac}(p_{0},p_{0'})]_{00'} = 
\\
& =[R^{ca}(-p_{0'},-p_{0})]_{00'}^{op}[T_{-}^{ab}(p_{0},p_{B})]_{0B}[R^{ac}(p_{0},-p_{0'})]_{00'}[T_{-}^{cb}(p_{0'},p_{B})]_{0'B} \,\, .
\end{aligned}
\end{equation}
The double-row monodromy is then constructed by letting an auxiliary excitation move towards a right-wall, bounce off it and finally move towards the left, back to its original position. In this process, the auxiliary excitation also scatters against a series of $N$ physical excitations, which are thought of as being located at fixed positions and hence encountered in reversed orders when first moving towards the right and then towards the left. This translates into the following expression
\begin{equation}\label{monodromy-Tminus}
[T_{-}^{ab}(p_{0},p_{B})]_{0B}:= \Biggl( \prod_{i=N}^{1}[R^{d_{i}a}(p_{i},-p_{0})]^{op}_{0i} \Biggr) [K^{ab}(p_{0},p_{B})]_{0B} \Biggl( \prod_{i=1}^{N} [R^{ad_{i}}(p_{0},p_{i})]_{0i} \Biggr) \,\, ,
\end{equation}
with $K^{ab}(p_{0},p_{B})$ satisfying \eqref{general-right-BYBE} and denoting one of the $K$-matrices given in section \ref{sec:boundary-theory} for the vector boundary. This only acts non-trivially on one auxiliary space, in this case $0$, and the boundary, hence falling in the category \eqref{matrices-M-and-N}. The auxiliary particle and the boundary are respectively taken in generic representations $a$ and $b$ (these being either $\text{L}$ or $\text{R}$, or the tilde versions), which appear as labels on both $K$ and the monodromy itself. The $R$-matrices (and their $op$) act on the same auxiliary space as $K$ and only one physical space, hence exhibiting the structure \eqref{matrices-M-and-N}. The labels $d_i$ refer to the representation acting on the physical space $i$, which can also be either $\text{L}$ or $\text{R}$ or the tilde versions. They do not explicitly appear on the left hand side of \eqref{monodromy-Tminus} because, as we will now show, the BYBE is not sensitive to such information and is satisfied for any collection $\{d_{1}, d_{2},...,d_{N} \}$. 

The proof that \eqref{monodromy-Tminus} satisfies \eqref{BYBE-appendix} generalises the one given in appendix C of \cite{Bielli:2024bve}, restricted to the case of a singlet wall and massless representations of the type $\text{L}$, and goes along similar lines of reasoning. We will hence show satisfaction of \eqref{BYBE-appendix} for $N=1$, with the case of higher-$N$ following by induction. The $N=2$ case has furthermore been checked via computer algebra.
To begin, consider two double-row monodromies, acting on different auxiliary spaces $0,0'$ and one boundary
\begin{equation}
\begin{aligned}
T_{-}^{ab}(p_{0},p_{B})&=[R^{da}(p_{1},-p_{0})]_{01}^{op}[K^{ab}(p_{0},p_{B})]_{0B}[R^{ad}(p_{0},p_{1})]_{01}
\\
T_{-}^{cb}(p_{0'},p_{B})&=[R^{dc}(p_{1},-p_{0'})]_{0'1}^{op}[K^{cb}(p_{0'},p_{B})]_{0'B}[R^{cd}(p_{0'},p_{1})]_{0'1} \,\, .
\end{aligned}
\end{equation}
The left hand side of \eqref{BYBE-appendix} then explicitly reads
\begin{align}
&[T_{-}^{cb}(p_{0'},p_{B})]_{0'B}[R^{ca}(p_{0'},-p_{0})]_{00'}^{op} [T_{-}^{ab}(p_{0},p_{B})]_{0B}[R^{ac}(p_{0},p_{0'})]_{00'} =
\notag \\[1.em]
& = [R^{dc}(p_{1},-p_{0'})]_{0'1}^{op}[K^{cb}(p_{0'},p_{B})]_{0'B} \,\,\,\, \cdot 
\notag \\
& \quad \cdot
\underbrace{[R^{cd}(p_{0'},p_{1})]_{0'1}[R^{ca}(p_{0'},-p_{0})]^{op}_{00'} [R^{da}(p_{1},-p_{0})]_{01}^{op}}_{\text{YBE \eqref{op-YBE-1}}} [K^{ab}(p_{0},p_{B})]_{0B}[R^{ad}(p_{0},p_{1})]_{01}[R^{ac}(p_{0},p_{0'})]_{00'}
\notag \\[1.em]
& = [R^{dc}(p_{1},-p_{0'})]_{0'1}^{op}[K^{cb}(p_{0'},p_{B})]_{0'B} \,\,\,\, \cdot 
\notag \\
& \quad \cdot [R^{da}(p_{1},-p_{0})]_{01}^{op} [R^{ca}(p_{0'},-p_{0})]^{op}_{00'} \underbrace{[R^{cd}(p_{0'},p_{1})]_{0'1} [K^{ab}(p_{0},p_{B})]_{0B}}_{\text{Commutativity \eqref{identity-MN-commutator}}}[R^{ad}(p_{0},p_{1})]_{01}[R^{ac}(p_{0},p_{0'})]_{00'}
\notag
\end{align}
\begin{align}
& = [R^{dc}(p_{1},-p_{0'})]_{0'1}^{op}[K^{cb}(p_{0'},p_{B})]_{0'B} \,\,\,\, \cdot 
\notag \\
& \quad \cdot [R^{da}(p_{1},-p_{0})]_{01}^{op} [R^{ca}(p_{0'},-p_{0})]^{op}_{00'} [K^{ab}(p_{0},p_{B})]_{0B} \underbrace{[R^{cd}(p_{0'},p_{1})]_{0'1} [R^{ad}(p_{0},p_{1})]_{01}[R^{ac}(p_{0},p_{0'})]_{00'}}_{\text{YBE} \eqref{YBE-appendix}}
\notag \\[1.em]
& = [R^{dc}(p_{1},-p_{0'})]_{0'1}^{op} \underbrace{[K^{cb}(p_{0'},p_{B})]_{0'B} [R^{da}(p_{1},-p_{0})]_{01}^{op}}_{\text{Commutativity} \eqref{identity-MN-commutator}} \,\,\,\, \cdot 
\notag \\
& \quad \cdot [R^{ca}(p_{0'},-p_{0})]^{op}_{00'} [K^{ab}(p_{0},p_{B})]_{0B} [R^{ac}(p_{0},p_{0'})]_{00'} [R^{ad}(p_{0},p_{1})]_{01} [R^{cd}(p_{0'},p_{1})]_{0'1}
\notag \\[2em]
& = [R^{dc}(p_{1},-p_{0'})]_{0'1}^{op} [R^{da}(p_{1},-p_{0})]_{01}^{op} \,\,\,\, \cdot 
\notag \\
& \quad \cdot \underbrace{[K^{cb}(p_{0'},p_{B})]_{0'B} [R^{ca}(p_{0'},-p_{0})]^{op}_{00'} [K^{ab}(p_{0},p_{B})]_{0B} [R^{ac}(p_{0},p_{0'})]_{00'}}_{\text{BYBE} \eqref{BYBE-appendix}} [R^{ad}(p_{0},p_{1})]_{01} [R^{cd}(p_{0'},p_{1})]_{0'1}
\notag \\[1.em]
& = \underbrace{[R^{dc}(p_{1},-p_{0'})]_{0'1}^{op} [R^{da}(p_{1},-p_{0})]_{01}^{op} [R^{ca}(-p_{0'},-p_{0})]^{op}_{00'}}_{\text{YBE} \eqref{op-YBE-2}} \,\,\,\, \cdot 
\notag \\
& \quad \cdot  [K^{ab}(p_{0},p_{B})]_{0B} [R^{ac}(p_{0},-p_{0'})]_{00'} [K^{cb}(p_{0'},p_{B})]_{0'B} [R^{ad}(p_{0},p_{1})]_{01} [R^{cd}(p_{0'},p_{1})]_{0'1}
\notag \\[2.em]
& =  [R^{ca}(-p_{0'},-p_{0})]^{op}_{00'} [R^{da}(p_{1},-p_{0})]_{01}^{op} \underbrace{[R^{dc}(p_{1},-p_{0'})]_{0'1}^{op} [K^{ab}(p_{0},p_{B})]_{0B}}_{\text{Commutativity} \eqref{identity-MN-commutator}} \,\,\,\, \cdot 
\notag \\
& \quad \cdot  [R^{ac}(p_{0},-p_{0'})]_{00'} \underbrace{[K^{cb}(p_{0'},p_{B})]_{0'B} [R^{ad}(p_{0},p_{1})]_{01}}_{\text{Commutativity} \eqref{identity-MN-commutator}} [R^{cd}(p_{0'},p_{1})]_{0'1}
\notag \\[1.em]
& =  [R^{ca}(-p_{0'},-p_{0})]^{op}_{00'} [R^{da}(p_{1},-p_{0})]_{01}^{op} [K^{ab}(p_{0},p_{B})]_{0B}   \,\,\,\, \cdot 
\notag \\
& \quad \cdot  \underbrace{[R^{dc}(p_{1},-p_{0'})]_{0'1}^{op} [R^{ac}(p_{0},-p_{0'})]_{00'} [R^{ad}(p_{0},p_{1})]_{01}}_{\text{YBE} \eqref{op-YBE-3}} [K^{cb}(p_{0'},p_{B})]_{0'B} [R^{cd}(p_{0'},p_{1})]_{0'1}
\notag \\[1.em]
& =  [R^{ca}(-p_{0'},-p_{0})]^{op}_{00'} [R^{da}(p_{1},-p_{0})]_{01}^{op} [K^{ab}(p_{0},p_{B})]_{0B}   \,\,\,\, \cdot 
\notag \\
& \quad \cdot   [R^{ad}(p_{0},p_{1})]_{01} [R^{ac}(p_{0},-p_{0'})]_{00'} [R^{dc}(p_{1},-p_{0'})]_{0'1}^{op}  [K^{cb}(p_{0'},p_{B})]_{0'B} [R^{cd}(p_{0'},p_{1})]_{0'1}
\notag \\[1.em]
& = [R^{ca}(-p_{0'},-p_{0})]^{op}_{00'} [T_{-}^{ab}(p_{0},p_{B})]_{0B} [R^{ac}(p_{0},-p_{0'})]_{00'} [T_{-}^{cb}(p_{0'},p_{B})]_{0'B} \,\, ,
\notag
\end{align}
and indeed after some manipulations coincides with the right hand side of \eqref{BYBE-appendix}, as desired.

\section{$T_{+}^{cb}$ commuting with $T_{-}^{ab}$}\label{app:commuting-monodromies}
In this appendix we would like to show the existence of a dual monodromy $T_{+}^{cb}$, satisfying the dual equation \eqref{dual_BYBE}, which commutes with the double-row monodromy $T_{-}^{ab}$ satisfying the BYBE, namely 
\begin{equation}\label{T-plus-T-minus-commutativity}
[T_{-}^{ab}(p_{0},p_{B})]_{0B}[T_{+}^{cb}(p_{0'},p_{B})]_{0'B}=[T_{+}^{cb}(p_{0'},p_{B})]_{0'B}[T_{-}^{ab}(p_{0},p_{B})]_{0B} \,\, .
\end{equation}
To this aim, we begin by considering the explicit definition \eqref{monodromy-Tminus} of the double-row monodromy $T_{-}^{ab}$, and the following generic expansion for a dual monodromy $T_{+}^{cb}$ which only acts non-trivially on the auxiliary space $0'$, with representation $c$, and the boundary:
\begin{equation}\label{monodromy-Tplus}
[T_{+}^{cb}(p_{0'},p_{B})]_{0'B} = t_{+}^{\alpha\beta\gamma\delta} \,\, \mathds{1} \otimes \mathbb{E}_{\alpha\beta} \otimes \mathds{1}_{N} \otimes \mathbb{E}_{\gamma\delta} \,\, .
\end{equation}
It is then immediate to notice that $[T_{+}^{cb}(p_{0'},p_{B})]_{0'B}$ has precisely the structure of $[N(p_{0'},p_{B})]_{0'B}$ defined in \eqref{matrices-M-and-N}, while all the $R$-matrices appearing in the definition \eqref{monodromy-Tminus} of $T_{-}^{ab}$ are of the form $[M(p_{0},p_{i})]_{0i}$ in \eqref{matrices-M-and-N}: this means that thanks to \eqref{identity-MN-commutator}, any choice of $T_{+}^{cb}$ of the form \eqref{monodromy-Tplus} freely goes through any $R$-matrix in the definition \eqref{monodromy-Tminus} and commutativity of $T_{+}^{cb}$ with $T_{-}^{ab}$ is satisfied if
\begin{equation}\label{condition}
[K^{ab}(p_{0},p_{B})]_{0B}[T_{+}^{cb}(p_{0'},p_{B})]_{0'B} = [T_{+}^{cb}(p_{0'},p_{B})]_{0'B} [K^{ab}(p_{0},p_{B})]_{0B} \,\, .
\end{equation}
At this point, noting that both matrices in \eqref{condition} are of the form given in \eqref{matrices-M-and-N}, one finds that
\begin{equation}
[K^{ab}(p_{0},p_{B})]_{0B}[T_{+}^{cb}(p_{0'},p_{B})]_{0'B} = k^{\mu\nu\rho\sigma} \,\, t_{+}^{\alpha\beta\gamma\delta } \,\, (-1)^{(\rho+\sigma)(\alpha+\beta)}\delta_{\sigma\gamma} \,\, \mathbb{E}_{\mu\nu} \otimes \mathbb{E}_{\alpha\beta} \otimes \mathds{1}_{N} \otimes \mathbb{E}_{\rho\delta} \,\, ,
\end{equation}
while
\begin{equation}\label{RHS}
\begin{aligned}
[T_{+}^{cb}(p_{0'},p_{B})]_{0'B} [K^{ab}(p_{0},p_{B})]_{0B} & = \!k^{\mu\nu\rho\sigma} \,\, t_{+}^{\alpha\beta\gamma\delta}\,\, (-1)^{(\mu+\nu)(\alpha+\beta+\gamma+\delta)}\delta_{\delta\rho} \,\, \mathbb{E}_{\mu\nu}\otimes \mathbb{E}_{\alpha\beta} \otimes \mathds{1}_{N} \otimes \mathbb{E}_{\gamma\sigma} 
\\
& = \! k^{\mu\nu\gamma\delta} \,\, t_{+}^{\alpha\beta\rho\sigma} \,\, (-1)^{(\mu+\nu)(\alpha+\beta+\rho+\sigma)}\delta_{\sigma\gamma} \,\, \mathbb{E}_{\mu\nu}\otimes \mathbb{E}_{\alpha\beta} \otimes \mathds{1}_{N} \otimes \mathbb{E}_{\rho\delta} \,\, ,
\end{aligned}
\end{equation}
after using on both sides that
\begin{equation}
\mathbb{E}_{\alpha\beta}\mathbb{E}_{\gamma\delta} = \delta_{\beta\gamma} \mathbb{E}_{\alpha\delta} \,\, ,
\end{equation}
and relabelling $\gamma \leftrightarrow\rho$, $\sigma \leftrightarrow \delta$ in the second line of \eqref{RHS}. The two expressions agree if
\begin{equation}
\sum_{\gamma=1}^{2} (-1)^{(\rho+\gamma)(\alpha + \beta)} \,\, k^{\mu\nu\rho\gamma} \,\, t_{+}^{\alpha\beta\gamma\delta} = \sum_{\gamma=1}^{2} k^{\mu\nu\gamma\delta} \,\, t_{+}^{\alpha\beta\rho\gamma} \qquad \qquad \forall \, \mu,\nu,\rho,\alpha,\beta,\delta \,\, \in \{1,2\} \,\, .
\end{equation}
Notice that in the last step we also exploited the requirement that $T_{+}^{cb}$ should respect fermion number conservation, such that each component $t_{+}^{\alpha\beta\rho\sigma}$ satisfies $|\alpha|+|\beta|+|\rho|+|\sigma|=0$ mod 2. At this point, spelling out the conditions for any value of the free indices and keeping in mind fermion number conservation for both $T_{+}^{cb}$ and $K^{ab}$, one finds that all identities are satisfied provided that:
\begin{equation}
t_{+}^{1212}=t_{+}^{1221}=t_{+}^{2112}=t_{+}^{2121}=0
\qquad \qquad \qquad
t_{+}^{1122}=t_{+}^{1111} 
\qquad\qquad\qquad
t_{+}^{2222}=t_{+}^{2211} \,\, .
\end{equation}
This condition requires $T_{+}^{cb}$ to act trivially on the boundary space, taking the form
\begin{equation}\label{constrained-monodromy-Tplus}
[T_{+}^{cb}(p_{0'},p_{B})]_{0'B} = \sum_{\alpha =1}^{2} t^{cb}_{\alpha} \, \mathds{1} \otimes \mathbb{E}_{\alpha\alpha} \otimes \mathds{1}_{N} \otimes \mathds{1}_{B} \,\, ,
\end{equation}
and effectively behaving as the simplest extension of the singlet $T_{+}^{c}$, after normalising $t_{1}^{cb}$ entry to 1.

\begin{section}{The general dual BYBE}\label{app:dual-BYBE}

\subsubsection*{\ul{\it The supertransposition}}
In the dual boundary Yang-Baxter equation and throughout its derivation, we make use of a supertransposition defined on the $\mathbb{E}_{\alpha\beta}$ matrices as
\begin{equation}\label{st-definition}
\begin{aligned}
\mathbb{E}_{11}^{st}=\mathbb{E}_{11}\qquad\mathbb{E}_{22}^{st}=\mathbb{E}_{22}\qquad\mathbb{E}_{12}^{st}=\mathbb{E}_{21}\qquad\mathbb{E}_{21}^{st}=-\mathbb{E}_{12}
\end{aligned}\,.
\end{equation}
We will also need to define an inverse supertransposition, $ist$, such that for any matrix $\left(M^{st}\right)^{ist}=\left(M^{ist}\right)^{st}=M$, which is necessary since our supertransposition is not involutive.\footnote{In fact, it is of order 4, such that the inverse operation can be thought of as applying supertransposition thrice.} Obviously, the appropriate choice is
\begin{equation}\label{ist-definition}
\begin{aligned}
\mathbb{E}_{11}^{ist}=\mathbb{E}_{11}\qquad\mathbb{E}_{22}^{ist}=\mathbb{E}_{22}\qquad\mathbb{E}_{12}^{ist}=-\mathbb{E}_{21}\qquad\mathbb{E}_{21}^{ist}=\mathbb{E}_{12}
\end{aligned}\,.
\end{equation}
\subsubsection*{\ul{\it Writing down the equation}}
In this part, let us present the boundary YBE dual to the general BYBE \eqref{BYBE-appendix} and investigate it in the specific cases where the $R$-matrix is one of those relevant to $AdS_3$.

To determine what this dual equation needs to be, let us start with the ansatz\footnote{Alongside the unknown matrices, this ansatz includes the $R$-matrix. One could instead start by replacing it with a fourth unknown matrix $W(p,q)$ and then fixing it in the same way as the other three, namely by finding the relation it needs to satisfy for the transfer matrix to commute. That relation would turn out to be $\left[W^{ac}(p,q)\right]^{ist_0st_{0'}}\left[R^{ca}(-q,-p)\right]^{op}=f^{ac}(p,q)\mathds{1}\otimes\mathds{1}$ and from braiding unitarity \eqref{unitarity-general} it would then follow that $W^{ac}(p,q)=\left[R^{ac}(-p,-q)\right]^{st_0 ist_{0'}}$.}
\begin{align}\label{dual_BYBE_appendix}
\left[K_D^{cb}(p_{0'},p_{B})\right]^{ist_{0'}}_{0'B}&\left[X^{ac}(p_0,p_{0'})\right]_{00'}\left[K^{ab}_D(p_0,p_{B})\right]_{0B}^{st_0}\left[Y^{ac}(p_0,p_{0'})\right]_{00'}=\nonumber
\\&[R^{ac}(-p_0,-p_{0'})]_{00'}^{st_0ist_{0'}}\left[K^{ab}_D(p_0,p_{B})\right]_{0B}^{st_0}\left[Z^{ac}(p_0,p_{0'})\right]_{00'}\left[K_D^{cb}(p_{0'},p_{B})\right]^{ist_{0'}}_{0'B}
\end{align}
where $X,Y,Z$ are some unknown matrices, to be determined, and $K_{D}$ the dual reflection matrix. As shown in the next part of this appendix, the commutativity of the transfer matrix then requires that these matrices satisfy the following relations
\begin{align}
&\left(X^{ac}(p_0,p_{0'})\right)^{ist_{0}}\,\left(\left[R^{ca}(p_{0'},-p_0)\right]^{op}\right)^{ist_{0'}}=\mathds{1}\otimes\mathds{1}\label{general_rel_1}\\
&Y^{ac}(p_0,p_{0'})\,\left(R^{ac}(p_0,p_{0'})\right)^{st_0ist_{0'}}=f^{ac}(p_0,p_{0'})\mathds{1}\otimes\mathds{1}\label{general_rel_2}\\
&\left(Z^{ac}(p_0,p_{0'})\right)^{st_{0'}}\left(R^{ac}(p_0,-p_{0'})\right)^{st_0}=\mathds{1}\otimes\mathds{1},\label{general_rel_3}
\end{align}
Note that the functions $f^{ac}$ in \eqref{general_rel_2} are the same that appear in the unitarity relation \eqref{unitarity-general}. Each of the above equations is enough to fix these matrices in terms of the coefficients of the $R$-matrix. For convenience let us write the $X$-matrix as 
\begin{align}
X^{ac}(p,q)=\left(x_{pq}^{ac}\right)^{\alpha\beta\gamma\delta}\mathbb{E}_{\alpha\beta}\otimes \mathbb{E}_{\gamma\delta}
\end{align}
and similarly for $Y$ and $Z$. Let us also define the functions
\begin{align}
\varphi^{ac}_{p,q}=\left[(r^{ac}_{pq})^{1212}(r^{ac}_{pq})^{2121}-(r^{ac}_{pq})^{1122}(r^{ac}_{pq})^{2211}\right]^{-1}\\
\theta^{ac}_{p,q}=\left[(r^{ac}_{pq})^{1111}(r^{ac}_{pq})^{2222}-(r^{ac}_{pq})^{1221}(r^{ac}_{pq})^{2112}\right]^{-1}\\
\zeta^{ac}_{p,q}=\left[(r^{ac}_{pq})^{1212}(r^{ac}_{pq})^{2121}+(r^{ac}_{pq})^{1111}(r^{ac}_{pq})^{2222}\right]^{-1}\\
\lambda^{ac}_{p,q}=\left[(r^{ac}_{pq})^{1221}(r^{ac}_{pq})^{2112}+(r^{ac}_{pq})^{1122}(r^{ac}_{pq})^{2211}\right]^{-1} 
\end{align}
The coefficients and normalization functions of the $X$,$Y$,$Z$ matrices then turn out to be 
\begin{equation}
\begin{aligned}
(x_{pq}^{ac})^{1111}&\!=\!(r^{ca}_{q,-p})^{2222}\theta^{ca}_{q,-p}& (x_{pq}^{ac})^{1122}&\!=\!-(r^{ca}_{q,-p})^{1122}\varphi^{ca}_{q,-p}&
\quad(x_{pq}^{ac})^{2211}&\!=\!-(r^{ca}_{q,-p})^{2211}\varphi^{ac}_{q,-p}\!&\\[0.5em](x_{pq}^{ac})^{2222}&\!=\!(r^{ca}_{q,-p})^{1111}\theta^{ca}_{q,-p}& (x_{pq}^{ac})^{1212}&\!=\!(r^{ca}_{q,-p})^{2121}\varphi^{ca}_{q,-p}&
(x_{pq}^{ac})^{2121}&\!=\!-(r^{ca}_{q,-p})^{1212}\theta^{ca}_{q,-p}\!&\\[0.5em]
(x_{pq}^{ac})^{1221}&\!=\!(r^{ca}_{q,-p})^{1221}\theta^{ca}_{q,-p}&(x_{pq}^{ac})^{2112}&\!=\!(r^{ca}_{q,-p})^{2112}\theta^{ca}_{q,-p} \!
\end{aligned}
\end{equation}
\vspace{3mm}
\begin{equation}
\begin{aligned}(y_{pq}^{ac})^{1111}&\!=\!f^{ac}_{pq}(r^{ac}_{pq})^{2222}\zeta^{ac}_{pq}&(y_{pq}^{ac})^{1122}&\!=\!f^{ac}_{pq}(r^{ac}_{p,q})^{2211}\lambda^{ac}_{p,q}&
(y_{pq}^{ac})^{2211}&\!=\!f^{ac}_{pq}(r^{ac}_{p,q})^{1122}\lambda^{ac}_{p,q}&\\[0.5em]
(y_{pq}^{ac})^{2222}&\!=\!f^{ac}_{pq}(r^{ac}_{p,q})^{1111}\zeta^{ac}_{p,q}& (y_{pq}^{ac})^{1212}&\!=\!f^{ac}_{pq}(r^{ac}_{p,q})^{2121}\zeta^{ac}_{p,q}&
(y_{pq}^{ac})^{2121}&\!=\!f^{ac}_{pq}(r^{ac}_{p,q})^{1212}\zeta^{ac}_{p,q}&\\[0.5em]
(y_{pq}^{ac})^{1221}&\!=\!-f^{ac}_{pq}(r^{ac}_{p,q})^{2112}\lambda^{ac}_{p,q}& (y_{pq}^{ac})^{2112}&\!=\!-f^{ac}_{pq}(r^{ca}_{q,-p})^{1221}\lambda^{ac}_{p,q}
\end{aligned}
\end{equation}
\vspace{3mm}
\begin{equation}
\begin{aligned}
(z_{pq}^{ac})^{1111}&=(r^{ac}_{p,-q})^{2222}\theta^{ac}_{p,-q}&(z_{pq}^{ac})^{1122}&=-(r^{ac}_{p,-q})^{2211}\varphi^{ac}_{p,-q}&
(z_{pq}^{ac})^{2211}&=-(r^{ca}_{q,-p})^{1122}\varphi^{ac}_{p,-q}&\\[0.5em]
(z_{pq}^{ac})^{2222}&=(r^{ac}_{p,-q})^{1111}\theta^{ac}_{p,-q}& (z_{pq}^{ac})^{1212}&=-(r^{ac}_{p,-q})^{2121}\varphi^{ac}_{p,-q}&
(z_{pq}^{ac})^{2121}&=-(r^{ac}_{p,-q})^{1212}\varphi^{ac}_{p,-q}&\\[0.5em]
(z_{pq}^{ac})^{1221}&=-(r^{ac}_{p,-q})^{2112}\theta^{ac}_{p,-q}& (z_{pq}^{ac})^{2112}&=-(r^{ac}_{p,-q})^{1221}\theta^{ac}_{p,-q}
\end{aligned}
\end{equation}
In an attempt to bring \eqref{dual_BYBE_appendix} to a more traditional form, we have tried to express the above matrices in terms of the $R$-matrices of the theory. For all the explicit $AdS_3$ $R$-matrices, i.e. when the $r^{\alpha\beta\gamma\delta}_{pq}$ correspond to the components of the matrices in \eqref{LtildeLtilde+RtildeRtilde-R-matrices}-\eqref{LRtilde+LtildeR+LtildeRtilde-Rmatrices}, the $X$ and $Y$ matrices simplify to \footnote{The second equality in these equations results from $\left(R^{ac}(p,q)\right)^{st_0ist_{0'}}=\left(R^{ac}(p,q)\right)^{ist_0st_{0'}}$ for all of our $R$-matrices.}
\begin{align}
\frac{1}{(x^{ac}_{pq})^{1111}}X^{ac}(p,q)&=\frac{1}{(r_{-p,q}^{ac})^{1111}} (R^{ac}(-p,q))^{st_0ist_{0'}}=\frac{1}{(r_{-p,q}^{ac})^{1111}}(R^{ac}(-p,q))^{ist_0st_{0'}}\label{X-as-R}\\
\nonumber\\
Y^{ac}(p,q)&= \left([R^{ca}(q,p)]^{op}\right)^{ist_0st_{0'}}=\left([R^{ca}(q,p)]^{op}\right)^{st_0ist_{0'}}\label{Y-as-R}
\end{align}
where the normalizations of $X$ and $R$ have been adjusted in the first equation.
We have been unable to extract a similar general relation for the matrix $Z$, however when considering only representations for which the $R$-matrix is of the true 6-vertex type, in the sense of appendix \ref{app:6vandfake-6v}, it turns out that
\begin{align}\label{Z-as-R}
\frac{1}{(z_{pq}^{(6v)})^{1111}}Z^{(6v)}(p,q)\!=\!\frac{1}{(r_{-p,q}^{(6v)})^{1111}} \left(\!\!\left(R^{(6v)}(-p,q)\right)^{st_0}\!\right)^{st_0}\!\!\!=\!\frac{1}{(r_{-p,q}^{(6v)})^{1111}}\!\!\left(\!\!\left(R^{(6v)}(-p,q)\right)^{st_{0'}}\!\right)^{st_{0'}} \,\, .
\end{align}
This relation \textit{does not} hold for the fake 6-vertex cases, therefore the dual equation \eqref{dual_BYBE_appendix} can be fully rewritten in terms of $R$-matrices
\begin{align}\label{general-dual-BYBE}
\left[K_D^{cb}(p_{0'},p_{B})\right]^{ist_{0'}}_{0'B}&\left[R^{ac}(-p_0,p_{0'})\right]_{00'}^{st_0ist_{0'}}\left[K_D^{ab}(p_0,p_{B})\right]_{0B}^{st_0}\left([R^{ca}(p_{0'},p_0)]^{op}_{00'}\right)^{ist_0st_{0'}}=
\\
\hphantom{(r^{ac}_{p_0,p_{0'}})^{1111}}&\left[R^{ac}(-p_0,-p_{0'})\right]_{00'}^{st_0ist_{0'}}\left[K^{ab}_D(p_0,p_{B})\right]_{0B}^{st_0}\left[R^{ac}(-p_0,p_{0'})\right]_{00'}^{st_0 st_{0}}\left[K_D^{cb}(p_{0'},p_{B})\right]^{ist_{0'}}_{0'B}\nonumber
\end{align}
only in those cases where all of the scalar factors that arise from \eqref{X-as-R} and \eqref{Z-as-R} conveniently cancel out when the explicit $R$-matrix coefficients are substituted, as it turns out that $(x_{pq}^{ac})^{1111}=(z_{pq}^{ac})^{1111}$.

\subsubsection*{\ul{\it Singlet boundary dual K-matrices}}
For singlet boundaries, the dual BYBE is satisfied by a dual $K$-matrix $K_{D}^{a}(p)$, for $a\in\{\text{L},\text{R},\tilde{\text{L}},\tilde{\text{R}}\}$, with
\begin{equation}
\begin{aligned}
K_D^{a}(p)\!=\!\mathbb{E}_{11}+g_D^{a}(p) \, \mathbb{E}_{22} \quad \text{with} \quad g_D^{\text{a}}(p)\!=\!\frac{s^{\text{a}}_D-x^-_{p}}{s^{\text{a}}_D + x^+_{p}} 
\quad , \quad 
g_D^{\tilde{\text{a}}}(p)\!=\!\frac{s^{\tilde{\text{a}}}_D + x^+_{p}}{s^{\tilde{\text{a}}}_D-x^-_{p}} \quad \text{for} \quad \text{a}\in\{\text{L},\text{R}\}  \,\, .
\end{aligned}
\end{equation}
In particular, each of the four homogeneous dual BYBEs, where both vector spaces $V_0,V_{0'}$ carry the same representation, is solved for the corresponding $K^a_D$-matrix shown above for any value of its constant $s_D^a\in\mathbb{C}$. The various mixed dual BYBEs, where the two spaces carry different representations, then require these constants to be related in the following way
\begin{align}
s_D^{L}=s_D^{\tilde{L}}=-\frac{1}{s^R_D}=-\frac{1}{s^{\tilde{R}}_D} \,\, .
\end{align}

\subsubsection*{\ul{\it Proof of transfer matrix commutativity}}
Let us now prove that the transfer matrix defined as the supertrace of (\ref{double_row_monodromy}) is commutative, provided that the partial monodromies $T_+(p),T_-(p)$ satisfy equations (\ref{dual_BYBE}) and (\ref{general-right-BYBE}) respectively. As already discussed at various stages above, while the double-row monodromy $T_{-}$ is of the form \eqref{vector-monodromy} and exhibits a non-trivial action on the physical spaces, the dual monodromy $T_{+}$ is taken to coincide with the solution to the dual BYBE as in \eqref{explicit-dual-monodromy} and acts trivially on the physical spaces. To simplify notation, we omit the momentum arguments in $T_{\pm}(p_i),\,i=0,0'$, which always matches the space in which the monodromy acts. We also omit denoting the momenta in the $X,Y,Z$ matrices as they are always ordered as  $X^{ac}(p_0,p_{0'})\equiv X^{ac}$ etc. Let us start from
\begin{equation}
\begin{aligned}
    \tau^{ab}(p_0) \tau^{cb}(p_{0'})&=\text{str}_0([T^{ab}_+]_{0B}[T^{ab}_-]_{0B})\text{str}_{0'}([T^{cb}_+]_{0'B}[T^{cb}_-]_{0'B})\\
                &=\text{str}_0\left([T_+^{ab}]_{0B}^{st_0}[T_+^{ab}]_{0B}^{st_0}\right)\text{str}_{0'}([T_+^{cb}]_{0'B}[T_+^{cb}]_{0'B})\\
                &=\text{str}_{00'}\left([T_+^{ab}]_{0B}^{st_0}[T_-^{ab}]_{0B}^{st_0}[T_+^{cb}]_{0'B}[T_-^{cb}]_{0'B}\right) \,\, .
\end{aligned}
\end{equation}
The first non-trivial step is exchanging $[T_-^{ab}]_{0B}$ and $[T_+^{cb}]_{0'B}$. In the case of a singlet boundary, where there is no boundary space, these two matrices act non-trivially in different spaces entirely and therefore commute trivially. However, when dealing with a vector boundary they generally do not commute as they both act in the boundary space and a specific choice of $T_+$ or $T_-$ needs to be made. In this work, as discussed in appendix \ref{app:commuting-monodromies}, we fix $K$ to be a solution of the vector-boundary BYBE and $K_D$ to be given by (\ref{vector-K-plus}). With this choice, dressing $K$ with R-matrices leads to a $T_-$ that acts on the boundary, while $T_+ (p)=K_D (p)$ does not, i.e. equation \eqref{T-plus-T-minus-commutativity} is satisfied. Therefore, in both the singlet and vector boundary cases, we can rewrite the above expression as 
\begin{align}
     \tau^{ab}(p_0) \tau^{cb}(p_{0'}) &=\text{str}_{00'}\left([T_+^{ab}]_{0B}^{st_0}[T_+^{cb}]_{0'B}[T_-^{ab}]_{0B}^{st_0}[T_-^{cb}]_{0'B}\right).
\end{align}
In order to obtain $\tau^{cb}(p_{0'})\tau^{ab}(p_0)$, we would like to also exchange $[T_+^{ab}]_{0B}^{st_0},[T_+^{cb}]_{0'B}$ and $[T_-^{ab}]_{0B}^{st_0},[T_-^{cb}]_{0'B}$. Since these pairs generally do not commute, particularly $[T_-^{ab}]_{0B}^{st_0}$ and $[T_-^{cb}]_{0'B}$ which both act on the physical spaces, this can only be achieved via the two reflection equations. To utilize these relations, let us use (\ref{general_rel_3}) to introduce a pair of $R$-matrices and rewrite the resulting expression using properties of the supertrace and supertransposition.
\begin{equation}
\resizebox{0.9\hsize}{!}{$
\begin{aligned}
      \tau^{ab}(p_0) \tau^{cb}(p_{0'})&=\!\text{str}_{00'}\!\left([T_+^{ab}]_{0B}^{st_0}[T_+^{cb}]_{0'B} \left[Z^{ac}\right]_{00'}^{st_{0'}}\left[R^{ac}(p_0,-p_{0'})\right]_{00'}^{st_0}[T_-^{ab}]_{0B}^{st_0}[T_-^{cb}]_{0'B}\right)\\
    &=\!\text{str}_{00'}\left(\left([T_+^{ab}]_{0B}^{st_0}\left[Z^{ac}\right]_{00'}[T_+^{cb}]_{0'B}^{ist_{0'}}\right)^{st_{0'}}\! \left([T_-^{ab}]_{0B}\left[R^{ac}(p_0,-p_{0'})\right]_{00'}\![T_-^{cb}]_{0'B}\right)^{st_0}\!\!\right)\\
    &=\!\text{str}_{00'}\!\left(\left([T_+^{ab}]_{0B}^{st_0}\!\left[Z^{ac}\right]_{00'}[T_+^{cb}]_{0'B}^{ist_{0'}}\right)^{ist_0 \, st_{0'}}[T_-^{ab}]_{0B}\left[R^{ac}(p_0,-p_{0'})\right]_{00'}[T_-^{cb}]_{0'B}\right)\, ,
\end{aligned} $}
\end{equation}
where $ist_{0,0'}$ denotes the inverse supertransposition in the respective space. Similarly, using braiding unitarity \eqref{unitarity-general} we have
\begin{equation}
\begin{aligned}
     & \tau^{ab}(p_0) \tau^{cb}(p_{0'})=\\
     &=\!\text{str}_{00'}\!\!\left[\left([T_+^{ab}]_{0B}^{st_0}\left[Z^{ac}\right]_{00'}[T_+^{cb}]_{0'B}^{ist_{0'}}\right)^{ist_0st_{0'}}[T_-^{ab}]_{0B}\left[R^{ac}(p_0,-p_{0'})\right]_{00'}[T_-^{cb}]_{0'B}\right]\\
     &=\!f^{ac}_{p_0p_{0'}}\text{str}_{00'}\!\!\left[\!\left([T_+^{ab}]_{0B}^{st_0}\left[Z^{ac}\right]_{00'}[T_+^{c}]_{0'B}^{ist_{0'}}\right)^{ist_0st_{0'}}[R^{ac}(-p_0,-p_{0'})]_{00'}\right.\\
     &\!\!\!\!\!\!\hphantom{\left([T_+^{a}]_{0B}^{st_0}\left[Z^{ac}\right]_{00'}([T_+^{c}]_{0'B})^{ist_{0'}}\right)^{ist_{0}}}\left[R^{ca}(-p_{0'},-p_0)\right]_{00'}^{op}[T_-^{ab}]_{0B}\left[R^{ac}(p_0,-p_{0'})\right]_{00'}[T_-^{cb}]_{0'B}\!\Bigr]\\
     &=\!f^{ac}_{p_0p_{0'}}\text{str}_{00'}\left[\!\left(\left[R^{ac}(-p_0,-p_{0'})\right]^{st_0ist_{0'}}_{00'}[T_+^{ab}]_{0B}^{st_0}\left[Z^{ac}\right]_{00'}[T_+^{cb}]_{0'B}^{ist_{0'}}\right)^{ist_0st_{0'}}\right.\\
     &\hphantom{\left(\left[R^{ac}(-p_0,-p_{0'})\right]^{st_0ist_{0'}}_{00'}aaa\right)}\left[R^{ca}(-p_{0'},-p_0)\right]_{00'}^{op}[T_-^{ab}]_{0B}\left[R^{ac}(p_0,-p_{0'})\right]_{00'}[T_-^{cb}]_{0'B}\!\biggr]
\end{aligned}
\end{equation}
The expression inside the trace is now a product of two terms, each being one side of one of the two reflection equations (\ref{general-right-BYBE}),(\ref{dual_BYBE_appendix}). We apply them both and follow the previous steps in reverse:
\begin{align}
     & \tau^{ab}(p_0) \tau^{cb}(p_{0'})\nonumber\\
     &\!\!=\!\!f^{ac}_{p_{0}p_{0'}}\text{str}_{00'}\!\left[\!\left(\! [T_+^{cb}]_{0'B}^{ist_{0'}}\left[X^{ac}\right]_{00'}[T_+^{ab}]_{0B}^{st_0}\left[Y^{ac}\right]_{00'}\!\right)^{ist_0st_{0'}}\![T_-^{cb}]_{0'B}\left[R^{ca}(p_{0'},-p_0)\right]^{op}_{00'}[T_-^{ab}]_{0B}\left[R^{ac}(p_0,p_{0'})\right]_{00'}\!\right]\nonumber\\
     &\!\!=\!\!f^{ac}_{p_0p_{0'}}\text{str}_{00'}\!\left[\! [T_+^{cb}]_{0'B}^{ist_{0'}}\left[X^{ac}\right]_{00'}[T_+^{ab}]_{0B}^{st_0}\left[Y^{ac}\right]_{00'}\!\left([T_-^{cb}]_{0'B}\left[R^{ca}(p_{0'},-p_0)\right]^{op}_{00'}[T_-^{ab}]_{0B}\left[R^{ac}(p_0,p_{0'})\right]_{00'}\right)^{st_0ist_{0'}}\!\right]\nonumber\\
     &\!\!=\!\!f^{ac}_{p_0p_{0'}}\text{str}_{00'}\!\left[\! [T_+^{cb}]_{0'B}^{ist_{0'}}\left[X^{ac}\right]_{00'}[T_+^{ab}]_{0B}^{st_0}\left[Y^{ac}\right]_{00'}\!\left([T_-^{cb}]_{0'B}\left[R^{ca}(p_{0'},-p_0)\right]^{op}_{00'}[T_-^{ab}]_{0B}\left[R^{ac}(p_0,p_{0'})\right]_{00'}\right)^{st_0ist_{0'}}\!\right]\nonumber\\
     &\!\!=\!\!f^{ac}_{p_0p_{0'}}\text{str}_{00'}\!\left[\! [T_+^{cb}]_{0'B}^{ist_{0'}}\left[X^{ac}\right]_{00'}[T_+^{ab}]_{0B}^{st_0}\left[Y^{ac}\right]_{00'} \!\!\left[R^{ac}(p_0,p_{0'})\right]_{00'}^{st_0ist_{0'}}\!\!\left(\![T_-^{cb}]_{0'B}\left[R^{ca}(p_{0'},p_0)\right]^{op}_{00'}[T_-^{ab}]_{0B}\!\right)^{st_0ist_{0'}}\!\right]\nonumber\\
     &\!\!=\!\text{str}_{00'}\!\left[\! \left([T_+^{cb}]_{0'B}^{ist_{0'}}\left[X^{ac}\right]_{00'}[T_+^{ab}]_{0B}^{st_0}\right)^{ist_{0}}\left([T_-^{cb}]_{0'B}\left[R^{ca}(p_{0'},-p_0)\right]_{00'}^{op}[T_-^{ab}]_{0B}\right)^{ist_{0'}}\!\right]\\
     &\!\!=\!\text{str}_{00'}\!\left[\! \left([T_+^{cb}]_{0'B}^{ist_{0'}}\left[X^{ac}\right]_{00'}[T_+^{ab}]_{0B}^{st_0}\right)^{ist_{0}}\left([T_-^{cb}]_{0'B}\left[R^{ca}(p_{0'},-p_0)\right]^{op}_{00'}[T_-^{ab}]_{0B}\right)^{ist_{0'}}\!\right]\nonumber\\
     &\!\!=\!\text{str}_{00'}\!\left[ [T_+^{cb}]_{0'B}^{ist_{0'}}[T_+^{ab}]_{0B}\left[X^{ac}\right]_{00'}^{ist_{0}}\left(\left[R^{ca}(p_{0'},-p_0)\right]^{op}_{00'}\right)^{ist_{0'}}[T_-^{cb}]_{0'B}^{ist_{0'}}[T_-^{ab}]_{0B}\right]\nonumber\\
     &\!\!=\!\text{str}_{00'}\!\left[ [T_+^{cb}]_{0'B}^{ist_{0'}}[T_+^{ab}]_{0B}[T_-^{cb}]_{0'B}^{ist_{0'}}[T_-^{ab}]_{0B}\right]\nonumber
\end{align}
where relations (\ref{general_rel_1}),(\ref{general_rel_2}) were used. Assuming, as in the beginning, that $[T_+^{a}]_{0B}$ and $[T_-^{c}]_{0'B}$ commute, we finally have
\begin{equation}
\begin{aligned}
      \tau^{ab}(p_0) \tau^{cb}(p_{0'})&=\text{str}_{00'}\left[ [T_+^{cb}]_{0'B}^{ist_{0'}}[T_-^{cb}]_{0'B}^{ist_{0'}}[T_+^{ab}]_{0B}[T_-^{ab}]_{0B}\right]\\
     &=\text{str}_{0'}\left[ [T_+^{cb}]_{0'B}^{ist_{0'}}[T_-^{cb}]_{0'B}^{ist_{0'}}\right]\text{str}_0\left[[T_+^{ab}]_{0B}[T_-^{ab}]_{0B}\right]\\
     &=\text{str}_{0'}\left[ [T_+^{cb}]_{0'B}[T_-^{cb}]_{0'B}\right]\text{str}_0\left[[T_+^{ab}]_{0B}[T_-^{ab}]_{0B}\right]\\
     &= \tau^{cb}(p_{0'}) \tau^{ab}(p_0) \,\, .
\end{aligned}
\end{equation}
\end{section}

\section{Mixed auxiliary representations}\label{app:6vandfake-6v}
In this appendix we reconsider the construction of the exchange relations for the operators $\mathcal{A},\mathcal{B},\mathcal{C},\mathcal{D}$ which characterise the boundary algebraic Bethe ansatz. While in section \ref{subsec:singlet-boundary} and \ref{subsubsec:vector-all-L-reps} we focused for simplicity on the case of auxiliary excitations in the $\text{L}$ representation, here we will attempt to generalise the construction to vector boundaries in representation $b$ and auxiliary excitations in representations $a,c$, for any $ a,b,c \in  \{ \text{L},\text{R},\tilde{\text{L}},\tilde{\text{R}} \}$. Due to the result of appendix \ref{app:Tminus-satisfies-BYBE} this extension of the formalism seems like a natural step, however, as we will now describe, while the construction remains completely unaffected by the extension to a vector boundary, it turns out to exhibit obstructions when considering certain combinations of representations for the auxiliary excitations. We will clarify this result, highlighting how its origin lies in the structure of certain $R$-matrices of the theory, which even exhibiting only 6 non-vanishing entries still carry some crucial information of a typical 8-vertex setting. 

The starting point of the construction is the BYBE \eqref{BYBE-appendix}, which we would like to write in a form that generalises \eqref{indices} to any possible set of representations. To this aim we write the monodromies as
\begin{equation}
[T_{-}^{ab}(p_{0},p_{B})]_{0B} := \mathbb{E}_{\mu\nu} \otimes \mathds{1} \otimes \mathcal{A}_{\mu\nu}^{ab}(p_{0},p_{B})
\qquad \qquad 
[T_{-}^{cb}(p_{0'},p_{B})]_{0B} :=  \mathds{1} \otimes \mathbb{E}_{\rho\sigma} \otimes \mathcal{A}_{\rho\sigma}^{cb}(p_{0'},p_{B}) \,\, ,
\end{equation}
with $\mathcal{A}^{ab}_{\mu\nu}$ quantum operators analogous to the ones in \eqref{operators-1} and \eqref{operators-2}, now acting on the physical spaces and a non-trivial boundary space $V_{1}\otimes ...\otimes V_{N}\otimes V_{B}$. In a similar way, we also write the $R$-matrices as
\begin{equation}
\begin{aligned}
[R^{ac}(p,q)]_{00'} &:=(r_{pq}^{ac})^{\alpha\beta\gamma\delta} \, \mathbb{E}_{\alpha\beta} \otimes \mathbb{E}_{\gamma\delta} \otimes \mathds{1}_{N+B}
\\
[R^{ca}(p,q)]_{00'}^{op} &:= (-1)^{(\lambda+\omega)(\tau+\eta)} \,(r_{pq}^{ca})^{\lambda\omega\tau\eta} \, \mathbb{E}_{\tau\eta} \otimes \mathbb{E}_{\lambda\omega} \otimes \mathds{1}_{N+B} \,\, ,
\end{aligned}
\end{equation}
with the generic momenta $(p,q)$ being $(p_{0},p_{0'})$, $(p_{0'},-p_{0})$ etc. as from equation \eqref{BYBE-appendix}, and the symbol $\mathds{1}_{N+B}$ meaning $\mathds{1}_{N}\otimes \mathds{1}_{B}$, having introduced $\mathds{1}_{N}$ in appendix \ref{app:notation-and-conventions}. Taking into account the grading of the tensor product, and the fact that $A_{\mu\nu}^{ab}$ carry a grading equal to $\mu+\nu$, one finally arrives at 
\begin{equation}\label{operators-relations-appendix}
\begin{aligned}
&\!\!(-1)^{(\gamma+\rho)(\tau+\beta)+(\rho+\sigma)(\mu+\beta+\gamma+\delta)}(r_{p_{0'} \, -p_{0}}^{ca})^{\sigma\gamma\tau\mu}(r_{p_{0}p_{0'}}^{ac})^{\nu\beta\gamma\delta}\mathbb{E}_{\tau\beta}\otimes \mathbb{E}_{\rho\delta}\otimes \mathcal{A}_{\rho\sigma}^{cb}(p_{0'},p_{B})\mathcal{A}_{\mu\nu}^{ab}(p_{0},p_{B})=
\\
&\!\!(-1)^{(\gamma+\rho)(\tau+\beta)+(\mu+\nu)(\nu+\beta+\gamma+\delta)}(r_{-p_{0'} \, -p_{0}}^{ca})^{\sigma\gamma\tau\mu}(r_{p_{0} \, -p_{0'}}^{ac})^{\nu\beta\gamma\delta}\mathbb{E}_{\tau\beta}\otimes \mathbb{E}_{\rho\delta}\otimes \mathcal{A}_{\mu\nu}^{ab}(p_{0},p_{B})\mathcal{A}_{\sigma\delta}^{cb}(p_{0'},p_{B}) ,
\end{aligned}
\end{equation}
which encodes a set of 16 relations, one for each combination of indices $\{\tau,\beta,\rho,\delta\}\in\{1,2\}$, between products of quantum operators. In each relation, the entries of generic $R$-matrices listed in section \ref{sec:bulk-theory} feature as weighting factors for the products of quantum operators, ultimately determining the relations amon them for each choice of representations $a,c$. It is here crucial to notice that while all $R$-matrices in section \ref{sec:bulk-theory} exhibit non-vanishing diagonal entries (encoded in the matrix elements $\mathbb{E}_{11}\otimes\mathbb{E}_{11},\mathbb{E}_{11}\otimes\mathbb{E}_{22},\mathbb{E}_{22}\otimes\mathbb{E}_{11},\mathbb{E}_{22}\otimes\mathbb{E}_{22}$), not all of them exhibit the same structure for the off-diagonal terms, which happen to fall into two main categories.
\subsection{Genuine 6-vertex case}\label{app:genuine-6-vertex-subsection}
This first class of $R$-matrices takes the following general structure
\begin{equation}
R\simeq 
\begin{pmatrix}
\# & 0 & 0 & 0
\\
0 & \# & \# & 0
\\
0 & \# & \# & 0
\\
0 & 0 & 0 & \#
\end{pmatrix} \,\, ,
\end{equation}
and is characterised by non-vanishing "internal" off-diagonal elements $\mathbb{E}_{12}\otimes\mathbb{E}_{21},\mathbb{E}_{21}\otimes\mathbb{E}_{12}$ and vanishing "external" off-diagonal elements $\mathbb{E}_{12}\otimes\mathbb{E}_{12},\mathbb{E}_{21}\otimes\mathbb{E}_{21}$. This is typical of models of the \textit{6-vertex type} and from a quick look at section \ref{sec:bulk-theory} one can realise that $R^{\text{L}\text{L}},R^{\text{R}\text{R}},R^{\tilde{\text{L}}\tilde{\text{L}}},R^{\tilde{\text{R}}\tilde{\text{R}}},R^{\text{L}\tilde{\text{R}}},R^{\text{R}\tilde{\text{L}}},R^{\tilde{\text{L}}\text{R}},R^{\tilde{\text{R}}\text{L}}$ fall in this set. The identities \eqref{operators-relations-appendix} can in this setting be conveniently disentangled in a fashion which generalises the steps described in section \ref{sec:Algebraic-Bethe-Ansatz}, leading to the following set of relations.
\begin{align}
\mathcal{B}^{ab}(p_{0},p_{B})\mathcal{B}^{cb}(p_{0'},p_{B}) = & \varphi^{ac}(p_{0},p_{0'}) \mathcal{B}^{cb}(p_{0'},p_{B})\mathcal{B}^{ab}(p_{0},p_{B}) \,\, ,
\label{BB-commutator-general} \\[2em]
\mathcal{A}^{ab}(p_{0},p_{B})\mathcal{B}^{cb}(p_{0'},p_{B}) = &  \hat{\varphi}^{ac}(p_{0},p_{0'}) \mathcal{B}^{cb}(p_{0'},p_{B}) \mathcal{A}^{ab}(p_{0},p_{B}) +
\notag \\[1em]
+ & \hat{\chi}^{ac}(p_{0},p_{0'})\mathcal{B}^{ab}(p_{0},p_{B})\mathcal{A}^{cb}(p_{0'},p_{B}) +
\notag \\[1em]
+ & \hat{\psi}^{ac}(p_{0},p_{0'})\mathcal{B}^{ab}(p_{0},p_{B})\mathcal{D}^{cb}(p_{0'},p_{B}) \,\, ,
\label{AB-commutator-general} \\[2em]
\mathcal{D}^{ab}(p_{0},p_{B})\mathcal{B}^{cb}(p_{0'},p_{B}) = &  \tilde{\varphi}^{ac}(p_{0},p_{0'}) \mathcal{B}^{cb}(p_{0'},p_{B}) \mathcal{D}^{ab}(p_{0},p_{B}) +
\notag\\[1em]
+ & \tilde{\chi}^{ac}(p_{0},p_{0'})\mathcal{B}^{ab}(p_{0},p_{B})\mathcal{D}^{cb}(p_{0'},p_{B}) +
\notag \\[1em]
+ & \tilde{\psi}^{ac}(p_{0},p_{0'})\mathcal{B}^{ab}(p_{0},p_{B})\mathcal{A}^{cb}(p_{0'},p_{B}) + 
\notag \\[1em]
+ & \tilde{\xi}^{ac}(p_{0},p_{0'})\mathcal{B}^{cb}(p_{0'},p_{B})\mathcal{A}^{ab}(p_{0},p_{B}) \,\, ,
\label{DB-commutator-general}
\end{align}
where for simplicity we introduced coefficients
\begin{align}
\varphi^{ac}(p_{0},p_{0'},p_{B})&:=- \frac{(r^{ca}_{p_{0'}\,-p_{0}})^{2211}(r^{ac}_{p_{0}p_{0'}})^{2222}}{(r^{ca}_{-p_{0'}\,-p_{0}})^{1111}(r^{ac}_{p_{0} \, -p_{0'}})^{2211}} \,\, ,
\label{phi-coefficient}\\[3em]
\hat{\varphi}^{ac}(p_{0},p_{0'},p_{B})&:=\frac{(r^{ca}_{p_{0'}\,-p_{0}})^{2211}\Bigl[(r^{ac}_{p_{0}p_{0'}})^{1221}(r^{ac}_{p_{0} p_{0'}})^{2112}+(r^{ac}_{p_{0}p_{0'}})^{1122}(r^{ac}_{p_{0} p_{0'}})^{2211}\Bigr]}{(r^{ac}_{p_{0}\,-p_{0'}})^{1111}(r^{ca}_{-p_{0'} \,  -p_{0}})^{1111}(r^{ac}_{p_{0} \,  p_{0'}})^{2211}} \,\, ,
\notag\\
\label{phi-tilde+phi-hat-coefficients}\\
\tilde{\varphi}^{ac}(p_{0},p_{0'},p_{B})&:= \frac{(r^{ac}_{p_{0}p_{0'}})^{2222}\Bigl[(r^{ca}_{p_{0'}\,-p_{0}})^{1111}(r^{ca}_{p_{0'}\, -p_{0}})^{2222}-(r^{ca}_{p_{0'}\,-p_{0}})^{1221}(r^{ca}_{p_{0'}\,- p_{0}})^{2112}\Bigr]}{(r^{ca}_{p_{0'}\,-p_{0}})^{1111}(r^{ca}_{-p_{0'} \,  -p_{0}})^{1122}(r^{ac}_{p_{0} \,  -p_{0'}})^{2211}} \,\, ,
\notag \\[3em]
\hat{\chi}^{ac}(p_{0},p_{0'},p_{B})&:=\frac{(r^{ac}_{p_{0}p_{0'}})^{2112}(r^{ac}_{p_{0} \, -p_{0'}})^{2211}}{(r^{ac}_{p_{0}\,-p_{0'}})^{1111}(r^{ac}_{p_{0} p_{0'}})^{2211}} \,\, ,
\notag \\
\label{chi-tilde+chi-hat-coefficients}\\
\tilde{\chi}^{ac}(p_{0},p_{0'},p_{B})&:= \frac{(r^{ca}_{-p_{0'} \, -p_{0}})^{1221}\Bigl[(r^{ac}_{p_{0}\,-p_{0'}})^{1221}(r^{ac}_{p_{0}\, -p_{0'}})^{2112}-(r^{ac}_{p_{0}\,-p_{0'}})^{1111}(r^{ac}_{p_{0}\,- p_{0'}})^{2222}\Bigr]}{(r^{ac}_{p_{0}\,-p_{0'}})^{1111}(r^{ca}_{-p_{0'} \,  -p_{0}})^{1122}(r^{ac}_{p_{0} \,  -p_{0'}})^{2211}} \,\, ,
\notag \\[3em]
\hat{\psi}^{ac}(p_{0},p_{0'},p_{B})&:=-\frac{(r^{ac}_{p_{0} \, -p_{0'}})^{2112}}{(r^{ac}_{p_{0} \, -p_{0'}})^{1111}} \,\, ,
\notag\\
\label{phi-tilde+psi-hat-coefficients}\\
\tilde{\psi}^{ac}(p_{0},p_{0'},p_{B})&:= \Biggl[ \frac{(r^{ca}_{-p_{0'}\, -p_{0}})^{1111}(r^{ca}_{p_{0'}\, -p_{0}})^{1221}(r^{ac}_{p_{0}p_{0'}})^{2222}}{(r^{ca}_{p_{0'}\, -p_{0}})^{1111}(r^{ca}_{-p_{0'}\, -p_{0}})^{1122}(r^{ac}_{p_{0}p_{0'}})^{2211}}- \frac{(r^{ac}_{p_{0}\, -p_{0'}})^{1221}(r^{ca}_{-p_{0'}\, -p_{0}})^{1221}(r^{ac}_{p_{0}\, p_{0'}})^{2112}}{(r^{ac}_{p_{0}\, -p_{0'}})^{1111}(r^{ca}_{-p_{0'}\, -p_{0}})^{1122}(r^{ac}_{p_{0}p_{0'}})^{2211}} \Biggr] \,\, ,
\notag \\[3em]
\tilde{\xi}^{ac}(p_{0},p_{0'},p_{B})&:=\Biggl[ \frac{(r^{ca}_{p_{0'}\,-p_{0}})^{2211}(r^{ac}_{p_{0}p_{0'}})^{1221}(r^{ca}_{p_{0'}\, -p_{0}})^{1221} (r^{ac}_{p_{0} p_{0'}})^{2222}}{(r^{ca}_{p_{0'}\, -p_{0}})^{1111}(r^{ca}_{-p_{0'}\, -p_{0}})^{1122}(r^{ac}_{p_{0}\, -p_{0'}})^{2211}(r^{ac}_{p_{0}p_{0'}})^{2211}} +
\notag\\[1em]
& \quad -\frac{(r^{ca}_{p_{0'}\,-p_{0}})^{2211}(r^{ac}_{p_{0}\, -p_{0'}})^{1221}(r^{ca}_{-p_{0'}\, -p_{0}})^{1221}(r^{ac}_{p_{0}p_{0'}})^{1221} (r^{ac}_{p_{0}p_{0'}})^{2112}}{(r^{ac}_{p_{0}\, -p_{0'}})^{1111}(r^{ca}_{-p_{0'}\, -p_{0}})^{1111}(r^{ca}_{-p_{0'}\, -p_{0}})^{1122}(r^{ac}_{p_{0}\, -p_{0'}})^{2211}(r^{ac}_{p_{0}p_{0'}})^{2211}}+
\notag\\[1em]
& \quad -\frac{(r^{ca}_{p_{0'}\,-p_{0}})^{2211}(r^{ac}_{p_{0}\, -p_{0'}})^{1221}(r^{ca}_{-p_{0'}\, -p_{0}})^{1221}(r^{ac}_{p_{0}p_{0'}})^{1122}}{(r^{ac}_{p_{0}\, -p_{0'}})^{1111}(r^{ca}_{-p_{0'}\, -p_{0}})^{1111}(r^{ca}_{-p_{0'}\, -p_{0}})^{1122}(r^{ac}_{p_{0}p_{0'}})^{2211}}\Biggr] \,\, .
\label{xi-tilde-coefficient}
\end{align}
Relation \eqref{BB-commutator-general} shows that the braided statistics found for the creation operators in equation \eqref{exch} for the singlet case with $a=c=\text{L}$ (correctly recovered from this more general setup), is valid for any interaction of auxiliary representations involving 6-vertex $R$-matrices, up to details encoded in the braiding factor. Some potential deviations from the argument in section \ref{sec:Algebraic-Bethe-Ansatz} seem however to arise from the other two relations, \eqref{AB-commutator-general} and \eqref{DB-commutator-general}, which exhibit two new features with respect to section \ref{sec:Algebraic-Bethe-Ansatz}.
\begin{itemize}
\item The coefficients $\hat{\varphi}^{ac}(p_{0},p_{0'})$ and $\tilde{\varphi}^{ac}(p_{0},p_{0'})$ in \eqref{phi-tilde+phi-hat-coefficients}, appearing in front of
$\mathcal{B}^{cb}(p_{0'},p_{B})\mathcal{A}^{ab}(p_{0},p_{B})$ on the right hand side of \eqref{AB-commutator-general} and $\mathcal{B}^{cb}(p_{0'},p_{B})\mathcal{D}^{ab}(p_{0},p_{B})$ on the right hand side of \eqref{DB-commutator-general}, seem to be quite different. For the boundary Bethe ansatz to work properly, these two terms should be the same, since their contributions should sum up as an eigenvalue term for the transfer matrix. Fortunately, substituting the explicit entries for any of the 6-vertex $R$-matrices listed above, one can indeed check that these two coefficients are always the same, even though their specific form differs for each choice of representations.
\item The huge coefficient $\tilde{\xi}^{ac}(p_{0},p_{0'})$ in \eqref{xi-tilde-coefficient}, appearing in front of $\mathcal{B}^{cb}(p_{0'},p_{B})\mathcal{A}^{ab}(p_{0},p_{B})$ on the right hand side of \eqref{DB-commutator-general}, should vanish, for the structure of such relation to match the one found in section \ref{sec:Algebraic-Bethe-Ansatz}. Once again, substituting the explicit entries for any of the 6-vertex $R$-matrices reveals the correctness of this expectation.
\end{itemize}
We conclude that the structure of the exchange relations \eqref{BB-commutator-general}, \eqref{AB-commutator-general} and \eqref{DB-commutator-general} is exactly the same as the one found in section \ref{sec:Algebraic-Bethe-Ansatz}, for either singlet or vector boundaries and any choice of 6-vertex interaction in the auxiliary spaces - up to details encoded in the above representation-dependent coefficients. 

\subsection{Fake 6-vertex case}
This second class of $R$-matrices takes the following general structure
\begin{equation}
R\simeq 
\begin{pmatrix}
\# & 0 & 0 & \#
\\
0 & \# & 0 & 0
\\
0 & 0 & \# & 0
\\
\# & 0 & 0 & \#
\end{pmatrix} \,\, ,
\end{equation}
and is characterised by vanishing "internal" off-diagonal elements $\mathbb{E}_{12}\otimes\mathbb{E}_{21},\mathbb{E}_{21}\otimes\mathbb{E}_{12}$ and non-vanishing "external" off-diagonal elements $\mathbb{E}_{12}\otimes\mathbb{E}_{12},\mathbb{E}_{21}\otimes\mathbb{E}_{21}$. From the summary in section \ref{sec:bulk-theory} it is clear that $R^{\text{L}\text{R}},R^{\text{R}\text{L}},R^{\text{L}\tilde{\text{L}}},R^{\text{R}\tilde{\text{R}}},R^{\tilde{\text{L}}\text{L}},R^{\tilde{\text{R}}\text{R}},R^{\tilde{\text{L}}\tilde{\text{R}}},R^{\tilde{\text{R}}\tilde{\text{L}}}$ fall in this second set of matrices. Despite exhibiting 6 non-vanishing entries, as one would typically expect for 6-vertex models, the location of the non vanishing off-diagonal entries is in this case reminiscent of a model of the \textit{8-vertex type}. We are thus dealing with a mixed structure lying somewhat in between a 6 and an 8 vertex model and it is a priori not clear whether this would allow for the application of the algebraic Bethe ansatz, as for a typical 6-vertex case, or would on the other hand prevent it, as for a typical 8-vertex case. Also in this setting the relations \eqref{operators-relations-appendix} can be disentangled in a fashion which generalises the steps followed in section \ref{sec:Algebraic-Bethe-Ansatz}, however this time the result appears quite different from the previous case, as the following schematic expressions arise
\begin{align}
\mathcal{B}^{ab}&(p_{0},p_{B})\mathcal{B}^{cb}(p_{0'},p_{B}) = 
\notag\\[0.5em]
&=\varphi^{ac} \mathcal{B}^{cb}(p_{0'},p_{B})\mathcal{B}^{ab}(p_{0},p_{B})  + \omega^{ac} \mathcal{C}^{cb}(p_{0'},p_{B})\mathcal{C}^{ab}(p_{0},p_{B}) + \theta^{ac} \mathcal{C}^{ab}(p_{0},p_{B})\mathcal{C}^{cb}(p_{0'},p_{B}) \,\, ,
\label{BB-commutator-general-8v} \\[1em]
\mathcal{A}^{ab}&(p_{0},p_{B})\mathcal{B}^{cb}(p_{0'},p_{B}) =
\notag\\[0.5em]
&=\hat{\varphi}^{ac} \mathcal{B}^{cb}(p_{0'},p_{B}) \mathcal{A}^{ab}(p_{0},p_{B}) + \hat{\omega}^{ac}\mathcal{A}^{cb}(p_{0'},p_{B})\mathcal{C}^{ab}(p_{0},p_{B}) + \hat{\theta}^{ac}\mathcal{D}^{cb}(p_{0'},p_{B})\mathcal{C}^{ab}(p_{0},p_{B}) \,\, ,
\label{AB-commutator-general-8v} \\[1em]
\mathcal{D}^{ab}&(p_{0},p_{B})\mathcal{B}^{cb}(p_{0'},p_{B}) = 
\notag\\[0.5em]
&=\tilde{\varphi}^{ac} \mathcal{B}^{cb}(p_{0'},p_{B}) \mathcal{D}^{ab}(p_{0},p_{B}) + \tilde{\omega}^{ac}\mathcal{A}^{cb}(p_{0'},p_{B})\mathcal{C}^{ab}(p_{0},p_{B}) + \tilde{\theta}^{ac}\mathcal{D}^{cb}(p_{0'},p_{B})\mathcal{C}^{ab}(p_{0},p_{B}) \,\, ,
\label{DB-commutator-general-8v}
\end{align}
with all coefficients being functions of $(p_{0},p_{0'})$. A first obvious difference between the above equations and \eqref{BB-commutator-general}, \eqref{AB-commutator-general}, \eqref{DB-commutator-general}, can be found in the exchange relation for the $\mathcal{B}$-operators, which in the spirit of section \ref{sec:Algebraic-Bethe-Ansatz} should be interpreted as creation operators. While in \eqref{BB-commutator-general} the creation of two excitations in reversed orders is the same up to a phase factor, equation \eqref{BB-commutator-general-8v} combines this information with the annihilation of two excitations, encoded in the $\mathcal{C}$-operators. A similar problem arises in the relations \eqref{AB-commutator-general-8v} and \eqref{DB-commutator-general-8v}, where once again the creation of an excitation via $\mathcal{B}$, followed by the diagonal action of $\mathcal{A}$ or $\mathcal{D}$ which neither create nor annihilate, is directly related to annihilations via $\mathcal{C}$. It is thus clear that in this setting the Bethe ansatz machinery of section \ref{sec:Algebraic-Bethe-Ansatz} does not apply in the standard way and the mixed structure of the $R$-matrices in this class turns out to reveal a closer connection to the 8-vertex case, for which problems in defining the pseudo-vacuum are well established, rather than to the 6-vertex one. There exists the possibility that the retained similarity with a 6-vertex model - namely the vanishing of the inner off-diagonal elements - might still allow to resolve the above obstructions by an appropriate redefinition of the creation and annihilation operators as (possibly very complicated) combinations of the $\mathcal{B}$ and $\mathcal{C}$ - something perhaps in the spirit of the transformations of \cite{DeLeeuw:2020ahx}. While very intriguing, this option remains outside the scope of the present analysis and is left as a possible future investigations.

Notice that a mathematical way to distinguish the true from the fake $6$-vertex case is that only the true $6$-vertex $R$-matrix preserves the {\it standard} $U(1)$ symmetry:
\begin{eqnarray}
\Delta^{op}(b) R = R\, \Delta(b) \qquad \text{with} \qquad \Delta(b) = b \otimes \mathds{1} + \mathds{1} \otimes b = \Delta^{op}(b) \qquad \text{and} \qquad b = (-1)^F,  
\end{eqnarray}
which means that for a true $6$-vertex $R$-matrix we have $[\Delta(b),R]=0$.
This is opposed to the fermionic number $b\otimes b$, which is always preserved for both true and fake $6$-vertex.

It is probably possible to perform a suitable map from the fake to the true $6$-vertex case: this map will most likely propagate through the formalism and update all the formulas. We have not found ourselves compelled to pursue this here.

\end{appendix}


\begin{thebibliography}{99}

\bibitem{Bogdan}
A.~Babichenko, B.~Stefa{\'n}ski, jr. and K.~Zarembo,
{\em Integrability and the {$AdS_3/CFT_2$} correspondence,} 
{JHEP {\bf 1003} (2010) 058}
  [\arXivlink{0912.1723}].
  
  
 
\bibitem{rev3}
  A.~Sfondrini,
  {\em Towards integrability for $AdS_3/CFT_2$,} 
  J.\ Phys.\ A {\bf 48} (2015)  023001
  [\arXivlink{1406.2971}]. 
 
\bibitem{Borsato:2016hud}
  R.~Borsato,
  {\em Integrable strings for AdS/CFT} 
  [\arXivlink{1605.03173}].

\bibitem{Seibold:2024qkh}
F.~K.~Seibold and A.~Sfondrini,
{\it $AdS_3$ Integrability, Tensionless Limits, and Deformations: A Review,}
[\arXivlink{2408.08414}].
 
\bibitem{Beisertreview}
G.~Arutyunov and S.~Frolov,
\emph{Foundations of the $AdS_5 \times S^5$ Superstring. Part I,}
J. Phys. A \textbf{42} (2009) 254003
[\arXivlink{0901.4937}].
N.~Beisert, C.~Ahn, L.~F.~Alday, Z.~Bajnok, J.~M.~Drummond, L.~Freyhult, N.~Gromov, R.~A.~Janik, V.~Kazakov and T.~Klose, \textit{et al.}
\emph{Review of AdS/CFT Integrability: An Overview,}
Lett. Math. Phys. \textbf{99} (2012) 3-32
[\arXivlink{1012.3982}].


\bibitem{OhlssonSax:2011ms}
O.~Ohlsson Sax and B.~Stefa{\'n}ski, jr.,
{\em Integrability, spin-chains, and the {$AdS_3/CFT_2$} correspondence,} 
{JHEP {\bf 1108} (2011) 029}
  [\arXivlink{1106.2558}].

\bibitem{seealso3}
R.~Borsato, O.~Ohlsson Sax, and A.~Sfondrini,
{\em {A dynamic $\mathfrak{su}(1|1)^2$ $S$-matrix for $AdS_3/CFT_2$},} 
{JHEP {\bf 1304} (2013) 113}
  [\arXivlink{1211.5119}].

\bibitem{Borsato:2012ss}
R.~Borsato, O.~Ohlsson Sax, and A.~Sfondrini,
{\em {All-loop Bethe ansatz equations for $AdS_3/CFT_2$},} 
{JHEP {\bf 1304} (2013) 116}
  [\arXivlink{1212.0505}].

\bibitem{Borsato:2013qpa}
R.~Borsato, O.~Ohlsson Sax, A.~Sfondrini, B.~Stefa\'nski, jr. and A.~Torrielli,
{\em {The all-loop integrable spin-chain for strings on AdS$_3 \times S^3
  \times T^4$: the massive sector},} 
{JHEP {\bf 1308} (2013) 043}
  [\arXivlink{1303.5995}].

\bibitem{Borsato:2014hja}
  R.~Borsato, O.~Ohlsson Sax, A.~Sfondrini and B.~Stefa\'nski, jr., 
{\it The complete AdS$_{3} \times$ S$^3 \times$ T$^4$ worldsheet S matrix,}
JHEP \textbf{10} (2014) 066
[\arXivlink{1406.0453}].

\bibitem{Borsato:2013hoa}
M.~Beccaria, F.~Levkovich-Maslyuk, G.~Macorini, and A.~Tseytlin,
{\em {Quantum corrections to spinning superstrings in $AdS_3\times S^3 \times
  M^4$: determining the dressing phase},} 
{JHEP {\bf 1304} (2013) 006}
  [\arXivlink{1211.6090}].
P.~Sundin and L.~Wulff,
{\em {World-sheet scattering in $AdS_3/CFT_2$},} 
{JHEP {\bf 1307} (2013) 007}
  [\arXivlink{1302.5349}].
L.~Bianchi, V.~Forini, and B.~Hoare,
{\em {Two-dimensional $S$-matrices from unitarity cuts},} 
{JHEP {\bf 1307} (2013) 088}
  [\arXivlink{1304.1798}].
O.~T.~Engelund, R.~W.~McKeown and R.~Roiban,
{\em Generalised unitarity and the worldsheet $S$-matrix in $AdS_n \times S^n \times M^{10-2n}$,} 
JHEP {\bf 1308} (2013) 023
[\arXivlink{1304.4281}].
R.~Borsato, O.~Ohlsson Sax, A.~Sfondrini, B.~Stefa\'nski, jr. and A.~Torrielli,
  {\em Dressing phases of $AdS_3/CFT_2$,} 
  Phys.\ Rev.\ D {\bf 88} (2013) 066004
  [\arXivlink{1306.2512}].
  L.~Bianchi and B.~Hoare,
  {\em $AdS_3 \times S^3 \times M^4$ string $S$-matrices from unitarity cuts,} 
  JHEP {\bf 1408} (2014) 097
  [\arXivlink{1405.7947}].
  
  [\arXivlink{1406.0453}].
  P.~Sundin and L.~Wulff,
  {\em The complete one-loop BMN $S$-matrix in $AdS_{3}\times  S^{3}\times T^{4}$,} 
  JHEP {\bf 1606} (2016) 062
[\arXivlink{1605.01632}].
  



 \bibitem{Sax:2012jv}
O.~Ohlsson Sax, B.~Stefa\'nski, jr. and A.~Torrielli,
{\em {On the massless modes of the $AdS_3/CFT_2$ integrable systems},} 
{JHEP {\bf 1303} (2013) 109}
  [\arXivlink{1211.1952}].

 
  
  O.~Ohlsson Sax, A.~Sfondrini and B.~Stefa\'nski, jr.,
  {\em Integrability and the Conformal Field Theory of the Higgs branch,} 
  JHEP {\bf 1506} (2015) 103
  [\arXivlink{1411.3676}].
  R.~Borsato, O.~Ohlsson Sax, A.~Sfondrini, B.~Stefa\'nski, jr. and A.~Torrielli,
  \emph{On the dressing factors, Bethe equations and Yangian symmetry of strings on $AdS_3 \times S^3 \times T^4$,} 
  J.\ Phys.\ A {\bf 50} (2017) 024004
  [\arXivlink{1607.00914}].
 
  M.~Baggio, O.~Ohlsson Sax, A.~Sfondrini, B.~Stefa\'nski jr. and A.~Torrielli,
  {\em Protected string spectrum in $AdS_3/CFT_2$ from worldsheet integrability,} 
  JHEP \textbf{04} (2017) 091
[\arXivlink{1701.03501}].
  O.~Ohlsson Sax and B.~Stefa\'nski, jr.,
  \emph{Closed strings and moduli in AdS$_{3}$/CFT$_{2}$,}
  JHEP {\bf 1805} (2018) 101
  [\arXivlink{1804.02023}].


 \bibitem{Zamol2}
A.~B.~Zamolodchikov and A.~B.~Zamolodchikov,
	{\em Massless factorized scattering and sigma models with topological terms,}
Nucl.\ Phys.\ B {\bf 379} (1992) 602.
P.~Fendley, H.~Saleur and A.~B.~Zamolodchikov,
\emph{Massless flows, 2. The Exact S-matrix approach,}
Int.\ J.\ Mod.\ Phys.\ A {\bf 8} (1993) 5751
[\arXivlink{hep-th/9304051}].
P.~Fendley and K.~A.~Intriligator,
\emph{Exact $N=2$ Landau-Ginzburg flows,}
Nucl. Phys. B \textbf{413} (1994) 653-674
[\arXivlink{hep-th/9307166}].

\bibitem{Fendley:1993jh}
  P.~Fendley and H.~Saleur,
  \emph{Massless integrable quantum field theories and massless scattering in $(1+1)$-dimensions}
  [\arXivlink{hep-th/9310058}].


\bibitem{DiegoBogdanAle}
  D.~Bombardelli, B.~Stefa\'nski, jr. and A.~Torrielli,
  \emph{The low-energy limit of $AdS_3/CFT_2$ and its TBA,} 
  JHEP {\bf 1810} (2018) 177
  [\arXivlink{1807.07775}].
 
    \bibitem{Lloyd:2013wza}
T.~Lloyd and B.~Stefa\'nski,
{\em {$AdS_3/CFT_2$, finite-gap equations and massless modes},} 
{JHEP {\bf 1404} (2014) 179}
  [\arXivlink{1312.3268}].
A.~Pittelli, A.~Torrielli and M.~Wolf,
  {\em Secret symmetries of type IIB superstring theory on $AdS_3 \times S^3 \times M^4$,} 
  J.\ Phys.\ A {\bf 47} (2014)   455402
  [\arXivlink{1406.2840}].
  A.~Prinsloo,
{\it D1 and D5-brane giant gravitons on $AdS_3 \times S^3 \times S^3 \times S^1$,}
JHEP \textbf{12} (2014) 094
[\arXivlink{1406.6134}].
  M.~C.~Abbott and I.~Aniceto,
  {\em Macroscopic (and Microscopic) Massless Modes,} 
  Nucl.\ Phys.\ B {\bf 894} (2015) 75
  [\arXivlink{1412.6380}].
    V.~Regelskis,
  {\em Yangian of $AdS_3/CFT_2$ and its deformation,} 
  J.\ Geom.\ Phys.\  {\bf 106} (2016) 213
  [\arXivlink{1503.03799}].

 
  


  L.~Wulff,
  {\em On integrability of strings on symmetric spaces,} 
  JHEP {\bf 1509} (2015) 115
  [\arXivlink{1505.03525}].
M. C. Abbott, J. Murugan, S. Penati, A. Pittelli, D. Sorokin, P. Sundin, J. Tarrant, M. Wolf and L. Wulff,
  {\em T-duality of Green-Schwarz superstrings on 
  $AdS_d \times S^d \times M^{10-2d}$,} 
  JHEP {\bf 1512} (2015) 104
  [\arXivlink{1509.07678}].
  
  M.~C.~Abbott and I.~Aniceto,
  \emph{Massless L\"uscher terms and the limitations of the AdS$_3$ asymptotic Bethe ansatz,}
  Phys.\ Rev.\ D {\bf 93} (2016)   106006
  [\arXivlink{1512.08761}].


 
   
  
M.~R.~Gaberdiel, R.~Gopakumar and C.~Hull,
  \emph{Stringy AdS$_{3}$ from the worldsheet,} 
  JHEP {\bf 1707} (2017) 090
    [\arXivlink{1704.08665}].


M.~Baggio and A.~Sfondrini,
  \emph{Strings on NS-NS backgrounds as integrable deformations,}
  Phys. \ Rev. \ D {\bf 98} (2018) 021902
   [\arXivlink{1804.01998}].
{}
  A.~Dei and A.~Sfondrini,
  \emph{Integrable spin chain for stringy Wess-Zumino-Witten models,} 
  JHEP {\bf 1807} (2018) 109
  [\arXivlink{1806.00422}].
  B.~Hoare, N.~Levine and A.~A.~Tseytlin,
  {\em On the massless tree-level $S$-matrix in 2d sigma models,} 
  J.\ Phys.\ A {\bf 52} (2019)   144005
  [\arXivlink{1812.02549}].


A.~Dei, L.~Eberhardt and M.~R.~Gaberdiel,
  {\em Three-point functions in AdS$_3$/CFT$_2$ holography,}
  JHEP {\bf 12} (2019) 012
  [\arXivlink{1907.13144}].
O.~Ohlsson Sax and B.~Stefa\'nski, jr.,
\emph{On the singularities of the RR $AdS_3 \times S^3 \times T^4$ S matrix,}
J. Phys. A \textbf{53} (2020)  155402
[\arXivlink{1912.04320}].
M.~C.~Abbott and I.~Aniceto,
  \emph{Integrable Field Theories with an Interacting Massless Sector,}
  Phys.Rev.D 103 (2021) 8, 086017
 [\arXivlink{2002.12060}].

A.~Dei, M.~R.~Gaberdiel, R.~Gopakumar and B.~Knighton,
\emph{Free field world-sheet correlators for $AdS_3$,}
JHEP \textbf{02} (2021) 081
[\arXivlink{2009.11306}].
  
  
  





  







 
  
\bibitem{Ben}
  B.~Hoare and A.~A.~Tseytlin,
  \emph{Towards the quantum S-matrix of the Pohlmeyer reduced version of $AdS_5 \times S^5$ superstring theory,}
  Nucl.\ Phys.\ B {\bf 851} (2011) 161
  [\arXivlink{1104.2423}].
{}
  B.~Hoare,
  \emph{Towards a two-parameter q-deformation of AdS$_3 \times S^3 \times M^4$ superstrings,}
  Nucl.\ Phys.\ B {\bf 891} (2015) 259
  [\arXivlink{1411.1266}].
{}
  G.~Giribet, C.~Hull, M.~Kleban, M.~Porrati and E.~Rabinovici,
  \emph{Superstrings on AdS$_{3}$ at $k =1$,}
  JHEP {\bf 1808} (2018) 204
  [\arXivlink{1803.04420}].
{}
  M.~R.~Gaberdiel and R.~Gopakumar,
  \emph{Tensionless string spectra on AdS$_{3}$,}
  JHEP {\bf 1805} (2018) 085
  [\arXivlink{1803.04423}].
{}
  L.~Eberhardt, M.~R.~Gaberdiel and R.~Gopakumar,
  \emph{The Worldsheet Dual of the Symmetric Product CFT,}
  JHEP {\bf 1904} (2019) 103
  [\arXivlink{1812.01007}].

  
J.~M.~Nieto Garc\'ia and A.~Torrielli,
{\em Norms and scalar products for $AdS_3$,}
J. \ Phys. \ A {\bf 53} (2020) 145401
[\arXivlink{1911.06590}].


 \bibitem{QSC}
A.~Cavagli\`a, N.~Gromov, B.~Stefa\'nski, jr., and A.~Torrielli,
\emph{Quantum Spectral Curve for $AdS_3/CFT_2$: a proposal,}
JHEP \textbf{12} (2021) 048
[\arXivlink{2109.05500}].
S.~Ekhammar and D.~Volin,
\emph{Monodromy Bootstrap for $\alg{su}(2|2)$ Quantum Spectral Curves: From Hubbard model to $AdS_3/CFT_2$,}
JHEP \textbf{03} (2022) 192
[\arXivlink{2109.06164}].

\bibitem{Cavaglia:2022xld}
A.~Cavagli\`a, S.~Ekhammar, N.~Gromov and P.~Ryan,
{\it Exploring the Quantum Spectral Curve for $AdS_3/CFT_2$,}
JHEP {\bf 12} (2023) 089
[\arXivlink{2211.07810}].
S.~Ekhammar,
{\it An Exploration of Q-Systems: From Spin Chains to Low-Dimensional AdS/CFT,} Uppsala thesis.
  
\bibitem{Ekhammar:2024kzp}
S.~Ekhammar, N.~Gromov and B.~Stefa\'nski, jr.,
{\it Demystifying the Massless Sector in $AdS_3$ Quantum Spectral Curve,}
[\arXivlink{2412.11915}].

\bibitem{AleSSergey}
S.~Frolov and A.~Sfondrini,
\emph{Massless S matrices for $AdS_3/CFT_2$,} JHEP \textbf{04} (2022) 067
[\arXivlink{2112.08895}].
S.~Frolov and A.~Sfondrini,
\emph{New Dressing Factors for $AdS_3/CFT_2$,} JHEP \textbf{04} (2022) 162
[\arXivlink{2112.08896}].
S.~Frolov and A.~Sfondrini,
\emph{Mirror Thermodynamic Bethe Ansatz for $AdS_3/CFT_2$,} JHEP \textbf{03} (2022) 138
[\arXivlink{2112.08898}].



\bibitem{Seibold:2022mgg}
F.~K.~Seibold and A.~Sfondrini,
\emph{Transfer matrices for $AdS_3/CFT_2$,} JHEP \textbf{05} (2022), 089
[\arXivlink{2202.11058}].
A.~Brollo, D.~le Plat, A.~Sfondrini and R.~Suzuki,
{\it Tensionless Limit of Pure\textendash{}Ramond-Ramond Strings and $AdS_3/CFT_2$,}
Phys. Rev. Lett. \textbf{131} (2023) 161604
[\arXivlink{2303.02120}].
S.~Frolov, A.~Pribytok and A.~Sfondrini,
{\it Ground state energy of twisted $AdS_3 \times S^3 \times T^4$ superstring and the TBA,}
JHEP \textbf{09} (2023) 027
[\arXivlink{2305.17128}].
A.~Brollo, D.~le Plat, A.~Sfondrini and R.~Suzuki,
{\it More on the tensionless limit of pure-Ramond-Ramond $AdS_3/CFT_2$,}
JHEP {\bf 12} (2023) 160
[\arXivlink{2308.11576}].

\bibitem{Fabri:2025rok}
M.~R.~Gaberdiel and V.~Sriprachyakul,
{\it Tensionless strings on $AdS_3 \times S^3 \times S^3 \times S^1$,}
[\arXivlink{2411.16848}].
M.~R.~Gaberdiel, F.~Lichtner and B.~Nairz,
{\it Anomalous dimensions in the symmetric orbifold,}
[\arXivlink{2411.17612}].
M.~R.~Gaberdiel, D.~Kempel and B.~Nairz,
{\it $AdS_3\times S^3$ magnons in the symmetric orbifold,}
[\arXivlink{2412.02741}].
M.~Fabri, A.~Sfondrini and T.~Skrzypek,
{\it Perturbed symmetric-product orbifold: first-order mixing and puzzles for integrability,}
[\arXivlink{2504.13091}].

\bibitem{recent}
S.~Frolov, D.~Polvara and A.~Sfondrini,
{\em On mixed-flux worldsheet scattering in AdS$_{3}$/CFT$_{2}$,}
JHEP \textbf{11} (2023) 055
[\arXivlink{2306.17553}].
N.~Baglioni, D.~Polvara, A.~Pone and A.~Sfondrini,
{\it On the worldsheet S matrix of the $AdS_3/CFT_2$ mixed-flux mirror model,} JHEP \textbf{05} (2024), 237
[\arXivlink{2308.15927}].
\bibitem{riab}
O. Ohlsson Sax, D. Riabchenko and B.~Stefa{\'n}ski, jr.,
{\it Worldsheet kinematics, dressing factors and odd
crossing in mixed-flux $AdS_3$ backgrounds}
[\arXivlink{2312.09288}].
\bibitem{gaber2}
M.~R.~Gaberdiel, R.~Gopakumar and B.~Nairz,
{\it Beyond the Tensionless Limit: Integrability in the Symmetric Orbifold,} JHEP \textbf{06} (2024), 030
[\arXivlink{2312.13288}].
S.~Frolov and A.~Sfondrini,
{\it Comments on Integrability in the Symmetric Orbifold}
[\arXivlink{2312.14114}].
B.~Hoare, A.~L.~Retore and F.~K.~Seibold,
{\it Elliptic deformations of the $AdS_3 \times S^3 \times T^4$ string,} JHEP \textbf{04} (2024), 042
[\arXivlink{2312.14031}].
R. Borsato, S. Driezen, B. Hoare, A. L. Retore, F. K. Seibold,
{\it Inequivalent light-cone gauge-fixings of strings on $AdS_n \times S^n$ backgrounds,} Phys. Rev. D \textbf{109} (2024) 106023
[\arXivlink{2312.17056}].
N.~Beisert and E.~Im,
{\it Affine Classical Lie Bialgebras
for AdS/CFT Integrability,}
[\arXivlink{2401.10327}]. S.~Frolov, D.~Polvara and A.~Sfondrini,
{\it Dressing Factors for Mixed-Flux $AdS_3\times S^3\times T^4$ Superstrings,}
[\arXivlink{2402.11732}].
A.~Dei, B.~Knighton, K.~Naderi and S.~Sethi,
{\it Tensionless AdS$_{3}$/CFT$_{2}$ and single trace $ T\overline{T} $,}
JHEP \textbf{11} (2024), 145
[\arXivlink{2408.00823}].
\bibitem{frolo}
S.~Frolov, D.~Polvara and A.~Sfondrini,
{\it Massive dressing factors for mixed-flux AdS$_3$/CFT$_2$,}
[\arXivlink{2501.05995}].
\bibitem{sei2}
F.~K.~Seibold and A.~Sfondrini,
{\it Interpolating families of integrable $AdS_3$ backgrounds,}
[\arXivlink{2502.07103}].
X.~Kervyn, D.~Polvara and A.~Sfondrini,
{\it Thermodynamics of integrable $N=2$ theories, squared,}
[\arXivlink{2502.10356}].
S.~Frolov, D.~Polvara and A.~Sfondrini,
{\it Exchange relations and crossing,}
[\arXivlink{2506.04096}].

\bibitem{Bielli:2024xuv}
D.~Bielli, V.~Gautam, V.~Moustakis, A.~Prinsloo and A.~Torrielli,
{\it Boundary scattering in massless AdS$_{3}$,}
JHEP \textbf{07} (2024), 266
[\arXivlink{2403.18594}].



  \bibitem{GhoshalZamo}
S.~Ghoshal and A.~B.~Zamolodchikov,
{\it Boundary S matrix and boundary state in two-dimensional integrable quantum field theory,}
Int. J. Mod. Phys. A \textbf{9} (1994) 3841
[erratum: Int. J. Mod. Phys. A \textbf{9} (1994) 4353]
[\arXivlink{hep-th/9306002}].
  
  
\bibitem{Prinsloo:2015apa}
A.~Prinsloo, V.~Regelskis and A.~Torrielli,
{\it Integrable open spin-chains in AdS3/CFT2 correspondences,}
Phys. Rev. D \textbf{92} (2015)  106006
[\arXivlink{1505.06767}].










\bibitem{gamma1}
  A.~Fontanella and A.~Torrielli,
  \emph{Geometry of Massless Scattering in Integrable Superstring,}
  JHEP {\bf 1906} (2019) 116
  [\arXivlink{1903.10759}].
  
\bibitem{gamma2}
  A.~Fontanella, O.~Ohlsson Sax, B.~Stefa\'nski and A.~Torrielli,
  \emph{The effectiveness of relativistic invariance in AdS$_{3}$,} 
  JHEP {\bf 1907} (2019) 105
  [\arXivlink{1905.00757}].

\bibitem{Bielli:2024bve}
D.~Bielli, V.~Moustakis and A.~Torrielli,
{\it Boundary Bethe ansatz in massless AdS$_{3}$,}
J. Phys. A \textbf{58} (2025) no.4, 045402
[\arXivlink{2408.15674}].

\bibitem{Bedu}
  G.~Bed\"urftig and H.~Frahm,
  \emph{Open $t-J$ chain with boundary impurities,} 
  J. Phys. A {\bf 32.25} (1999) 4584.

\bibitem{Bracken:1997zww}
A.~J.~Bracken, X.~Y.~Ge, Y.~Z.~Zhang and H.~Q.~Zhou,
{\it Integrable open-boundary conditions for the $q$-deformed supersymmetric $U$ model of strongly correlated electrons,}
Nucl. Phys. B \textbf{516} (1998) 588
[\arXivlink{cond-mat/9710141}].

\bibitem{AppelVlaar}
A.~Appel and B.~Vlaar,
{\it Boundary transfer matrices arising from quantum symmetric pairs,} \arXivlink{2410.21654}.

\bibitem{Angela}
A.~Foerster,
{\it Quantum group invariant supersymmetric $t-J$ model with periodic boundary conditions,}
J. Phys. A \textbf{29.23} (1996) 7625.

\bibitem{Foerster:1993fp}
A.~Foerster and M.~Karowski,
{\it The Supersymmetric $t - J$ model with quantum group invariance,}
Nucl. Phys. B \textbf{408} (1993) 512.

\bibitem{Slav}
H.~Frahm and N.~A.~Slavnov,
{\it New solutions to the reflection equation and the projecting method,}
J. Phys. A \textbf{32.9} (1999) 1547.

\bibitem{Ge}
X.~Y.~Ge,
{\it Integrable open-boundary conditions for the $q$-deformed extended Hubbard model,}
Mod. Phys. Lett. \textbf{B13} (1999) 499.

\bibitem{Gonzalez-Ruiz:1994loj}
A.~Gonz\'alez-Ruiz,
{\it Integrable open boundary conditions for the supersymmetric t-J model. The Quantum group invariant case,}
Nucl. Phys. B \textbf{424} (1994) 468
[\arXivlink{hep-th/9401118}].

\bibitem{Links}
J.~Links and A.~Foerster,
{\it On the construction of integrable closed chains with quantum supersymmetry,}
J. Phys. A \textbf{30.7} (1997) 2483.

\bibitem{Grab}
A.~M.~Grabinski and H.~Frahm,
{\it Non-diagonal boundary conditions for super spin chains,}
J. Phys. A \textbf{43.4} (2010) 045207.

\bibitem{Gould}
J.~R.~Links and M.~D.~Gould,
{\it Integrable systems on open chains with quantum supersymmetry,}
Int. J. Mod. Phys.  \textbf{B10.25} (1996) 3461.

\bibitem{Martins:1999jbx}
M.~J.~Martins and X.~W.~Guan,
{\it Integrable supersymmetric correlated electron chain with open boundaries,}
Nucl. Phys. B \textbf{562} (1999) 433
[\arXivlink{solv-int/9907006}].

\bibitem{Shiro}
M.~Shiroishi and M.~Wadati,
{\it Integrable boundary conditions for the one-dimensional Hubbard model,}
Journal of the Physical Society of Japan \textbf{66.8} (1997) 2288.

\bibitem{Yao}
Y.~Zhang,
{\it On the graded quantum Yang-Baxter and reflection equations,}
Communications in theoretical physics \textbf{29.3} (1998) 377.

\bibitem{Zhou}
Y.~Zhang and H.~Zhou,
{\it Quantum integrability and exact solution of the supersymmetric $U$ model with boundary terms,}
Phys. Rev. D \textbf{58.1} (1998) 51.

\bibitem{Zhou:1998joy}
H.~Q.~Zhou, X.~Y.~Ge, J.~Links and M.~D.~Gould,
{\it Graded reflection equation algebras and integrable Kondo impurities in the one-dimensional $t - J$ model,}
Nucl. Phys. B \textbf{546} (1999) 779
[\arXivlink{cond-mat/9809056}].
T.~Gombor, C.~Kristjansen, V.~Moustakis and X.~Qian,
{\it On exact overlaps of integrable matrix product states: inhomogeneities, twists and dressing formulas,}
JHEP \textbf{02} (2025), 100
[\arXivlink{2410.23117}].

\bibitem{Sklyanin:1988yz}
E.~K.~Sklyanin,
{\it Boundary Conditions for Integrable Quantum Systems,}
J. Phys. A \textbf{21} (1988), 2375-2389
 
\bibitem{Majumder:2021zkr}
S.~Majumder, O.~O.~Sax, B.~Stefa\'nski and A.~Torrielli,
{\it Protected states in $AdS_3$ backgrounds from integrability,}
J. Phys. A \textbf{54} (2021) 415401
[\arXivlink{2103.16972}].




\bibitem{Torrielli:2021hnd}
A.~Torrielli,
{\it A study of integrable form factors in massless relativistic $AdS_3$,}
J. Phys. A \textbf{55} (2022) no.17, 175401
[\arXivlink{2106.06874}].


\bibitem{Fedor}
F.~Levkovich-Maslyuk,
{\it The Bethe ansatz,}
J. Phys. A \textbf{49} (2016), 323004
[\arXivlink{1606.02950}].



\bibitem{case}
Z.~Bajnok,
{\it Equivalences between spin models induced by defects,}
J. Stat. Mech. \textbf{0606} (2006), P06010
[\arXivlink{hep-th/0601107}].
R.~I.~Nepomechie and A.~L.~Retore,
{\it Surveying the quantum group symmetries of integrable open spin chains,}
Nucl. Phys. B \textbf{930} (2018), 91-134
[\arXivlink{1802.04864}].
M.~Vanicat,
{\it Integrable Floquet dynamics, generalized exclusion processes and ``fused" matrix ansatz,}
Nucl. Phys. B \textbf{929} (2018), 298-329
[\arXivlink{1711.08884}].
C.~Paletta and T.~Prosen,
{\it Integrability of open boundary driven quantum circuits,}
[\arXivlink{2406.12695}]. T.~Gombor and Z.~Bajnok,
{\it Dual overlaps and finite coupling 't Hooft loops,}
[\arXivlink{2408.14901}].

\bibitem{Doikou:2004qd}
A.~Doikou,
{\it From affine Hecke algebras to boundary symmetries,}
Nucl. Phys. B \textbf{725} (2005), 493-530
[\arXivlink{math-ph/0409060}].

\bibitem{Nepomechie:2008ab}
R.~I.~Nepomechie and E.~Ragoucy,
{\it Analytical Bethe ansatz for the open AdS/CFT $SU(1|1)$ spin chain,}
JHEP \textbf{12} (2008), 025
[\arXivlink{0810.5015}].



\bibitem{Lloyd:2014bsa}
T.~Lloyd, O.~Ohlsson Sax, A.~Sfondrini and B.~Stefa\'nski, jr.,
{\it The complete worldsheet S matrix of superstrings on AdS$_3 \times$ S$^3 \times$ T$^4$ with mixed three-form flux,}
Nucl. Phys. B \textbf{891} (2015), 570-612
doi:10.1016/j.nuclphysb.2014.12.019
[\arXivlink{1410.0866}].


\bibitem{Torrielli:2023uam}
A.~Torrielli,
{\it A study of form factors in relativistic mixed-flux AdS$_{3}$,}
JHEP \textbf{03} (2024), 082
[\arXivlink{2312.17557}].










\bibitem{Cava}
A.~Cavagli\`a, N.~Gromov, J.~Julius, M.~Preti and N.~S.~Sokolova,
{\it Probing Line Defect CFT with Mixed-Correlator Bootstrability,}
[\arXivlink{2412.07624}].
F.~Coronado, S.~Komatsu and K.~Zarembo,
{\it Coulomb Branch and Integrability,}
[\arXivlink{2506.07222}].
R.~Demjaha and K.~Zarembo,
{\it String Integrability on the Coulomb Branch,}
[\arXivlink{2506.17955}].
\bibitem{DeLeeuw:2020ahx}
M.~De Leeuw, C.~Paletta, A.~Pribytok, A.~L.~Retore and A.~Torrielli,
{\it Free Fermions, vertex Hamiltonians, and lower-dimensional AdS/CFT,}
JHEP \textbf{02} (2021), 191
[\arXivlink{2011.08217}].

\end{thebibliography}
\end{document}